\documentclass[11pt]{article}
\usepackage{amsmath, amssymb, amsfonts,euscript,mathrsfs}
\usepackage{bm,bbm,latexsym,amsthm}
\usepackage{psfig, epsfig, verbatim}
\usepackage{natbib}
\usepackage{times}
\usepackage{setspace}
\usepackage{multirow}
\usepackage{color}
\usepackage[normalem]{ulem}
\usepackage{graphicx}
\usepackage{float}
\usepackage{subfig}
\usepackage{pslatex}
\usepackage{rotating}
\usepackage{geometry}
\usepackage{soul}
\RequirePackage{amsthm,amsmath}
\RequirePackage{natbib}
\RequirePackage[colorlinks,citecolor=blue,urlcolor=blue]{hyperref}
\usepackage{graphicx}
\usepackage{float}
\usepackage{subfig}
\usepackage{multirow}
\usepackage{ulem}
\usepackage{soul,xcolor,colortbl}

\newcommand{\PMTwo}{PM$_{2.5}$}
\newcommand{\SOTwo}{SO$_2$}
\newcommand{\NOx}{NO$_x$}
\newcommand{\COTwo}{CO$_2$}
\newcommand\independent{\protect\mathpalette{\protect\independenT}{\perp}}
\def\independenT#1#2{\mathrel{\rlap{$#1#2$}\mkern2mu{#1#2}}}

\makeatletter
\newcommand*{\rom}[1]{\expandafter\@slowromancap\romannumeral #1@}
\makeatother
\newcommand*{\LargerCdot}{\raisebox{-0.25ex}{\scalebox{1.2}{$\cdot$}}}

\title{\Large Bayesian Methods for Multiple Mediators: Relating Principal Stratification and Causal Mediation in the Analysis of Power Plant Emission Controls}

\author{ Chanmin Kim$^{1}$, 
Michael J. Daniels$^{2}$, Joseph W. Hogan$^{3}$, Christine Choirat$^{4}$ and Corwin M. Zigler$^5$\\
$^{1}$Department of Biostatistics, Boston University School of Public Health\\
$^{2}$Department of Statistics, University of Florida\\
$^{3}$Department of Biostatistics, Brown University\\
$^{4}$Swiss Data Science Center\\
$^{5}$Department of Statistics and Data Sciences, University of Texas at Austin
}

\begin{document}

\date{}


\newcommand{\nc}{\newcommand}
\nc{\nn}{\nonumber}
\nc{\fns}{\footnotesize}

\nc{\revisionline}{\vspace{.1in} \today \vspace{.1in} \hrule\hrule\hrule\vspace{.1in}}
\nc{\newpp}{\vspace{.1in} \noindent}

\nc{\slideline}{\smallskip \hrule\hrule \smallskip}

\nc{\wh}{\widehat}

\def\boxit#1{\vbox{\hrule\hbox{\vrule\kern6pt
          \vbox{\kern6pt#1\kern6pt}\kern6pt\vrule}\hrule}}

\newcommand{\se}{\text{se} }
\newcommand{\spec}{\text{sp} }
\newcommand{\fpr}{\text{{\sc fpr}}}
\newcommand{\fnr}{\text{{\sc fnr}}}

\nc{\Ef}{ {\rm E}_{\infty} }
\nc{\Ex}{ {\rm E} }
\nc{\Ec}{ {\rm E}_1 }
\nc{\Pf}{ {\rm P}_{\infty} }
\nc{\Pc}{ {\rm P}_{1} }
\nc{\Prb}{ {\rm P} }
\nc{\sd}{\pm \hat{\sigma} }

\nc{\indep}{{\, \perp \! \! \! \perp  \,} }
\nc{\tsps}{^{ {\rm T} } }

\nc{\pu}{\pi_{\rm U}}
\nc{\pbi}{\pi_{\rm B}}
\nc{\pnb}{\pi_{\rm NB}}
\nc{\prp}{\propto}
\nc{\pr}{ {\rm pr} }

\newcommand{\Poi}{\text{Poi}}
\newcommand{\Bin}{\text{Bin}}
\newcommand{\Ber}{\text{Ber}}
\newcommand{\Mult}{\text{Mult}}

\nc{\al}{\alpha}
\nc{\dl}{\delta}
\nc{\la}{\lambda}
\nc{\vep}{\varepsilon}
\nc{\eps}{\epsilon}

\nc{\snf}{\sum_{n=1}^{\infty}}
\nc{\skf}{\sum_{k=1}^{\infty}}
\nc{\sner}{\sum_{n=1}^{86}}
\nc{\sjn}{\sum_{j=1}^{n}}
\nc{\skn}{\sum_{k=1}^{n}}
\nc{\sumim}{\sum_{i=1}^m}
\nc{\sumjn}{\sum_{j=1}^n}
\nc{\sumlL}{\sum_{l=1}^{L}}
\nc{\sumL}{\sum_{l=1}^{L}}
\nc{\sumkK}{\sum_{k=1}^{K_i}}
\nc{\sumrR}{\sum_{r=1}^R}

\nc{\hivp}{\sum_{ {\rm HIV}^+ } }
\nc{\sumiN}{ \sum_{i=1}^N }
\nc{\summM}{ \sum_{m=1}^M }
\nc{\sumjM}{ \sum_{j=1}^M }

\nc{\lsq}{\left[}
\nc{\rsq}{\right]}
\nc{\lbc}{\left \{ }
\nc{\rbc}{\right \} }
\nc{\lp}{\left(}
\nc{\rp}{\right)}

\nc{\imp}{\Rightarrow}
\nc{\lbf}{\lim_{b \rightarrow \infty}}
\nc{\limNinf}{\lim_{N \rightarrow \infty}}
\nc{\limminf}{\lim_{m \rightarrow \infty}}
\nc{\limninf}{\lim_{n \rightarrow \infty}}
\nc{\convd}{\stackrel{D}{\longrightarrow}}
\nc{\convp}{\stackrel{P}{\longrightarrow}}
\nc{\eqd}{\stackrel{{\EuScript D}}{=}}

\nc{\trans}{^{\text T}}
\nc{\ol}{\overline}
\nc{\logit}{\text{logit}\,}

\nc{\rl}{ {\rm {\bf R} } }
\nc{\zah}{ {\rm {\bf Z} } }

\nc{\lkn}{\Lambda^n_k}
\nc{\stp}{ {\cal C}_b }
\nc{\istp}{ {\cal I}_A }
\nc{\snb}{S_{N_b}}
\nc{\stb}{S_{T_b}}
\nc{\ixlog}{I_{ \{ 0 \leq x \leq \log \al \} } }
\nc{\iulog}{I_{ \{ 0 \leq u  \leq \log \al \} } }
\nc{\rgn}{ \Upsilon_n }
\nc{\var}{{\rm var}}
\nc{\cov}{{\rm cov}}
\nc{\corr}{{\rm corr}}
\nc{\dpl}{\partial}
\nc{\half}{ {\textstyle \frac{1}{2}} }
\nc{\tr}{{\rm trace}}
\nc{\real}{\mathbb{R}}
\nc{\bbC}{\mathbb{C}}
\nc{\bbR}{\mathbb{R}}
\nc{\bbone}{\mathbb{1}}
\nc{\bbP}{\mathbb{P}}

\def\boxit#1{\vbox{\hrule\hbox{\vrule\kern6pt\vbox{\kern6pt#1\kern6pt}\kern6pt\vrule}\hrule}}

\nc{\calb}{ {\cal B} }
\nc{\calc}{ {\cal C} }
\nc{\bcalc}{ \mbox{\boldmath{${\cal C}$}}}
\nc{\cald}{ {\cal D} }
\nc{\cale}{ {\cal E} }
\nc{\cali}{ {\cal I} }
\nc{\call}{ {\cal L} }
\nc{\calm}{ {\cal M} }
\nc{\caln}{ {\cal N} }
\nc{\cals}{ {\cal S} }
\nc{\calo}{ {\cal O} }
\nc{\bcalo}{ \mbox{\boldmath{${\cal O}$}}}
\nc{\calt}{ {\cal T} }
\nc{\calv}{ {\cal V} }
\nc{\bcalu}{ \mbox{\boldmath{${\cal U}$}}}
\nc{\calu}{ {\cal U} }
\nc{\calw}{ {\cal W} }
\nc{\calx}{ {\cal X} }

\nc{\sca}{ {\EuScript A} }
\nc{\scb}{ {\EuScript B} }
\nc{\scc}{ {\EuScript C} }
\nc{\scd}{ {\EuScript D} }
\nc{\sce}{ {\EuScript E} }
\nc{\scf}{ {\EuScript F} }
\nc{\scF}{ {\EuScript f} }
\nc{\scg}{ {\EuScript G} }
\nc{\sch}{ {\EuScript H} }
\nc{\sci}{ {\EuScript I} }
\nc{\scj}{ {\EuScript J} }
\nc{\sck}{ {\EuScript K} }
\nc{\scl}{ {\EuScript L} }
\nc{\sclic}{ \scl_i^{\rm c} }
\nc{\scm}{ {\EuScript M} }
\nc{\scn}{ {\EuScript N} }
\nc{\sco}{ {\EuScript O} }
\nc{\scp}{ {\EuScript P} }
\nc{\scq}{ {\EuScript Q} }
\nc{\scr}{ {\EuScript R} }
\nc{\scs}{ {\EuScript S} }
\nc{\sct}{ {\EuScript T} }
\nc{\scu}{ {\EuScript U} }
\nc{\scv}{ {\EuScript V} }
\nc{\scw}{ {\EuScript W} }
\nc{\scx}{ {\EuScript X} }
\nc{\scy}{ {\EuScript Y} }
\nc{\scz}{ {\EuScript Z} }
\nc{\scxo}{ {\EuScript X}_{\rm obs} }
\nc{\Xobs}{ \pmb{\scx}_{\rm obs} }
\nc{\Xcom}{ \pmb{\scx} }
\nc{\Xmis}{ \pmb{\scx}_{\rm mis} }

\nc{\bsci}{ \mbox{\boldmath{$\sci$}}}  
\nc{\bscj}{ \mbox{\boldmath{$\scj$}}}  

\nc{\sumlic}{\sum_{l \in sclic}}

\nc{\scyo}{ {\EuScript Y}_{\rm obs} }

%
%
%

\nc{\bga}{\begin{array}{c}}
\nc{\ena}{\end{array}}

\nc{\mhat}{ {\hat{p}}_M }
\nc{\fhat}{ {\hat{p}}_F }
\nc{\ph} { \hat{p} }

\nc{\ta}{ {\tilde{a}} }
\nc{\tc}{ {\tilde{c}} }

\nc{\bi}{\mbox{\boldmath{$i$}}} 
\nc{\bal}{\mbox{\boldmath{$\alpha$}}} 
\nc{\balpha}{\mbox{\boldmath{$\alpha$}}} 
\nc{\bone}{\mbox{\boldmath{$1$}}} 
\nc{\bbet}{\mbox{\boldmath{$\beta$}}} 
\nc{\bbeta}{\mbox{\boldmath{$\beta$}}} 
\nc{\bDel}{\mbox{\boldmath{$\Delta$}}} 
\nc{\bDelta}{\mbox{\boldmath{$\Delta$}}} 
\nc{\bdel}{\mbox{\boldmath{$\delta$}}}
\nc{\bdelta}{\mbox{\boldmath{$\delta$}}}
\nc{\bet}{\mbox{\boldmath{$\eta$}}}
\nc{\beps}{\mbox{\boldmath{$\epsilon$}}}
\nc{\bvep}{\mbox{\boldmath{$\vep$}}} 
\nc{\bgam}{\mbox{\boldmath{$\gamma$}}} 
\nc{\bgamma}{\mbox{\boldmath{$\gamma$}}} 
\nc{\bGamma}{\mbox{\boldmath{$\Gamma$}}} 
\nc{\boldeta}{\mbox{\boldmath{$\eta$}}} 
\nc{\bLam}{\mbox{\boldmath{$\Lambda$}}} 
\nc{\bLambda}{\mbox{\boldmath{$\Lambda$}}} 
\nc{\blambda}{\mbox{\boldmath{$\lambda$}}} 
\nc{\bmu}{ \mbox{\boldmath{$\mu$}}} 
\nc{\boldnu}{ \mbox{\boldmath{$\nu$}}} 
\nc{\bOm}{ \mbox{\boldmath{$\Omega$}}} 
\nc{\bOmega}{ \mbox{\boldmath{$\Omega$}}} 
\nc{\bom}{ \mbox{\boldmath{$\omega$}}} 
\nc{\bomega}{ \mbox{\boldmath{$\omega$}}} 
\nc{\bpi}{ \mbox{\boldmath{$\pi$}}} 
\nc{\bPi}{ \mbox{\boldmath{$\Pi$}}} 
\nc{\bpsi}{ \mbox{\boldmath{$\psi$}}} 
\nc{\bPsi}{ \mbox{\boldmath{$\Psi$}}} 
\nc{\bphi}{ \mbox{\boldmath{$\phi$}}} 
\nc{\bPhi}{ \mbox{\boldmath{$\Phi$}}} 
\nc{\bxi}{ \mbox{\boldmath{$\xi$}}} 
\nc{\bXi}{ \mbox{\boldmath{$\Xi$}}} 
\nc{\bSig}{\mbox{\boldmath{$\Sigma$}}}
\nc{\bSigma}{\mbox{\boldmath{$\Sigma$}}}
\nc{\bsig}{\mbox{\boldmath{$\sigma$}}}
\nc{\bsigma}{\mbox{\boldmath{$\sigma$}}}
\nc{\btau}{\mbox{\boldmath{$\tau$}}}
\nc{\bThe}{\mbox{\boldmath{$\Theta$}}}
\nc{\bTheta}{\mbox{\boldmath{$\Theta$}}}
\nc{\bthe}{\mbox{\boldmath{$\theta$}}}
\nc{\btheta}{\mbox{\boldmath{$\theta$}}}
\nc{\bzeta}{\mbox{\boldmath{$\zeta$}}}
\nc{\bIm}{\mbox{\boldmath{$\Im$}}}


\nc{\ba}{ { \bf a }}
\nc{\bA}{ { \bf A }}
\nc{\bB}{ { \bf B }}
\nc{\bb}{ { \bf b }}
\nc{\bc}{ { \bf c }}
\nc{\bC}{ { \bf C }}
\nc{\bD}{ { \bf D }}
\nc{\bd}{ { \bf d }}
\nc{\be}{ { \bf e }}
\nc{\bF}{ { \bf F }}
\nc{\boldf}{ { \bf f }}
\nc{\bG}{ { \bf G }}
\nc{\bh}{ { \bf h }}  
\nc{\bH}{ { \bf H }}  
\nc{\bI}{ { \bf I }}  
\nc{\bJ}{ { \bf J }}  
\nc{\bk}{ { \bf k }}  
\nc{\bK}{ { \bf K }}  
\nc{\bL}{ { \bf L }}
\nc{\bM}{ { \bf M }}
\nc{\bn}{ { \bf n }}
\nc{\bO}{ { \bf O }}
\nc{\bP}{ { \bf P }}
\nc{\bp}{ {\bf p }}
\nc{\br}{ { \bf r }}
\nc{\bR}{ { \bf R }}
\nc{\bolds}{ { \bf s }}  
\nc{\bS}{ { \bf S }} 
\nc{\bT}{ { \bf T }} 
\nc{\bt}{ { \bf t }} 
\nc{\bu}{ { \bf u }} 
\nc{\bU}{ { \bf U }}  
\nc{\bv}{ { \bf v }}
\nc{\bV}{ { \bf V }}  
\nc{\bW}{ { \bf W }}  
\nc{\bw}{ { \bf w }}  
\nc{\bx}{ { \bf x }}
\nc{\bX}{ { \bf X }}
\nc{\by}{ { \bf y }} 
\nc{\bY}{ { \bf Y }}  
\nc{\bz}{ { \bf z }} 
\nc{\bZ}{ { \bf Z }}  

\nc{\YR}{[\bY,R]}
\nc{\YgivenR}{[\bY \mid R]}
\nc{\RgivenY}{[R \mid \bY]}
\nc{\Y}{[\bY]}
\nc{\R}{[R]}

\nc{\dio}{d_i^o}
\nc{\timi}{t_{i,m_i}}
\nc{\betahat}{\hat{\bbet}}
\nc{\mui}{\bmu_{\rm I}}
\nc{\mue}{\bmu^{\rm E}}
\nc{\mup}{\bmu^{\rm P}}
\nc{\muihat}{\hat{\bmu}_{\rm I}}
\nc{\muehat}{\hat{\bmu}^{\rm E}}
\nc{\muphat}{\hat{\bmu}^{\rm P}}
\nc{\delhat}{\hat{\bdel}}
\nc{\muhat}{\hat{\bmu}}

\nc{\iid}{\stackrel{\rm iid}{\sim}}
\nc{\law}{\stackrel{\scl}{=}}

\nc{\phiij}{ \phi_{ij}( \Delta_0) }
\nc{\phiiprmj}{ \phi_{i'j}( \Delta_0) }
\nc{\phiijprm}{ \phi_{ij'}( \Delta_0) }
\nc{\phixy}{ \phi( X_i(S_{ik}), Y_j(T_{jl}) ) }
\nc{\phixydo}{ \phi( X_i(S_{ik}), Y_j(T_{jl})-\Delta_0 ) }
\nc{\phixyd}{ \phi( X_i(S_{ik}), Y_j(T_{jl})-\Delta) }
\nc{\phixydstar}{ \phi^*( X_i(S_{ik}), Y_j(T_{jl})-\Delta) }
\nc{\phixystdttil}{ \tilde{\phi}( X_i(s), Y_j(t)-\Delta, \theta) }
\nc{\phixydttil}{ \tilde{\phi}( X_i(S_{ik}), Y_j(T_{jl})-\Delta, \theta) }
\nc{\Nmn}{{\sqrt{N} \over mn}}

\nc{\Xis}{X_i(s)}
\nc{\Yjt}{Y_j(t)}


\nc{\bthehat}{\hat{\bthe}}

\nc{\Ritil}{\tilde{R}_i}

\nc{\Ybar}{\overline{Y}}
\nc{\Rbar}{\overline{R}}
\nc{\Nbar}{\overline{N}}
\nc{\intzeroinf}{\int_0^\infty}

\nc{\Fhat}{\hat{F}}
\nc{\Ghat}{\hat{G}}

\nc{\FhatS}{\hat{F}(S_{ik})}
\nc{\GhatT}{\hat{G}(T_{jl})}

\nc{\Fhatik}{\hat{F}_{ik}}
\nc{\Ghatjl}{\hat{G}_{jl}}
\nc{\Fik}{F_{ik}}
\nc{\Gjl}{G_{jl}}
\nc{\phiijkl}{\phi_{ik,jl}(\Delta)}
\nc{\phiijkltil}{\tilde{\phi}_{ik,jl}(\Delta_0,\theta_0)}
\nc{\ord}{N^{-3/2}}           
\nc{\sumijkl}{\sum_{ijkl}}

\nc{\Citil}{\tilde{C}_i}
\nc{\Crtil}{\tilde{C}_r}
\nc{\Djtil}{\tilde{D}_j}
\nc{\Ditil}{\tilde{D}_i}

\nc{\Cithe}{\tilde{C}^{\theta}_i}
\nc{\Djthe}{\tilde{D}^{\theta}_j}

\nc{\Sikthe}{S_{ik}^{\theta}}
\nc{\Tjlthe}{T_{jl}^{\theta}}

\nc{\Zi}{ \bZ_{-i}}
\nc{\zic}{ \lbc z(\bs_j) \: : \: i \neq j \rbc }
\nc{\zkap}{ \bz_{\kappa} }
\nc{\sumi}{ \sum_i }
\nc{\sumj}{ \sum_j }
\nc{\sumij}{ \sum_{i < j} }
\nc{\sumiandj}{ \sum_{i, j} }
\nc{\zsi}{ z(\bs_i) }
\nc{\Zsi}{ Z(\bs_i) }
\nc{\zsj}{ z(\bs_j) }
\nc{\zsn}{ z(\bs_n) }
\nc{\zsone}{ z(\bs_1) }
\nc{\pZ}{ \Pr \lbc \bZ \rbc }
\nc{\qz}{ Q( \bz ) }
\nc{\qZ}{ Q( \bZ ) }

\nc{\thetaYD}{\theta_{Y\mid D}}
\nc{\thetaD}{\theta_D}
\nc{\psiDY}{\psi_{D\mid Y}}
\nc{\psiY}{\psi_Y}

\nc{\tn}{\Theta^{\nu}}
\nc{\Etn}{E_{\theta^{\nu}}}
\nc{\tnone}{\Theta^{\nu+1}}
\nc{\Lm}{L_{\text{m}}}
\nc{\Lo}{L_{\text{o}}}
\nc{\Ym}{Y_{\text{m}}}
\nc{\Yo}{Y_{\text{o}}}
\nc{\ym}{y_{\text{m}}}
\nc{\yo}{y_{\text{o}}}

\nc{\vijb}{v_{ij} - \bX_{i(j)}  \bbet}
\nc{\vikb}{v_{ik} - \bX_{i(k)}  \bbet}
\nc{\vilb}{v_{il} - \bX_{i(l)}  \bbet}
\nc{\betart}{ \bbet^{(r)}_{t_i} }
\nc{\betarj}{ \bbet^{(r)}_j } 
\nc{\yij}{y_{ij}}
\nc{\Xmisi}{ {\bX_{ i{\rm (mis)} }} }
\nc{\Xobsi}{ {\bX_{ i{\rm (obs)} }} }
\nc{\Zobsi}{ {\bZ_{ i{\rm (obs)} }} }
\nc{\bSigobs}{ \bSig_{  {\rm obs} } }
\nc{\bSigmis}{ \bSig_{  {\rm mis} } }
\nc{\bSigmo}{ \bSig_{  {\rm mis,obs} } }
\nc{\bSigom}{ \bSig_{  {\rm obs,mis} } }
\nc{\Xil}{{\bX}_{il}}
\nc{\Zil}{{\bZ}_{il} }
\nc{\omilr}{\omega_{il}^{(r)}}
\nc{\delio}{\bdel_i^{{\rm obs}} }

\nc{\obs}{{\text{obs}}}
\nc{\mis}{{\text{mis}}}
\nc{\rep}{{\text{rep}}}

\nc{\yio}{ {y_i^{\rm o} }}
\nc{\Yio}{ {Y_i^{\rm o }} }
\nc{\Yim}{ {Y_i^{\rm m} }}
\nc{\yim}{ {y_i^{\rm m} }}
\nc{\Yc}{Y^{\rm c}}
\nc{\Yic}{Y_i^{\rm c}}
\nc{\yc}{y^{\rm c}}
\nc{\yic}{y_i^{\rm c}}

\nc{\yi}{y_i}
\nc{\Yi}{Y_i}

\nc{\fyic}{f ( \yic ; \; \psiY )}
\nc{\fyi}{f ( y_i ; \; \psiY ) }
\nc{\fdigivenyic}{f ( d_i  \mid  \yic ; \; \psiDY )}
\nc{\fditilgivenyic}{f ( \tilde{d}_i  \mid  \yic ; \; \psiDY )}
\nc{\fditilgivenyi}{f ( \tilde{d}_i  \mid  \yi ; \; \psiDY )}
\nc{\Fditilgivenyic}{F ( \tilde{d}_i  \mid  \yic ; \; \psiDY )}
\nc{\Fditilgivenyi}{F ( \tilde{d}_i  \mid  \yi ; \; \psiDY )}
\nc{\fdigivenyi}{f (d_i \mid y_i ; \;  \psiDY  )}
\nc{\fyicdi}{f \left( \yic, d_i \right)}
\nc{\fyidi}{f \left( \yi, d_i \right)}

\nc{\fymidr}{f_{Y \mid R}}
\nc{\fyr}{f_{Y,R}}
\nc{\frmidy}{f_{R \mid Y}}
\nc{\fy}{f_Y}
\nc{\fr}{f_R}

\nc{\fyicgivendi}{f (\yic \mid d_i; \; \thetaYD )}
\nc{\fyigivendi}{f (\yi \mid d_i; \; \thetaYD )}
\nc{\fyicgivens}{f (\yic \mid s; \; \thetaYD )}
\nc{\fyigivens}{f (\yi \mid s; \; \thetaYD )}
\nc{\fdi}{f ( d_i; \; \thetaD )}

\nc{\fyicX}{f ( \yic \mid X_i; \; \psiY )}
\nc{\fyiX}{f ( y_i \mid X_i; \; \psiY ) }
\nc{\fdigivenyicX}{f ( d_i  \mid  \yic, X_i ; \; \psiDY )}
\nc{\fdigivenyiX}{f (d_i \mid y_i, X_i ; \;  \psiDY  )}
\nc{\fyicdiX}{f \left( \yic, d_i \mid X_i \right)}

\nc{\fyicgivendiX}{f (\yic \mid d_i, X_i; \; \thetaYD )}
\nc{\fyigivendiX}{f (y_i \mid d_i, X_i; \; \thetaYD )}
\nc{\fdiX}{f ( d_i \mid X_i; \; \thetaD )}

\nc{\Yistar}{\bY_i^*}

\nc{\Dio}{D_i^{\rm obs}}
\nc{\bdelio}{\bdel_{ i \, {\rm (obs)}} }

\nc{\fygivend}{f_{Y \mid \delta}}
\nc{\fyd}{f_{Y, \delta}}
\nc{\fd}{f_\delta}
\nc{\FD}{F_D}
\nc{\fygivenbd}{f_{Y\mid b, \delta}}

\nc{\alphahat}{\hat{\bal}}
\nc{\phihat}{\hat{\bphi}}
\nc{\thetahat}{\hat{\bthe}}
\nc{\thetatilde}{\tilde{\bthe}}
\nc{\scoretheta}{\bS(\bthe; \, \scc)}
\nc{\hesstheta}{\bH(\bthe; \, \scc)}
\nc{\infotheta}{\sci(\bthe; \, \scc)}
\nc{\sitheta}{\bs_i(\bthe; \, \scc_i)}
\nc{\sithetahat}{\bs_i(\thetahat; \, \scc_i)}

\nc{\loglikobs}{\ell_{{\rm o}}(\bthe; \, \sco)}
\nc{\scoreobs}{\bS_{{\rm o}}(\bthe; \, \sco)}
\nc{\hessobs}{\bH_{{\rm o}}(\bthe; \, \sco)}
\nc{\infoobs}{\scj_{{\rm o}}(\bthe; \, \sco)}

\nc{\Cil}{\scc_{il}}
\nc{\olog}{\lambda^*(\bthe, \Xobs)}
\nc{\LthetaC}{\scl(\bthe; \, \scc)}
\nc{\LthetaCi}{\scl_i(\bthe; \, \scc_i)}
\nc{\LthetaCil}{\scl_i (\bthe; \, \scc_{il}) }
\nc{\lthetaC}{\ell(\bthe; \, \scc)}
\nc{\lthetaCi}{\ell_i(\bthe; \, \scc_i)}
\nc{\lthetaCil}{\ell_i (\bthe; \, \scc_{il}) }
\nc{\Qtheta}{\scq \left( \bthe \, \left| \,  \bthe^{(r)} \right. \right)}
\nc{\thetar}{\bthe^{(r)}}
\nc{\thetas}{\bthe^{(s)}}
\nc{\alphas}{\bal^{(s)}}
\nc{\psis}{\psi^{(s)}}
\nc{\alphasplusone}{\bal^{(s+1)}}
\nc{\psisplusone}{\bpsi^{(s+1)}}
\nc{\alphapsis}{\left( \alphas, \psis \right)}

\nc{\thetarplusone}{\bthe^{(r+1)}}
\nc{\ologi}{\lambda^*_i(\bthe, \Xobs)}
\nc{\llogi}{\lambda_i \left( \bthe, \tilde{\Xcom}_{il} \right) }

\nc{\scxil}{\tilde{\Xcom}_{il}} 
\nc{\siginv}{\bSig_i^{-1}}

\nc{\fofym}{ f \left( \by_i \mid \bbet_m, \bSig \right) }
\nc{\mphim}{ \phi_M \lsq \bSig^{-1/2}(\by_i - \bX_i \bbet_m) \rsq }
\nc{\mphit}{ \phi_M \lsq \bSig^{-1/2}(\by_i - \bX_i \bbet_{t_i}) \rsq }
\nc{\mphij}{ \phi_M \lsq \bSig^{-1/2}(\by_i - \bX_i \bbet_j) \rsq }
\nc{\mphik}{ \phi_M \lsq \bSig^{-1/2}(\by_i - \bX_i \bbet_k) \rsq }
\nc{\expkerm}{ \exp  \lbc -\half \bu_i(\bbet_m)' \bSig^{-1} \bu_i(\bbet_m)
  \rbc } 
\nc{\expkerk}{ \exp  \lbc -\half \bu_i(\bbet_k)' \bSig^{-1} \bu_i(\bbet_k)
  \rbc } 
\nc{\expkerj}{ \exp  \lbc -\half \bu_i(\bbet_j)' \bSig^{-1} \bu_i(\bbet_j)
  \rbc } 
\nc{\normscorem}{\left( \bX_i' \bSig^{-1} \bX_i \bbet_m - \bX_i' \bSig^{-1}
  \by_i \right) } 
\nc{\normscorej}{\left( \bX_i' \bSig^{-1} \bX_i \bbet_j - \bX_i' \bSig^{-1}
  \by_i \right) } 
\nc{\piti}{ \pi \left( t_i, \bal, \bZ_i\bgam \right) }
\nc{\omij}{ \om_{ij} \left( t_i, \bal, \bZ_i\bgam \right) }
\nc{\phibetak}{ \phi_M(\bbet_k) }
\nc{\phibetaj}{ \phi_M(\bbet_j) }
\nc{\dphidbetak}{ \left. \dpl \phibetak \right/ \dpl \bbet_k }
\nc{\dphidbetakf}{ \frac{ \dpl \phibetak }{ \dpl \bbet_k } }
\nc{\uik}{\bu_i \left( \bbet_k  \right)}
\nc{\mset}{ \{ 0, 1, \ldots, M \} }
\nc{\betasigma}{ \left( \lbc \bbet^{(r)}_t \rbc, \bSig^{(r)} \right) }
\nc{\Thetar}{ \bThe^{(r)} }

\nc{\shatkm}{\hat{S}_{\rm KM}}

\nc{\ds}{\displaystyle}

\nc{\beq}{\begin{eqnarray*}}
\nc{\eeq}{\end{eqnarray*}}

\nc{\beqna}{\begin{eqnarray}}
\nc{\eeqna}{\end{eqnarray}}

\nc{\bct}{\begin{center}}
\nc{\ect}{\end{center}}

\nc{\bds}{\begin{description}}
\nc{\eds}{\end{description}}

\nc{\bit}{\begin{itemize}}
\nc{\eit}{\end{itemize}}
 
\nc{\bnu}{\begin{enumerate}}
\nc{\enu}{\end{enumerate}}

\nc{\bgt}{\begin{table}}
\nc{\bgtb}{\begin{center} \begin{tabular}}
\nc{\entb}{\end{tabular} \end{center} }
\nc{\ent}{\end{table}}

\nc{\ts}{\textstyle}


\maketitle

\begin{abstract}
Emission control technologies installed on power plants are a key feature of many air pollution regulations in the US. While such regulations are predicated on the presumed relationships between emissions, ambient air pollution, and human health, many of these relationships have never been empirically verified.  The goal of this paper is to develop new statistical methods to quantify these relationships.  We frame this problem as one of mediation analysis to evaluate the extent to which the effect of a particular control technology on ambient pollution is mediated through causal effects on power plant emissions.  Since power plants emit various compounds that contribute to ambient pollution, we develop new methods for multiple intermediate variables that are measured contemporaneously, may interact with one another, and may exhibit joint mediating effects.  Specifically, we propose new methods leveraging two related frameworks for causal inference in the presence of mediating variables: principal stratification and causal mediation analysis. We define principal effects based on multiple mediators, and also introduce a new decomposition of the total effect of an intervention on ambient pollution into the natural direct effect and natural indirect effects for all combinations of mediators.  Both approaches are anchored to the same observed-data models, which we specify with Bayesian nonparametric techniques. We provide assumptions for estimating principal causal effects, then augment these with an additional assumption required for causal mediation analysis.  The two analyses, interpreted in tandem, provide the first empirical investigation of the presumed causal pathways that motivate important air quality regulatory policies. \\
\end{abstract}

\noindent%

\section{Introduction}
Motivated by a evidence of the association between ambient air pollution and human health outcomes, the US Environmental Protection Agency (EPA) oversees a vast program for air quality management designed to limit population exposure to harmful air pollution \citep{pope_iii_fine-particulate_2009, dominici_particulate_2014}.  Fine particulate matter of diameter 2.5 micrometers or less (\PMTwo) is of particular importance, with regulations to limit exposure to \PMTwo\, estimated to account for over half of the benefits and a substantial portion of the costs of all monetized federal regulations \citep{office_of_management_and_budget_2013_2013}.  
A large contributor to ambient \PMTwo\, in the US is the power generating sector, in particular coal-fired power plants.  These plants emit \PMTwo\, directly into the atmosphere, but are also major sources of sulfur dioxide (\SOTwo) and nitrogen oxides (\NOx) that, once emitted into the atmosphere, contribute to secondary formation of \PMTwo\, through chemical reaction, coagulation and other mechanisms. The amount \PMTwo\, formation initiated by emissions of \SOTwo\, and \NOx\, depends largely on atmospheric conditions such as temperature \citep{Hodan2004}.  Power plants are also major sources of \COTwo\, emissions.



A variety of regulatory programs under the purview of the Clean Air Act (e.g., the Acid Rain Program) are designed to reduce emissions from power plants, with one goal of reducing population exposure to ambient \PMTwo.  One key strategy for achieving this reduction is the installation of \SOTwo\, control technologies such as flue-gas desulfurization scrubbers (henceforth, ``scrubbers''), on power plant smokestacks to reduce \SOTwo\, emissions and, in turn PM$_{2.5}$. Estimates of the annualized human health benefits of regulatory polices such as the Acid Rain Program rely heavily on presumed relationships between such control strategies, emissions, ambient \PMTwo, and human health. While the underlying physical and chemical understanding of the link between power plant emissions and \PMTwo\, is well established, there remains considerable uncertainty about the effectiveness of specific strategies for reducing harmful pollution amid the realities of actual regulatory implementation. Accordingly, the EPA and other stakeholders have increasingly emphasized the need to provide evidence of which specific air pollution control strategies are most effective or efficient for reducing population exposures to \PMTwo \citep{hei_accountability_working_group_assessing_2003,u.s._epa_workshop_2013}.


The goal of this paper is to deploy newly-developed statistical methods to examine the causal effect of scrubbers installed at coal-fired power plants on the ambient concentration of ambient \PMTwo\, using observed data on power plant emissions and ambient pollution.  Physical and chemical understanding of these processes provide strong support for the expectation that scrubbers reduce ambient \PMTwo ``through'' reducing emissions of \SOTwo, but this relationship has never been empirically verified using observed data in the context of regulations that may simultaneously impact a variety of factors.  A key statistical challenge to verifying this relationship derives from the fact that \SOTwo\, emissions are highly correlated with emissions of \NOx\, and \COTwo\, and \NOx\, is known to play an important role in the formation of ambient \PMTwo, possibly through interactions with \SOTwo\,.  Thus, the question will be formally framed as one of mediation analysis: To what extent is the causal effect of a scrubber (the ``treatment'') on ambient \PMTwo\, (the ``outcome'') mediated through reduced emissions of \SOTwo, \NOx\, and \COTwo\, (the ``mediators'')?  Recovering a statistical answer to this question amid the problem of multiple highly correlated and possibly interacting mediators that are measured contemporaneously requires new methods development and would also serve to bolster the promise of statistical methods in studies of air pollution that have historically relied on physical and chemical knowledge and not on statistical analysis.

To answer this question, we develop new methods that draw from two frameworks for estimating causal effects in the presence of mediating variables: (1) principal stratification \citep{Fran:Rubi:2002} and (2) causal mediation analysis \citep{Robi:Gree:1992}.  The methodological contributions of this paper come in three areas. First, we develop new methods to accommodate multivariate mediating variables that are measured {\it contemporaneously} (not sequentially), are correlated, and may interact with each to impact the outcome (see Figure \ref{dag}. for a an illustrative directed acyclic graph).  This is essential for evaluating scrubbers because power plants simultaneously emit multiple pollutants that may interact through atmospheric processes to impact ambient \PMTwo.  Existing methods in the literature for both principal stratification and mediation analysis have primarily focused on settings with a single mediator (e.g., \cite{Baro:Kenn:1986,Fran:Rubi:2002,Vand:2009, Joff:Gree:2009, Dani:Roy:Kim:Hoga:Perr:2012}) and existing extensions to cases with multiple mediating variables cannot accommodate the setting of power plant emissions where mediators may simultaneously and jointly impact the outcome \citep{Wang:Nels:Albe:2013,Imai:2013,Vand:Vans:2014,daniel:2014}.  Our second methodological contribution is the use of Bayesian nonparametric approaches to model the observed distribution of emissions and pollution outcomes, making use of a multivariate Gaussian copula model to link flexibly-modeled marginal distributions of observed outcomes to a joint distribution of potential outcomes. Similar strategies with a single mediator have received recent attention in the principal stratification literature (\cite{bartolucci2011modeling,Ma:2011,schwartz2011bayesian,conlon2014surrogacy}) and are emerging for causal mediation analysis \citep{Dani:Roy:Kim:Hoga:Perr:2012,kim_bnp_2015}.  These approaches are important for confronting continuous mediators and infinitely many principal strata, and are deployed here in a novel way to address the problem of multiple mediators while flexibly modeling the observed-data distributions of both mediators and outcomes.
Finally, we provide a unification of principal stratification and causal mediation analysis. While the mathematical relationships between these two approaches are well understood \citep{mealli2003assumptions,Vand:2011,mattei2011augmented}, there has not been, to our knowledge, a comprehensive deployment of both perspectives in a complementary fashion to illuminate the scientific underpinnings of a specific problem.  \cite{Baccini:2015} made important progress in this direction using different observed-data models to estimate principal effects and mediation effects in a problem with a single mediator.  In contrast, the approach developed here uses the exact same observed-data models to ground both perspectives, proposes a common set of basic assumptions for estimating both principal effects and mediating effects, modularizes an additional assumption required to augment a principal stratification analysis in order to obtain estimates of natural direct and indirect effects, and considers settings with multiple mediating variables. Ultimately, we provide a new dimension of quantitative, statistical evidence for supporting air policy regulatory decisions. 


\section{Scrubber Installation and Linked Data Sources}\label{sec:ARP}
Title IV of the Clean Air Act established the Acid Rain Program (ARP), which required major emissions reductions of \SOTwo\, (and other emissions) by ten million tons relative to 1980 levels.  This reduction was achieved mostly through cutting emissions from power plants, or more formally, electricity-generating units (EGUs).  Impacts of the ARP have been evaluated extensively, and the program is generally lauded as a success due to marked national decreases in \SOTwo\, and \NOx\, coming at relatively low cost.  Estimates of the annualized human health benefits of the entire ARP range from \$50 billion to \$100 billion \citep{chestnut_fresh_2005}, but rely heavily on presumed relationships between power plant emissions, ambient \PMTwo, and human health. 

While power plants under the ARP had latitude to elect a variety of strategies to reduce emissions, one key strategy is the installation of a scrubber to reduce $\text{SO}_2$ emissions.   The precise extent to which installation of a scrubber reduces ambient $\text{PM}_{2.5}$ through reducing $\text{SO}_2$ emissions remains unknown, and has never been estimated empirically amid the realities of actual regulatory implementation where pollution controls may impact a variety of factors that are also related to the formation of $\text{PM}_{2.5}$.  Knowledge of these relationships is complicated by the fact that power plants emit more than just \SOTwo, and emissions of a variety of pollutants likely interact in the surrounding atmosphere to form ambient \PMTwo.    


To provide refined evidence of the extent to which scrubbers reduce emissions and cause improvements to ambient air quality, we assembled a national database of ambient air quality measures, weather conditions, and information on power plants.  Specifically, we assembled data on 258 coal-fired power plants from the EPA Air Markets Program Data and the Energy Information Administration, with information on plant characteristics, emissions control technologies installed (if any), and emissions of \SOTwo, \NOx, and \COTwo\, during 2005, five years after promulgation of an important phase of regulations under the Acid Rain Program. For each power plant, we augment the data set with annual average ambient \PMTwo\, concentrations in 2005 and baseline meteorologic conditions in 2004 measured at all monitoring stations in the EPA Air Quality System that are located within 150km.  The 150km range was chosen both to acknowledge that atmospheric processes carry power plant emissions across distances at least this great, but also to minimize the number of monitoring stations considered within range of more than one power plant.   We regard any power plant as ``treated'' with scrubbers in 2005 if at least 10\% of the plant's total heat input was attributed to a portion of the plant equipped with a scrubber as of January 2005. Note that this proportion was nearly 0\% or nearly 100\% for the vast majority of plants, indicating robustness to this 10\% cutoff.  Other power plant characteristics are listed in Table \ref{Data}. The data files and programs to assemble the analysis data set are available at \url{https://dataverse.harvard.edu/dataverse/mmediators} and \url{https://github.com/lit777/MultipleMediators}, respectively.

\begin{table}[h]
\centering
\caption{Summary statistics for covariates and outcomes available for the analysis of \SOTwo\, scrubbers.}\resizebox{\textwidth}{!}{  
\begin{tabular}{lcccc} \hline \hline
 & \multicolumn{2}{c}{\underline{Have scrubbers (n=59)}} & \multicolumn{2}{c}{\underline{Have no scrubber (n=190)}} \\
  & Median & IQR & Median & IQR \\ \hline
   \underline{Monitor Data} \\
 Average Ambient \PMTwo\, 2005 ($\mu g/m^3$)  & 12.4 & (7.8, 14.8) &13.7& (11.8, 15.2)\\
 Average Temperature 2004 ($^\circ$C)&11.5& (10.1, 15.0) &12.8& (10.4, 16.1)\\
 Average Barometric Pressure 2004 ($mmHg$) &737.8& (686.7, 752.4) &746.1& (739.1, 755.6)\\

 \\ \underline{Power Plant Level Data} \\
Total \SOTwo\, Emission 2005 (tons)&644.3& (257.3, 1819.9) &1267.1& (504.9, 2707.6)\\
Total \NOx\, Emission 2005 (tons)&852.1& (394.2, 1531.3) &442.5& (193.7, 878.2)\\
Total \COTwo\, Emission 2005 (1000 tons)& 505.3& (232.5, 960.7) &283.6& (117.7, 559.0)\\

 \\ \underline{Unit Level Data} \\
Average Heat Input 2004 (1000 MMBtu)&4653.3& (2266.4, 9363.9) & 2783.4& (1147.6, 5448.1) \\
Total Operating Time  2004 (hours $\times$ \# units) &7944.0& (7565.8, 8154.9) &7583.9& (7171.0, 7985.9) \\
Sulfur Content in Coal 2004 (lb/MMBtu)&1.0& (0.5, 2.2) &0.7& (0.3, 1.1)\\
Num. of \NOx\, Controls 2004 (\# units) &1.0& (1.0, 1.5) &1.0& (0.9, 1.3)\\
Pct. operating Capacity 2004 (MMBtu/MMBtu $\times$ 100) & 20.2& (10.0, 28.8) &16.4 & (9.3, 24.6)\\
Heat Rate 2004 (MMBtu/MWh) & 268.5 & (175.5, 436.9) & 254.3 & (152.6, 396.8)
\end{tabular}}
\label{Data}
\end{table}

\section{Causal Mediation Analysis and Principal Stratification}\label{sec:ps_and_med}
\subsection{Mediation Analysis with a Single Mediator}\label{sec:singlemediator}
To fix ideas, consider the single mediator case. Let $Z_i \in \{0,1\}$ indicate the presence of the intervention of interest, here, whether power plant $i$ has installed scrubbers  in January 2005 ($Z_i=1$) and let $\boldsymbol{Z} = (Z_1, \cdots, Z_n)$ be the vector of intervention indicators for power plants $i=1, \cdots, n$.  Using potential-outcomes notation \citep{rubi:1974}, let $M_i(\boldsymbol{Z})$ denote the potential emissions that the $i$-th power plant would be generated under the vector of scrubber assignments $\boldsymbol{Z}$, and let $Y_i(\boldsymbol{Z}; \boldsymbol{M})$  denote the potential ambient \PMTwo\, outcome that could, in principle, be defined for any scrubber assignment vector $\boldsymbol{Z}$ and any vector of intermediate emissions values $\boldsymbol{M}$.  Throughout the paper, we adopt the stable unit treatment value assumption (SUTVA; Rubin 1980) which implies 1) there is no ``interference'' in the sense that potential intermediate and outcome values from power plant $i$ do not depend on scrubber treatments and emissions intermediates of other power plants (i.e, $M_i(\boldsymbol{Z}) = M_i(Z_i)$ and $Y_i(\boldsymbol{Z}; \boldsymbol{M}) = Y_i(Z_i; M_i)$) and 2) there are ``no multiple versions'' of scrubber treatments such that whenever $Z_i = Z_i^\prime$, $M_i(Z_i) = M_i(Z_i^\prime)$ and $Y_i(Z_i;M_i(Z_i)) = Y_i(Z_i^\prime, M_i(Z_i^\prime))$.  For reasons that will become clear later, we augment the standard SUTVA to also assume ``no multiple versions'' of emissions intermediates which states, if $M_i=M_i^\prime$, then $Y_i(Z_i;M_i) = Y_i(Z_i; M_i^\prime)$ (Forastiere et al. 2016\nocite{Forastiere:2016}). We revisit possible violations of SUTVA in Section \ref{sec:discussion}, but note here that the linkage of power plants to monitors within 150km provides some justification for this assumption.


The {\it natural direct effect} \citep{Pear:2001} is defined by $\mbox{NDE}=E[Y_i(1;M_i(0)) - Y_i(0;M_i(0))]$, representing the effect of the intervention obtained when setting the mediator to its `natural' value $M_i(0)$; i.e., its realization in the absence of the intervention. The {\it natural indirect effect} is defined as
$\mbox{NIE}=E[Y_i(1;M_i(1)) - Y_i(1;M_i(0))]$, representing the effect of holding the intervention status fixed at $Z=1$ but changing the value of the mediator from $M(0)$ to $M(1)$.  The total causal effect of the intervention on the outcome can then be defined as $\mbox{TE}=\mbox{NDE}+\mbox{NIE}=E[Y_i(1;M_i(1)) - Y_i(0;M_i(0))]$.  Similar controlled effects could also be defined to represent causal effects at specific values of $M$ \citep{Pear:2001,Robi:Gree:1992}.  

Implicit in the definition of these effects is the conceptualization of hypothetical interventions that could independently manipulate values of both $Z$ and $M$ to, for example, ``block'' the effect on the mediator.  Thus, it is important to note that potential outcomes of the form $Y_i(Z_i; M_i(Z_i'))$ are purely hypothetical for $Z_i \ne Z_i^\prime$, and can never be observed for any observational unit.  Such unobservable potential outcomes have been referred to as {\it a priori counterfactuals} \citep{Robi:Gree:1992, rubin_direct_2004}. We revisit conceptualization of {\it a priori} counterfactuals in the context of the power plant study in Section \ref{sec:notation}, but note here the distinction between {\it a priori} counterfactuals and potential outcomes of the form $Y_i(Z_i; M_i(Z_i))$ that are {\it observable} and  actually observed for some units. 


\subsection{Principal Stratification}
A distinct but related framework for defining causal effects in the presence of intermediate variables is {\it principal stratification} \citep{Fran:Rubi:2002}. Continuing with the single-mediator case, principal stratification considers only a single intervention and relies on definition of two causal effects: the effect of $Z_i$ on $M_i$, defined as $M_i(1) - M_i(0)$, and the effect of $Z_i$ on $Y_i$, defined as $Y_i(1;M_i(1)) - Y_i(0;M_i(0))$.  The objective is to estimate {\it principal effects}, which are average causal effects of $Z_i$ on $Y_i$ within {\it principal strata} of the population defined by $(M_i(0), M_i(1))$.  

With principal stratification, {\it dissociative effects} are defined to quantify the extent to which the intervention causally affects outcomes when the intervention does not causally affect the mediator, for example, $E[Y_i(1;M_i(1)) - Y_i(0;M_i(0)) \,|\, M_i(1) = M_i(0)]$. Dissociative effects are similar to direct effects in a mediation analysis in that they represent causal effects of an intervention on the outcome among the subpopulation where there is no causal effect on the mediator, but they refer only to the specific subpopulation with $M(1)=M(0)$. \cite{Vand:2008} and \cite{mealli2012refreshing} show that dissociative effects represent a quantity that is only one contributor to the NDE, with the amount of contribution tied to the size of the subpopulation with $M(1)=M(0)$.

 {\it Associative effects} are defined to quantify the causal effect of the intervention on the outcome among those for which the intervention {\it does} causally affect the mediator, for example, $E[Y_i(1;M_i(1)) - Y_i(0;M_i(0)) \,|\, M_i(1) < M_i(0)]$.  An associative effect that is large in magnitude relative to the dissociative effect indicates that the causal effect of the intervention on the outcome is greater among those for which the mediator is causally affected, compared to those for which the mediator is not affected. This could be interpreted as suggestive of a causal pathway whereby the intervention impacts the outcome through changing the mediator, but note that associate effects are generally a combination of the NDE and NIE for a defined subpopulation.
  
Dissociative effects that are similar in magnitude to associative effects indicate that the intervention effect on the outcome is similar among observations that do and do not exhibit causal effects on the mediator, which could be interpreted as suggestive of other causal pathways through which $Z_i$ affects $Y_i$.  

A primary distinction between principal stratification and causal mediation analysis is that principal effects only pertain to population subgroups comprised of observations with particular values of $(M_i(0), M_i(1))$, whereas natural direct and indirect effects are defined for the whole population (as discussed in detail in  \cite{mealli2012refreshing}).
 Importantly, note that the {\it a priori} counterfactuals of the form $Y_i(Z_i, M_i(Z_i^\prime))$ for $Z_i \ne Z_i^\prime$ do not appear in the definition of principal effects, which rely only on the definition of {\it observable} potential outcomes $Y_i(Z_i, M_i(Z_i))$. Thus, there is no conception in principal stratification of a hypothetical intervention acting on $M_i$ independently from $Z_i$, and there is no definition of a causal effect of $Z_i$ on $Y_i$ that is mediated through $M_i$.  From a modeling perspective, principal effects can be estimated when an outcome model is specified conditional on both potential mediators (intermediate outcomes), $M_i(0)$ and $M_i(1)$ while causal mediation analysis has tended to rely on an outcome model that depends on the observed mediator. The differences in modeling strategies that are typically employed in principal stratification and causal mediation analysis complicate comparisons, as results of such analyses have typically been driven in part by different modeling assumptions. In Section 5, we will propose a new set of assumptions to build a common observed-data model for principal stratification and causal mediation analysis.

\subsection{Existing Considerations for Multiple Mediators}\label{sec:multiplemediators}
\begin{figure}[h]
\centering
\scalebox{0.45}
{\includegraphics{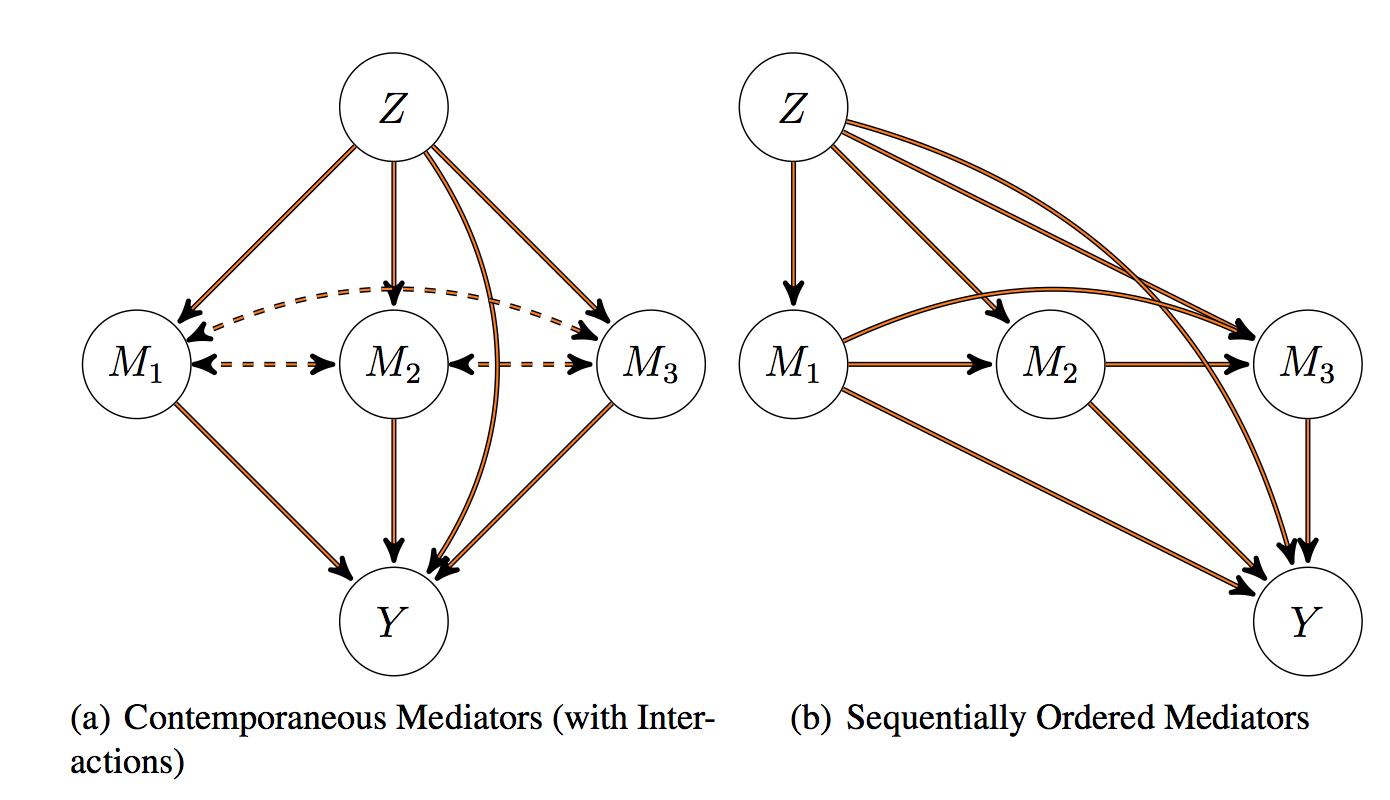}} 
\caption{Directed Acyclic Graphs : a) contemporaneous mediators with interactions (our case) and b) sequentially ordered mediators.}
\label{dag}
\end{figure}

Extensions of the causal mediation ideas outlined in Section \ref{sec:singlemediator} to settings of multiple mediating variables are emerging. For contemporaneously observed mediators, straightforward extensions of the Baron and Kenny (1986) regression-based structural equation model approach \citep{MacK:2008} have been proposed. For each of $K$ contemporaneous mediators $(M_{1}, M_{2}, \cdots, M_{K})$, a series of regression models is used to estimate mediator-specific NIEs in a manner that implies additivity of indirect effects:
\begin{equation}
\text{JNIE} = \sum_{k=1}^K \text{NIE}_k \quad \text{and} \quad \text{TE} = \text{NDE} + \text{JNIE},
\label{additivity}\end{equation}
where JNIE is used to denote the joint natural indirect effect due to changes in all $K$ mediators, and NIE$_k = E[Y_i(1;M_{k,i}(1))-Y_i(1;M_{k,i}(0))]$ represents the natural indirect effect of the $k$-th mediator.  These approaches assume that each $M_{k,i}$ mediates the treatment effect independently of the other mediators, without interactions among mediators (i.e., the mediators are {\it causally independent} or {\it parallel}). Figure \ref{dag}.a without dashed lines illustrates this case. \cite{Wang:Nels:Albe:2013} propose an alternative modeling approach under the setting of causally independent mediators.
If the mediators interact with each other in terms of their impact on the outcome, then additivity of indirect effects as in the above cannot hold; and estimation of multivariate mediated effects can then be further complicated by correlations among the mediators. 
Dependence among mediators has been considered when $M_k$ are observed sequentially (i.e., sequential mediators; Figure \ref{dag}.b), as in \cite{Imai:2013}. \cite{albert2011generalized}, and \cite{daniel:2014} propose approaches for either sequentially dependent mediators or mediators that do not affect nor interact with each other. These approaches offer a decomposition of the JNIE in the case of sequential dependence, and assume additivity of natural indirect effects otherwise. \cite{Vand:Vans:2014} discuss an approach to decompose the JNIE further when the mediators simultaneously affect each other; however, their approach does not evaluate the impact of each individual mediator (see Section 4.3).\cite{taguri2015causal} propose an approach for contemporaneous, non-ordered mediators, but rely on an assumption that the mediators are conditionally independent given observed covariates, which does not fully represent the possibility of contemporaneous interactions among the mediators, as may be the case with multiple emissions (in particular \SOTwo\, and \NOx\,) and the formation of ambient \PMTwo.  Section \ref{sec:sim} examines the possibility of contemporaneous interactions among (possibly correlated) mediators in the context of the scrubber study.

In summary, existing methods for multiple mediators rely on either assumed causal independence of (parallel) mediators and additivity of indirect effects, sequential dependence of mediators, or on restrictive assumptions of conditional independence among mediators. \cite{Vand:Vans:2014} point out that if there are interactions between the effects of (non-sequential) multiple mediators on the outcome, the joint indirect effect may not be the sum of all three indirect effects. They note that, in principle, an analysis could proceed with an outcome model including interactions $M_j M_k$ for all $\{j,k\}$ combinations combined with models for ($E(M_j,M_k)$).  However, this approach would lead to issues of model compatibility between the models for $M_j$ and $M_k$ and that for the product $M_j M_k$.  The lack of satisfactory methods for more general settings of multiple contemporaneously-measured mediators motivates the methods developed herein, where we offer a new decomposition of the joint natural indirect effect into individual indirect effects that may not affect the outcome additively.

\section{New Methods for Causal Mediation Analysis and Principal Stratification with Multiple Contemporaneous Mediators}\label{sec:estimands}
\subsection{Notation for Multiple Mediating Variables}\label{sec:notation}
Suppressing the $i$ subscript indexing power plants, let $\{ M_k(z); k=1,\ldots,K \}$ denote the potential emissions of $K$ pollutants that would occur if a power plant were to have scrubber status $Z=z$, for $z=0,1$.  While much of our development is general for any $K$, we focus on the case $K=3$ so that $M_k(z), k=1,2,3$ denotes the potential emissions of SO$_2$, NO$_x$, and CO$_2$, respectively.  The causal effect of the scrubber on emission $k$ can then be defined as a comparison between $M_k(1)$ and $M_k(0)$.  Let $\bM(z_1, z_2,z_3) \equiv \{M_1(z_1), M_2(z_2), M_3(z_3)\}$ denote potential emissions under a set of three scrubber statuses $\{z_1, z_2,z_3\}$.  

We similarly define potential \PMTwo\, outcomes, but extend the notation to define potential concentrations under different values of scrubber status, $Z$, and different possible values of emissions, $\bM(z_1, z_2,z_3)$.  Thus, in full generality, each power plant has a set of $2^{K+1}=16$ potential outcomes for \PMTwo, $Y(z;\bM(z_1,z_2,z_3))$, which denote potential values of \PMTwo\, that would be observed under intervention $Z=z$ with pollutant emissions set at values under
interventions $z_1, z_2, z_3$.  Definition of all $16$ potential \PMTwo\, concentrations is required for definition of natural direct and indirect effects and entails {\it a priori} counterfactuals. 
For example, $Y(1;\bM(0,0,1))$ would represent the potential ambient \PMTwo\, concentration near a plant under the hypothetical scenario where the plant installs a scrubber ($z=1$), but where emissions of \SOTwo\, and \NOx\, are set to what they would be without the scrubber ($z_1=z_2=0$) and emissions of \COTwo\, are set to what they would be with the scrubber ($z_3=1$).  This may be conceptualized as a setting where a power plant installs a scrubber, but offsets the cost of the technology by burning coal with a higher sulfur content and discontinuing use of a different \NOx\, control, thus ``blocking'' the intervention and maintaining \SOTwo\, and \NOx\, emissions at levels that would have occurred without the \SOTwo\, technology.  Principal stratification will only rely on potential outcomes with  $z=z_1=z_2=z_3$ that are observable from the data, such as  $\bM(1,1,1)$ and $Y(1;\bM(1,1,1))$ observed for any power plant that installs a scrubber. Finally, let $\boldsymbol{X}$ denote a vector of baseline covariates measured at the power plant or the surrounding area.

\subsection{Observable Outcomes: Principal Causal Effects for Multiple Mediators}
Extending principal stratification to settings where the intermediate variable is multivariate is conceptually straightforward.  Principal stratification defines a principal stratum for every combination of the joint vector $(\bM(0,0,0), \bM(1,1,1))$, and principal causal effects are defined as comparisons between $Y(0; \bM(0,0,0))$ and $Y(1; \bM(1,1,1))$ within principal strata. 

For any subset $\mathcal{K} \subseteq \{1,2,3\}$, let $|\bM(1,1,1) - \bM(0,0,0)|_\mathcal{K}$ denote the element-wise absolute differences between emissions of the subset of pollutants in $\mathcal{K}$, e.g.,  $|\bM(1,1,1) - \bM(0,0,0)|_{\mathcal{K} = \{1,3\}} = \{|M_1(1) - M_1(0)|, |M_3(1)-M_3(0)|\}$. Definitions of quantities such as average associative and dissociative effects can proceed following \cite{zigler_estimating_2012} by defining:
\begin{small}
\begin{eqnarray*}
\text{EDE}_\mathcal{K} &=& E[Y(1;\bM(1,1,1)) - Y(0;\bM(0,0,0)) \,\big\vert\,
|(\bM(1,1,1) - \bM(0,0,0) )|_{\mathcal{K}} < C _{\mathcal{K}}^D],\\
\text{EAE}_\mathcal{K} &=& E[Y(1;\bM(1,1,1)) - Y(0;\bM(0,0,0)) \,\big\vert\,
|(\bM(1,1,1) - \bM(0,0,0) )|_{\mathcal{K}} > C _{\mathcal{K}}^A],
\end{eqnarray*}
\end{small}where $C_\mathcal{K}^A$ denotes a vector of thresholds beyond which a change in each emission in $\mathcal{K}$ is considered meaningful, $C_\mathcal{K}^D$ is a vector of thresholds below which changes in these emissions are considered not meaningful, and $>$ and $<$ represent element-wise comparisons. Note that the dissociate effect is now defined on principal strata where potential changes (or differences) in the intermediate variables are less than some vector of thresholds $|(\bM(1,1,1) - \bM(0,0,0) )|_{\mathcal{K}} < C _{\mathcal{K}}^D$ instead of principal stratum with strict equality $|(\bM(1,1,1) - \bM(0,0,0) )|_{\mathcal{K}} = \{0,0,0\}_\mathcal{K}$ to accommodate continuous intermediate values. For example, $\mathcal{K} = \{1,3\}$ would be used to define the associative (dissociative) effect in the subpopulation exhibiting an effect on \SOTwo\, and \COTwo\, in excess of $C_\mathcal{K}^A$ (below $C_\mathcal{K}^D$), without regard to the effect on \NOx. For the data analysis in Section \ref{sec:dataanalysis}, we divide the EAE defined above into two parts: EAE$^+_\mathcal{K}$ will denote the average associative effects among power plants where all emissions in $\mathcal{K}$ are causally increased in excess of $C_\mathcal{K}^A$, while EAE$^-_\mathcal{K}$ will denote the average associative effect in power plants where all emissions in $\mathcal{K}$ were causally reduced in excess of $C_\mathcal{K}^A$.  Note that these summary quantities only consider a subset of principal strata that may be of interest.  For example, analogous average principal effects could be calculated among strata where some emissions are decreased and others are increased.  We avoid burdensome notation for such summaries, but will revisit estimates in additional principal strata in the context of the data analysis in Section \ref{sec:dataanalysis}. 

In addition to estimating average dissociative and associative effects for different $\mathcal{K}$ as defined above, interest may lie in entire surfaces of, for example, how the causal effect on \PMTwo\, varies as a function of the causal effect on each emission (``causal effect predictiveness'' surface \citep{gilbert_evaluating_2008}).

\subsection{Observable and {\it a priori} Counterfactual Outcomes: Natural Direct and Indirect Effects for Multiple Mediators}\label{sec:mediationeffects}
Extending definitions of natural direct and indirect effects to the multiple mediator setting is somewhat more complicated.  The natural direct effect is defined as $\text{NDE}$ = $E[Y(1;\bM(0,\cdots,0))-Y(0; \bM(0, \cdots, 0))]$, representing the causal effect of $Z$ on $Y$ that is ``direct'' in the sense that it is not attributable to changes in any of the $K$ emissions. The joint natural indirect effect of all $K$ mediators, $\text{JNIE}_{12\cdots K}$, is derived by subtracting the natural direct effect from the total effect, $\text{JNIE}_{12 \cdots K}=\text{TE}-\text{NDE} = E[Y(1;\bM(1,1,\cdots,1))-Y(1; \bM(0, 0, \cdots,0))]$.

In addition to $\text{JNIE}_{12\cdots K}$, we introduce a decomposition into the natural indirect effects attributable to changes in different combinations of the $K$ mediators.  Maintaining focus on the case where $K=3$, the $\text{JNIE}_{123}$ can be decomposed into emission-specific indirect effects and the joint indirect effects of all possible pairs of emissions.  See Figure 5 in the Web Appendix for a graphical representation.

We define the {\it mediator-specific} NIE for the $k$-th emission as a comparison between the potential \PMTwo\, outcome under scrubbers and the analogous outcome with the value of the $k$-th emission fixed to the natural potential value that would be observed without scrubbers. Specifically, for emissions of \SOTwo, \NOx, and \COTwo\, define:
\begin{eqnarray}
\text{NIE}_1 = E[Y(1; \bM(1,1,1)) - Y(1; \bM(0,1,1))], \nonumber\\
\text{NIE}_2 = E[Y(1; \bM(1,1,1)) - Y(1; \bM(1,0,1))], \label{niek}\\
\text{NIE}_3 = E[Y(1; \bM(1,1,1)) - Y(1; \bM(1,1,0))]. \nonumber
\end{eqnarray}

In a similar fashion we can define the joint natural indirect effect
attributable to subsets of mediators $j$ and $k$ for
$j\neq k$  as differences
between the observable potential \PMTwo\, outcomes under scrubbers and the analogous
{\it a priori} counterfactual with values of pollutants $j$ and $k$ set to their natural values that would be observed without scrubbers.  For example, $\text{JNIE}_{12}$ defines the joint natural indirect effects of mediators 1 (\SOTwo) and 2 (\NOx) as
\begin{equation*}
\text{JNIE}_{12}  =  E[Y(1; \bM(1,1,1)) - Y(1; \bM(0,0,1))].
\end{equation*}
Values of JNIE$_{jk}$ for other pairs of mediators can be defined analogously, and all such pairs correspond to the second row in Figure 5 in the Web Appendix.  Note that the joint natural indirect effect of each pair of mediators is not equal to the sum of corresponding mediator-specific NIEs unless there is no overlap between mediator-specific NIEs (additivity). For example, we can represent the relationship between $\text{JNIE}_{12}$ and the mediator-specific effects $\text{NIE}_1$ and $\text{NIE}_2$ as
\begin{small}
\begin{eqnarray*}
\lefteqn{(\text{NIE}_1 + \text{NIE}_2 ) - \text{JNIE}_{12}  }\\
&=&  E[Y(1;\bM(1,1,1)) - Y(1; \bM(0,1,1))  - Y(1; \bM(1,0,1)) + Y(1; \bM(0,0,1))].
\end{eqnarray*}
\end{small}
Thus, if this quantity is not equal to $0$, we argue that additivity of mediator-specific NIEs does not
hold. Note that the above decomposition of JNIE$_{123}$ differs from \cite{Vand:Vans:2014}, which considers the portion of the JNIE$_{123}$ mediated through $M_1$, then sequentially considers the additional contribution of each mediator in the presence of the others. This presumed ordering of mediators precludes estimation of the effect through different pairs of mediators such as $\text{JNIE}_{23}$ or $\text{JNIE}_{13}$, the availability of which is a benefit of our proposed decomposition.  Our decomposition also differs from Daniel et al. (2015) who only allow interacting overlap between mediator-specific NIEs when one mediator causally affects another.

Note that alternative definitions of NIE could use contrasts of the form: $\text{NIE}^*_{1} =  E[Y(0; \bM(1,1,1)) - Y(0; \bM(0,1,1))]$.  Such a strategy is also considered in Daniel et al. (2015), but defining NIE$_k^*$ in this way would rely entirely on {\it a priori} counterfactuals, whereas a benefit of using the definitions in (\ref{niek}) is that each definition uses the observable potential outcome $Y(1;\bM(1,1,1))$, comparing against only one {\it a priori} counterfactual (e.g., $Y(1; \bM(0,1,1))$).


\section{Flexible Bayesian Models Assumptions and Estimation}\label{sec:Models}
Under the assumptions developed in this section, Bayesian inference for the causal effects defined in Section 4 follows from specifying models for the joint distribution of all potential mediators (conditional on covariates) and the outcome model conditional on all potential mediators and covariates, and prior distributions for unknown parameters. Posterior distributions cannot be computed directly from observed data because potential outcomes are never jointly observed in both the presence and absence of a scrubber and {\it a priori} counterfactuals are never observed.  Our estimation strategy consists of three steps. First, we specify nonparametric models for the observed data.  The marginal distribution of each observed mediator (i.e.,$\bM(0,0,0)=\{M_1(0), M_2(0), M_3(0)\}$ observed for power plants that did not install scrubbers and  $\bM(1,1,1),=\{M_1(1), M_2(1), M_3(1)\}$ observed for those that did) is specified separately and then linked into a coherent joint distribution using a Gaussian copula model \citep{nelsen_introduction_1999}.  The models for the potential outcomes $Y(1;\bM(1,1,1))$ and $Y(0;\bM(0,0,0))$ are specified conditional on covariates and all potential mediators ($\bM(1,1,1)$ and $\bM(0,0,0)$) that are never observed simultaneously. Thus, the conditional outcome models are estimated via the data augmentation for unobserved potential mediators. Second, we introduce two assumptions for estimating the TE and the associative and dissociative effects.  Third, we employ an additional assumption to equate the distributions of {\it a priori} counterfactuals to those of the observed potential outcomes under intervention $Z=1$ to allow estimation of the natural direct and indirect effects.  We also provide optional modeling assumptions to sharpen posterior inference for the power plant evaluation. Throughout, we estimate the distribution of the covariates, $F_{\boldsymbol{X}}(\boldsymbol{x})$, using the empirical distribution.

\subsection{Models for the Observed Data}\label{sec:observedmodels}
We specify Dirichlet process mixtures for the marginal
distribution of  each mediator \citep{muller_bayesian_1996}.  For each
intervention $z=0,1$, $k=1,2,3$ and baseline covariates $\boldsymbol{X}=\boldsymbol{x}$, the conditional distribution of the $k$-th observed
mediator is specified as
\begin{eqnarray*}
M_{k,i} | Z_i=z, \boldsymbol{X}_i=\boldsymbol{x}_i & \sim &
N(\beta^z_{k0,i}+\boldsymbol{x}_i^\top \boldsymbol{\beta}^z_{k1},\,\,
\tau^z_{k,i}), \quad M_{k,i}\geq0; \,\, i=1,\cdots, n_z\\
\beta^z_{k0, i}, \tau^z_{k,i} & \sim & F_{k}^z,\\
F_{k}^z & \sim & DP (\lambda_{k}^z, \,\,\mathcal{F}_{k}^z),
\end{eqnarray*}
where $\{i=1,2,\ldots,n_z\}$ denotes the observations with $Z=z$ and $k$ indicates the $k$-th mediator.  We bound the mediator from below (0) using a truncated normal kernel (within the interval [0, $\infty$)). $\beta^z_{k0,i}$ and $\tau^z_{k,i}$ denote the intercept and precision parameters for the $k$-th emission at the $i$-th power plant that received intervention $z$. Here, $DP$ denotes the Dirichlet process with two parameters, a mass
parameter ($\lambda_k^z$) and a base measure ($\mathcal{F}_k^z$). To not overly complicate the model we only `mixed' over the intercept and precision parameters in the conditional distributions,
$\beta^z_{k0, i}$ and $\tau^z_{k,i}$. The base distribution $\mathcal{F}_k^z$ is taken to be the
normal-Gamma distribution,
$N(\mu_k^z, S_k^z) G(a_k^z,b_k^z)$.  Details including hyper prior specification are given in Section A of the Web Appendix.  

The marginal distributions of each mediator under each $z=0,1$ are linked to model the joint distribution of  $[M_1, M_2, M_3 | Z=z, X=x]$ with Gaussian copula models of the form:
\begin{equation*}
F_{\bM(z,z,z)} (\mathbf{m}_{z,z,z}) = \Phi_{3}[\Phi_{1}^{-1}\{F_{M_1(z)}(m_1)\},\Phi_{1}^{-1}\{F_{M_2(z)}(m_2)\},
 \Phi_{1}^{-1}\{F_{M_3(z)}(m_3)\}],
\end{equation*}
\noindent where $\mathbf{m}_{z,z,z}$ are values of potential mediators under intervention $Z=z$ and $\Phi_k$ is the $k$-variate standard normal CDF.  Note that we elect to model the marginal distribution of each univariate random variable separately, and then combine with the Gaussian copula model, rather than directly model the joint distributions of $[M_1, M_2, M_3 |Z = z, X = x]$. Thus, we allow full flexibility using DP mixtures of (truncated) normals for the marginal distributions (the fit of which can be checked empirically) and use the Gaussian copula to link them to construct the joint distribution of potential mediators. The Gaussian copula model implies some (correlation) structure to the joint distribution of all observable potential outcomes, without implying any specific causal structure. Flexibility of this structure derives from the fact that each marginal distribution is modeled as nonparametric with infinite dimensional parameter spaces.
The strategy is designed to coalesce with the modeling strategy in Section \ref{sec:assumption}.  Note that other potential alternatives to link the fixed marginal distributions such as mixtures of marginals (e.g. $H(x_1,x_2) = p F(x_1) + (1-p) G(x_2)$ or $H(x_1,x_2) = \sqrt{F(x_1)G(x_2)}$) do not specify the full joint distribution distribution of $(x,y)$ \citep{nelsen_introduction_1999}) and our method does not limit the number of the mediators in general. While the joint distribution of all potential mediators ($\bM(0,0,0)$ and $\bM(1,1,1)$) is also modeled via the same Gaussian copula model, this entails modeling unobserved potential mediators and will be discussed as a part of the assumptions in Section \ref{sec:assumption}

To model the distributions of the potential outcomes for each $z=0,1$ conditional on all potential mediators and covariates, we use a locally weighted mixture of normal regression models \citep{muller_bayesian_1996} that is induced by specifying a DP mixture of normals for the joint distribution of the outcome, all mediators and covariates. For each intervention $z=0,1$, potential values of all (counterfactual) mediators and baseline covariates $\boldsymbol{X}=\boldsymbol{x}$, the conditional distribution of the observed outcome $y_i$ is specified as
\begin{eqnarray*}
\lefteqn{f(y_i | \mathbf{m}_i(0,0,0), \mathbf{m}_i(1,1,1), \boldsymbol{x}_i,Z_i=z})\\
& = & \sum_{l=1}^{\infty} \omega_l^z
  N(y_i,\mathbf{m}_i(0,0,0),\mathbf{m}_i(1,1,1),\boldsymbol{x}_i \,|\,\boldsymbol{\mu}_l^z, \Sigma_l^z)
\end{eqnarray*}
where $\omega_l^z = \gamma_l^z / (\sum_{j=1}^\infty \gamma_j^z N(\mathbf{m}_i(0,0,0),\mathbf{m}_i(1,1,1),\boldsymbol{x}_i \,|\,\boldsymbol{\mu}_{j,\setminus 1}^z, \Sigma_{j, (\setminus 1, \setminus 1)}^z))$ and $\boldsymbol{\mu}_{j,\setminus 1}^z$ denotes all elements of mean parameters $\boldsymbol{\mu}_j^z$ except for $Y_i$. Similarly, $\Sigma_{j, (\setminus 1, \setminus 1)}^z$ denotes
a submatrix of covariance matrix $\Sigma_{j}^z$ formed by deleting the the first row and the first column. The weight involves the parameter $\gamma_j^z$  where $\gamma_j^z = \gamma_j^{\prime,z}\prod_{h<j} (1-\gamma_h^{\prime,z})$ and $\gamma_j^{\prime,z} \sim \text{Beta}(1,\alpha^z)$. This flexible conditional model specification is a necessary feature in our case since we allow the outcome model to capture nonlinear and/or interaction effects of the mediators. Note again that this outcome model is conditional on all potential mediators $\{\bM(0,0,0), \bM(1,1,1)\}$ which cannot be observed at the same time. We use a similar approach to that used in \cite{schwartz2011bayesian} to model the observed outcome distribution conditional on partly missing potential intermediate variables by constructing {\it complete intermediate data}. Here, we impute unobserved potential mediators for each unit with a data-augmentation approach based on the joint distribution of all potential mediators specified above. Details about hyper prior specification and posterior computation are given in the Web Appendix.

\subsection{Assumptions for Estimation of Causal Effects}\label{sec:assumption}
To estimate causal effects based on the model for the observed data specified in Section \ref{sec:observedmodels}, we formulate assumptions relating observed quantities to both observable outcomes and {\it a priori} counterfactuals.  Denote the conditional distribution $[Y(z; \bM(z_1, z_2,
z_3)) \,|\, \bM(0,0,0) = \mathbf{m}_{0,0,0}, \bM(1,1,1)=\mathbf{m}_{1,1,1}, \boldsymbol{X}=\boldsymbol{x}]$ with $f_{z, \bM(z_1,z_2,z_3)}(y\,|\,\mathbf{m}_{0,0,0}, \mathbf{m}_{1,1,1},
\boldsymbol{x})$ where $\mathbf{m}_{z_1, z_2, z_3}$ is a vector of
hypothetical values of the mediators under the interventions $z_1, z_2, z_3$. The conditional distribution $[\bM(z_1,z_2,z_3)|\boldsymbol{X}=\boldsymbol{x}]$
is denoted by $f_{\bM(z_1,z_2,z_3)}(\mathbf{m}_{z_1,z_2,z_3} |
\boldsymbol{x})$. Other conditional distributions are defined analogously, and we henceforth omit conditioning on covariates $\bX = \boldsymbol{x}$ to simplify notation. 

\subsubsection{Assumptions for principal causal effects}\label{sec:assumptions_ps}
We begin with an ignorability assumption stating that, conditional on covariates, ``assignment'' to scrubbers is unrelated to the observable potential outcomes:
\noindent Assumption 1 {\bf (Ignorable treatment assignment)} $$\{Y(z; \bM(z,z,z)),
  \bM(0,0,0), \bM(1,1,1) \}\independent Z | \boldsymbol{X}=\boldsymbol{x},$$ for
$z = 0,1$.  This assumption permits estimation of the distributions of potential outcomes under intervention $Z=z$ with observed data on ambient \PMTwo\, and emissions under the same intervention.

We adopt a Gaussian copula model to link the distributions of $( M_1(z), \allowbreak M_2(z), M_3(z))$ for $z=0,1$ into a single joint distribution of observable potential outcomes.

\noindent Assumption2:  The joint distribution of all potential mediators conditional on covariates follows a Gaussian copula model \citep{nelsen_introduction_1999}:
\begin{small}
\begin{eqnarray*}
 \lefteqn{F_{\bM(0,0,0),\bM(1,1,1)}(\mathbf{m}_{0,0,0}, \mathbf{m}_{1,1,1}) =} \\
 & &
 \Phi_{6}[\Phi_{1}^{-1}\{F_{M_1(0)}(m_1)\},\Phi_{1}^{-1}\{F_{M_2(0)}(m_2)\},
 \Phi_{1}^{-1}\{F_{M_3(0)}(m_3)\},\Phi_{1}^{-1}\{
 F_{M_1(1)}(m_1)\},\\
& &\Phi_{1}^{-1}\{F_{M_2(1)}(m_2)\},\Phi_{1}^{-1}\{F_{M_3(1)}(m_3)\}]
\end{eqnarray*}
\end{small}
where $\Phi_{6}$ is the multivariate normal CDF with mean $\mathbf{0}$ and a correlation matrix $\boldsymbol{R}$.

Assumption 2 implies a joint distribution of all observable potential mediators in a manner consistent with the models for  $[M_1, M_2, M_3 | Z=z, X=x]$ described in Section \ref{sec:observedmodels}. However, this entire joint distribution of potential mediators under both interventions is not fully identified from the data since potential mediators under different interventions are never jointly observed.  Specifically, entries of the correlation matrix $\boldsymbol{R}$ corresponding to, for example, the correlation between  $M_j(0)$ and $M_k(1)$, are not identifiable in the sense that no amount of data can estimate unique values for these parameters.  Nonetheless, proper prior distributions for these parameters can still permit inference from proper posterior distributions.  Such parameters are sometimes referred to as ``partially identifiable'' in the sense that increasing amounts of data may lead the supports of posterior distributions to converge to sets of values that are smaller than those specified in the prior distribution \citep{gustafson2010bayesian, mealli2013using}.  This can arise due to restrictions on the joint distribution implied by the models for the marginal distributions (e.g., the positive-definiteness restriction on $\boldsymbol{R}$ may exclude some possible values for its entries).  We discuss two prior specifications for the partially-identified parameters in $\boldsymbol{R}$, noting that further details of partial identifiability in the principal stratification context appear in \cite{schwartz2011bayesian}.


\subsubsection{Assumptions for Mediation Effects}\label{sec:assumptions_med}
Towards estimation of natural direct and indirect effects, we augment the assumptions of Section \ref{sec:assumptions_ps} with one relating observable outcomes to {\it a priori} counterfactual outcomes. 

\noindent Assumption 3: For intervention $Z=1$, the conditional distribution of the potential outcome given values of all potential mediators (and covariates) is the same regardless of whether the mediator values were induced by $Z=1$ or $Z=0$.

This assumption implies that the {\it a priori} counterfactual $Y(1; \bM(0, 0, 0))$ and  the observable potential outcomes $Y(1; \bM(1,1,1))$ have the same conditional distribution,
\begin{eqnarray*}
\lefteqn{f_{1,\bM(0,0,0)}(y\,|\,\bM(0,0,0)=\mathbf{m},
  \bM(1,1,1), \boldsymbol{x}) }\\
  & = & f_{1,\bM(1,1,1)}(y\,|\,\bM(0,0,0),
\bM(1,1,1) =\mathbf{m}, \boldsymbol{x}) . 
\end{eqnarray*}

This assumption also applies to any two mediators in the absence of the intervention. For
instance, the {\it a priori} counterfactual of $\text{PM}_{2.5}$, $Y(1; \bM(0, 1, 0))$, and $Y(1; \bM(1,
1, 1))$ have the same conditional distribution regardless of whether
corresponding emissions values arose under a scrubber
($Z=1$) or absent a scrubber ($Z=0$),
\begin{eqnarray*}
\lefteqn{f_{1,\bM(0,1,0)}(y\,|\,\bM(0,1,0)=\mathbf{m},
  \bM(1,0,1) , \boldsymbol{x}) }\\
& = &f_{1,\bM(1,1,1)}(y\,|\,\bM(0,0,0),\bM(1,1,1)=\mathbf{m}, \boldsymbol{x}).
\end{eqnarray*}



The key point is that the distribution of \PMTwo\, under a given (unobservable) combination of mediators ($\mathbf{m}$) only depends on the values of the mediators and not the intervention that led to those mediators.  Asserting this assumption in this case relies in part on what is known about the underlying chemistry relating \SOTwo, \NOx, and \COTwo\, emissions to \PMTwo.  Note that such an assumption may be more difficult to justify in, say, a clinical study where assumptions about {\it a priori} counterfactuals might pertain to choices of study participants.

The above assumption can be cast as two homogeneity assumptions of the form proposed in \cite{Forastiere:2016}. For example, one implication of Assumption 3 is that the {\it a priori} counterfactual $Y(1; M(0,0,0))$ is homogeneous across all principal strata with $M(0,0,0)=\mathbf{m}$, regardless of the value of $M(1,1,1)$.  Viewing Assumption 3 in terms of the implied homogeneity across principal strata aids interpretation and justification in the context of the power plant example.  Homogeneity across strata implies that the potential ambient air quality value in the area surrounding a power plant is related to (possibly counterfactual) emission levels only, and not to the power plant characteristics that govern effectiveness of scrubbers for reducing emissions (i.e., the power plant characteristics that determine the exact principal stratum membership). This underscores the importance of including covariates in $\boldsymbol{X}$ that capture characteristics of the monitoring locations (e.g., temperature and barometric pressure). Appendix D provides details of the relationship between Assumption 3 and assumptions of homogeneity across principal strata.  While Assumption 3 implies homogeneity assumptions, the converse is not true in the case of multiple mediators due to the connection of Assumption 3 to {\it a priori} counterfactuals defined to have mediator values induced by different interventions (e.g., $Y(1; M(0,1,0)$). We discuss a sensitivity analysis to this assumption in Web Appendix J.

\subsubsection{Optional Modeling Assumptions to Sharpen Posterior Inference}\label{sec:correlation}
With the above model specification, the partial identifiability of the model parameters in $\boldsymbol{R}$ warrants careful attention.  Proper but noninformative prior distributions for these parameters could be specified marginally for these parameters as $\text{Unif}(-1,1)$, or equivalently, as conditionally uniform on intervals satisfying positive definiteness restrictions for the correlation matrix.  In either case, posterior inference may exhibit large uncertainty. 

We consider in detail an alternative prior specification similar to that in \cite{zigler_estimating_2012} to sharpen posterior inference.  Specifically, the correlations between mediators under different interventions are specified as follows:
\begin{small}
\begin{equation*}
\text{Cor}(M_j(0), M_k(1)) = \frac{\text{Cor}(M_j(0), M_k(0))+\text{Cor}(M_j(1), M_k(1))}{2}\times
\rho, \,\, \text{ for   } \,\,  j, k=1, 2, 3, \label{corr1}
\end{equation*}
\end{small}
with $\rho$ a sensitivity parameter.  This strategy implies that
(a) the correlation between the same mediator $(j = k)$ under opposite
interventions is $\rho$, and (b) the correlation between different mediators $(j\neq k)$
under opposite interventions is an attenuated version of the correlation
observed separately under each intervention.  Section B of the Web Appendix provides a correlation matrix implied by this assumption in the case of 2 mediators. We assume a single $\rho$ and specify a uniform prior distribution, $\rho \sim \text{Unif}(0,1)$, but a different parameter could be specified for each mediator.   

As an additional assumption to sharpen posterior inference, we assume that the correlations between emissions (mediators) are all positive. Support for this assumption comes from observed-data estimates of these conditional correlations that are all positive.

In summary, assumptions 1-2 are sufficient to estimate the principal causal effects, and pertain only to observable potential outcomes.  Adding assumption 3 relating observed quantities to {\it a priori} counterfactuals permits estimation of direct and indirect effects for mediation analysis.  The optional assumptions here in Section \ref{sec:correlation} are designed to sharpen posterior inference in the power plant analysis.  

\subsubsection{Posterior Inference}
A Markov chain Monte Carlo (MCMC) algorithm is used to sample from this posterior distribution and estimate causal effects using the following steps: (1) sampling parameters from each marginal distribution for potential mediators and conditional distribution for potential outcomes defined in Section \ref{sec:observedmodels}; (2) sampling parameters from the correlation matrix $R$ of the Gaussian copula;  (3) sampling via data augmentation {\it a priori} counterfactual mediators from the joint distribution; (4) computing causal effects based on all potential mediators and outcomes including imputed {\it a priori} outcomes and mediators; (5) iterate Steps 1-4. The specifics of estimation (conditional on our specific model formulation) are based on the existing literature on Bayesian estimation of causal effects (and principal causal effects in particular), for example, in \cite{mattei2011augmented,zigler_estimating_2012,Dani:Roy:Kim:Hoga:Perr:2012}. 

The Web Appendix contains details of the MCMC procedure (Section F), prior specification for all other model hyper-parameters (Section A), and the procedure for computing the principal causal effects and  the mediation effects from the posterior distributions of model parameters (Section C). 

\section{Numerical Study}\label{sec:sim}
We examine the performance of the proposed model under combinations of the following two data generating scenarios: (1) correlations among the mediators (Case 1: uncorrelated mediators vs. Case 2: correlated mediators) and (2) interaction terms between the mediators in the outcome model (Case A: interaction term between $M_1$ and $M_2$ vs. Case B: interaction terms between $M_1$ and $M_2$, and between $M_2$ and $M_3$). Data sets of size $n=500$ are simulated for each of the four cases (1/A, 1/B, 2/A, 2/B), each with  three continuous confounders. In all cases, the three mediators are generated based on a multivariate normal distribution. See the Web Appendix (Section G) for the exact data generating mechanism.

We compare our method for estimating mediation effects to a regression-based model\citep{MacK:2008}:
\begin{eqnarray*}
M_1 &=& \alpha_{01}+\alpha_{11}Z+\boldsymbol{X}^\top  \boldsymbol{\alpha}_1+\epsilon_1\\
M_2 &=& \alpha_{02}+\alpha_{12}Z+\boldsymbol{X}^\top  \boldsymbol{\alpha}_2+\epsilon_2\\
M_3 &=& \alpha_{03}+\alpha_{13}Z+\boldsymbol{X}^\top  \boldsymbol{\alpha}_3+\epsilon_3\\
Y &=& \beta_{0}+\beta_{1}Z+\beta_{2}M_1+\beta_{3}M_2+\beta_{4}M_3+\boldsymbol{X}^\top  \boldsymbol{\beta}+\epsilon_Y
\end{eqnarray*}
where $\epsilon_1, \epsilon_2, \epsilon_3,$ and $\epsilon_Y$ are all independently distributed as $N(0,\sigma)$. 

Table \ref{table:sim} summarizes the results based on 400 replications for each of the four scenarios. It shows that our proposed model (BNP)
performs well in terms of bias and MSE for all cases. Note that the true effects change when the mediators are correlated in the presence of interaction term(s) in the outcome model. Thus, with any interaction effects of the mediators, it is desirable to capture the correlation structure of the mediators, which our method does by flexibly modeling the joint distribution of all potential mediators. Also, the flexible Bayesian nonparametric model can capture both complex relationships/interactions among the mediators and non-additive and nonlinear forms of mediators and/or confounders in the outcome model. In each scenario, interaction terms in the outcome model introduce non-additivity in the joint natural indirect effect (e.g., $\text{JNIE}\neq \text{NIE}_1+\text{NIE}_2+\text{NIE}_3$) and the traditional regression model has larger biases (and larger MSEs) for mediation effects.


\begin{table}[h]
\centering
\caption{Simulation results for point estimators of causal mediation and principal causal effects over 400 replications. The columns correspond to bias and MSE relative to the true values of the causal effects for each scenario (Cases 1 and 2, and Cases A and B)  under two different models; {\bf Parametric} : a regression based model for the causal mediation effects; {\bf BNP} : Our Bayesian nonparametric method.}
\resizebox{\textwidth}{!}{  
\begin{tabular}{c|c|c|rcrc|c|rcrc} \hline \hline
\multicolumn{1}{c}{} & \multicolumn{1}{c}{} & \multicolumn{5}{|c|}{\bf \large Case 1} & \multicolumn{5}{c}{\bf \large Case 2}\\ \cline{3-12} 
\multicolumn{1}{c}{} & \multicolumn{1}{c}{}& \multicolumn{1}{|c|}{} & \multicolumn{2}{c}{BNP} & \multicolumn{2}{c|}{Parametric} & \multicolumn{1}{|c|}{Truth} & \multicolumn{2}{c}{BNP} & \multicolumn{2}{c}{Parametric} \\
\multicolumn{1}{c}{} & \multicolumn{1}{c}{}& \multicolumn{1}{|c|}{Truth} & Bias & MSE & Bias & MSE & \multicolumn{1}{|c|}{Truth} & Bias & MSE & Bias & MSE \\
 \hline
 \multirow{9}{*}{\bf \large Case A} & TE & 0.73& \cellcolor{blue!25} 0.02& (0.09) & \cellcolor{red!25} -0.03 &(0.08)  & 0.92 & \cellcolor{blue!25} -0.04 & (0.08) & \cellcolor{red!25} \bf 0.20 & (0.33)\\
 & JNIE & 1.73& \cellcolor{blue!25} 0.06& (0.11) &\cellcolor{red!25} \bf 0.21 &(0.07) & 1.92& \cellcolor{blue!25} 0.04 & (0.08) & \cellcolor{red!25}0.02 & (0.47)\\
 & NDE & -1 &\cellcolor{blue!25} -0.04 &(0.01) & \cellcolor{red!25} \bf -0.25 &(0.15) & -1 &\cellcolor{blue!25} -0.08 & (0.01)  & \cellcolor{red!25} \bf -0.20 & (0.08)\\
 & NIE$_1$ & -0.16& \cellcolor{blue!25} 0.00 &(0.00) & \cellcolor{red!25} -0.01 &(0.01)& 0.03& \cellcolor{blue!25} -0.05& (0.00) & \cellcolor{red!25}\bf -0.38 & (0.26)\\
 & NIE$_2$ & 2.45&\cellcolor{blue!25} 0.02 &(0.10) &\cellcolor{red!25} -0.02 &(0.08)& 2.65&\cellcolor{blue!25} -0.05 & (0.08) &\cellcolor{red!25}\bf -0.39 & (0.31)\\
 & NIE$_3$ & -0.32& \cellcolor{blue!25}0.00 &(0.00) &\cellcolor{red!25} -0.01 &(0.01)& -0.32& \cellcolor{blue!25}0.01 & (0.00) & \cellcolor{red!25} -0.01 & (0.01)\\
 & JNIE$_{12}$ & 2.05& \cellcolor{blue!25} 0.05 &(0.10) & \cellcolor{red!25} \bf0.22 &(0.14)& 2.23&\cellcolor{blue!25} 0.03 & (0.08) & \cellcolor{red!25} \bf 0.21& (0.44)\\
 & JNIE$_{13}$ & -0.48& \cellcolor{blue!25} 0.01 &(0.01) &\cellcolor{red!25} -0.01 &(0.01)& -0.29&\cellcolor{blue!25} -0.04 & (0.00) & \cellcolor{red!25} \bf -0.38& (0.28)\\
 & JNIE$_{23}$ & 2.13& \cellcolor{blue!25} 0.02 &(0.10) &\cellcolor{red!25} -0.02 &(0.09)& 2.33& \cellcolor{blue!25} -0.04 & (0.08) & \cellcolor{red!25}\bf -0.39 & (0.33)\\ \cline{2-12}
 \hline 
 \multirow{9}{*}{\bf\large Case B} & TE & 1.08& \cellcolor{blue!25} -0.02 & (0.10) & \cellcolor{red!25} -0.01 & (0.08) & 1.33 &\cellcolor{blue!25} -0.09& (0.08)& \cellcolor{red!25}-0.01 & (0.08) \\
 & JNIE &2.08 &\cellcolor{blue!25} -0.00 & (0.10)& \cellcolor{red!25}\bf 0.16& (0.12) & 2.33 &\cellcolor{blue!25} -0.00 & (0.08) & \cellcolor{red!25}-0.08 &(0.11) \\
 & NDE & -1& \cellcolor{blue!25} -0.01 & (0.00)&\cellcolor{red!25} \bf -0.17 & (0.04)& -1&\cellcolor{blue!25} -0.09 & (0.01) & \cellcolor{red!25}0.08& (0.02)\\
 & NIE$_1$ & -0.16& \cellcolor{blue!25} -0.01 & (0.00) &\cellcolor{red!25} -0.01 & (0.01)& 0.03 &\cellcolor{blue!25} -0.05 & (0.01) &\cellcolor{red!25} \bf -0.20& (0.04) \\
 & NIE$_2$ & 2.51& \cellcolor{blue!25} -0.02 & (0.10) &\cellcolor{red!25} 0.02 & (0.09) &2.78&\cellcolor{blue!25} -0.08& (0.09)& \cellcolor{red!25}\bf -0.25 & (0.15) \\
 & NIE$_3$ & -0.13& \cellcolor{blue!25} 0.00 & (0.00) & \cellcolor{red!25}0.01 & (0.01)& -0.05 &\cellcolor{blue!25}-0.02 & (0.01)&\cellcolor{red!25} -0.08& (0.01) \\
 & JNIE$_{12}$ & 2.11 &\cellcolor{blue!25} 0.01& (0.10) & \cellcolor{red!25}\bf 0.25 & (0.16) &2.37& \cellcolor{blue!25}0.00 & (0.08)& \cellcolor{red!25}0.01 &(0.10) \\
 & JNIE$_{13}$ & -0.29&\cellcolor{blue!25} -0.00& (0.00) &\cellcolor{red!25} -0.01& (0.01)& -0.02&\cellcolor{blue!25} -0.07 & (0.01)& \cellcolor{red!25}\bf-0.27& (0.08) \\
 & JNIE$_{23}$ & 2.48&\cellcolor{blue!25} -0.04& (0.10) &\cellcolor{red!25} -0.08 & (0.09) & 2.75&\cellcolor{blue!25} -0.09 & (0.09)& \cellcolor{red!25}\bf-0.34& (0.21) \\ \cline{2-12}
 \hline 
\end{tabular}\label{table:sim}
}
\end{table}

\section{Analysis of Power Plant Scrubbers in the Acid Rain Program}\label{sec:dataanalysis}
Here we estimate causal effects of having scrubbers installed in January 2005 ($Z$) on annual average emissions of \SOTwo, \NOx, and \COTwo\, in 2005 ($M_1, M_2, M_3$) and on the 2005 annual average ambient \PMTwo\, concentration within 150km of a power plant ($Y$).  Emissions are log-transformed. 
Before reporting results, note that basic checks of the fit of marginal nonparametric models appear in Web Appendix I, indicating  fit that is clearly superior to simple parametric models.

A simple comparison of means indicates that the 150km area around power plants with scrubbers installed ($Z=1$) had average ambient \PMTwo\, that was lower, on average, than the areas surrounding power plants without scrubbers (12.4 vs. 13.7 $\mu g/m^3$).  Similarly, the power plants with scrubbers also emitted less \SOTwo, more \NOx, and more \COTwo\, than the plants without scrubbers.  Table \ref{Data} lists the covariates in $X$ to adjust for confounding and presents summary statistics for scrubber and non-scrubber power plants.

We present an analysis with the proposed method using the constrained prior specification in Section \ref{sec:correlation}. Analysis using uniform prior distributions on all elements of the correlation matrix appears in the Web Appendix.  All reported estimates are listed as posterior means (95\% posterior intervals).  The analysis estimates that having scrubbers installed causes \SOTwo\, emissions to be -1.17 (-1.86, 1.55) 1000 tons lower, on average, than they would be without the scrubber.  The analogous causal effects for \NOx\, and \COTwo\, emissions were  0.04 (0.00, 0.07) 1000 tons and 0.001 (-0.00, 0.004) million tons, respectively, indicating that scrubbers did not significantly affect these emissions, on average.  The total effect (TE) of having scrubbers installed on ambient \PMTwo\, within 150km is estimated to be -1.12 (-2.07, -0.29) $\mu g/m^3$, suggesting  a reduction amounting to approximately 10\% of the national annual regulatory standard for \PMTwo. 

\subsection{Principal Causal Effects}
For the $k$-th emission, let $\sigma_k$ denote the posterior standard deviation of the estimated individual-level causal effect of a scrubber on $M_k$, with posterior mean estimates $\hat{\sigma}_1 = 0.24$, $\hat{\sigma}_2 = 0.42$, $\hat{\sigma}_3 = 0.02$.  Let $\hat{\sigma}_\mathcal{K}$ denote the vector of $\hat{\sigma}_k$ for the emissions in $\mathcal{K}$.  To summarize dissociative effects, we set $C_{\mathcal{K}}^D = 0.25 \hat{\sigma}_\mathcal{K}$ to estimate EDE$_\mathcal{K}$ among power plants where the scrubber effect on emissions in $\mathcal{K}$ is within one-fourth of a standard deviation of the effect in the population. Similarly, we summarize associative effects with $C_{\mathcal{K}}^A = 0.25\hat{\sigma}_\mathcal{K}$ to estimate EAE$^-_\mathcal{K}$ (EAE$^+_\mathcal{K}$) among power plants where the scrubber causally reduces (increases) emissions in $\mathcal{K}$ more than one-fourth of a standard deviation of the effect in the population. 

\begin{figure}[p]
\subfloat[$k=1$ (\SOTwo)]{
  \includegraphics[width=.5\textwidth]{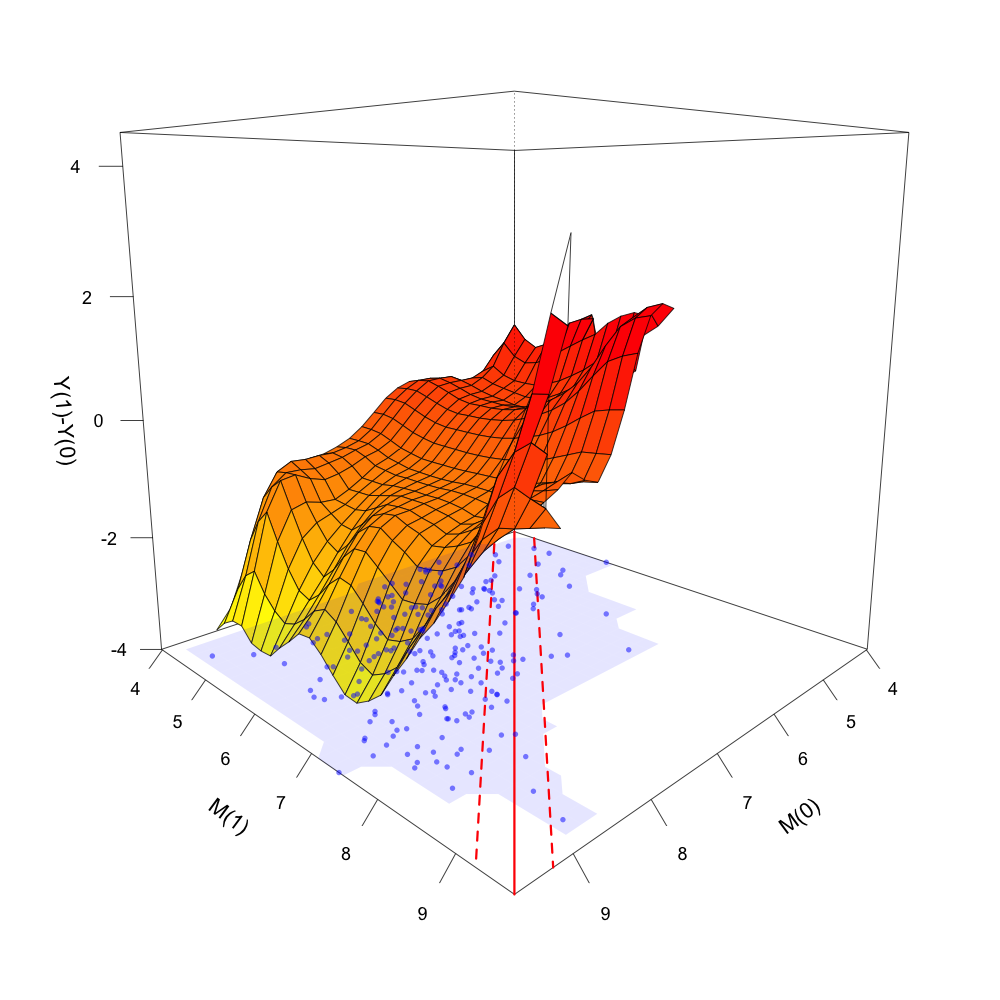}\label{so2}}
\subfloat[$k=2$ (\NOx)]{
  \includegraphics[width=.5\textwidth]{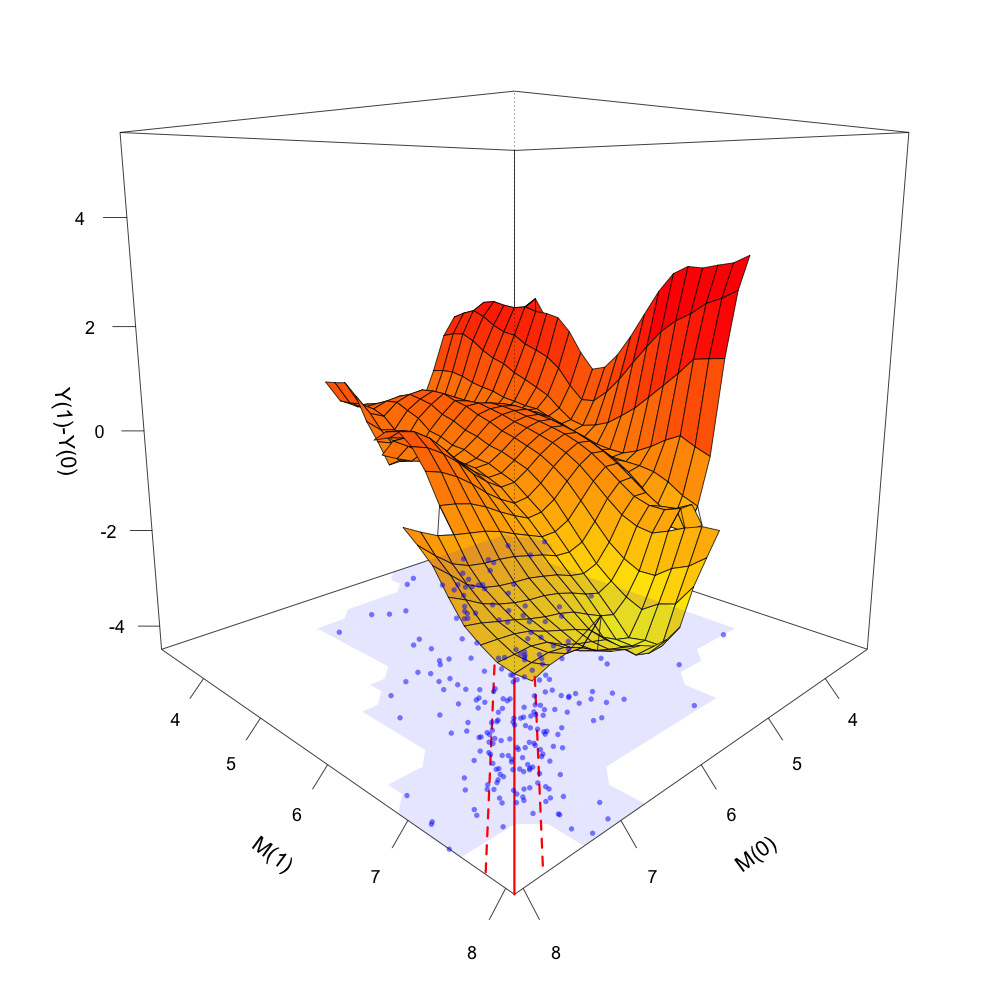}\label{nox}}\\
\subfloat[$k=3$ (\COTwo)]{
  \includegraphics[width=.5\textwidth]{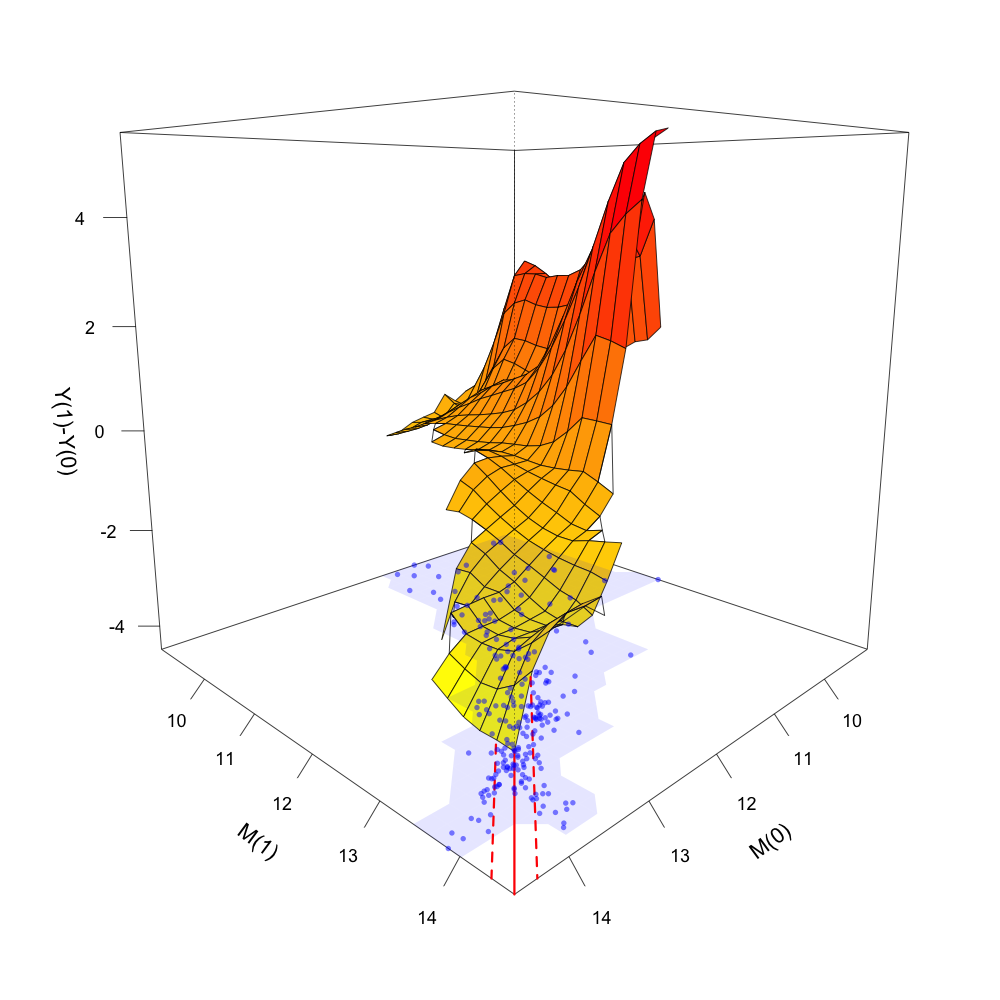}\label{co2}}
\caption{Average surface plots of the causal effect on \PMTwo\, for different values (log scales) of  $(M_k(0), M_k(1))$.  Values of $(M_k(0), M_k(1))$ are plotted on the $x$- and $y$- axes, and determine the causal effect of a scrubber on emission $k$.  The corresponding value of the causal effect of a scrubber on \PMTwo, $Y(1)-Y(0)$, is plotted on the $z$-axis.  The cloud of points in the $xy$-plane are one MCMC draw of 249 pairs of $(M_k(0), M_k(1))$.  Red lines are at $M_k(0) =  M_k(1)$ (solid line) and $+/- 0.25 \hat{\sigma}_k$ (dashed lines). }
\label{fig6}
\end{figure}

Before providing estimates of specific principal effects, we first examine 3-D surface plots in Figure \ref{fig6}.  For each emission separately ($k \in \{1,2,3\}$), Figure \ref{fig6} depicts estimated scrubber effects on \PMTwo\, across varying effects on emissions determined by values of $(M_k(0), M_k(1))$ simulated from the model.   Note the pattern for all emissions that the surfaces are sloped downward in the direction of increasing $M_k(0)$ and $M_k(1)$ (sloped towards the viewer), indicating larger effects on \PMTwo\, among plants with larger emissions values under both scrubber statuses, i.e., larger plants. 

In Figure \ref{fig6}\subref{so2} for \SOTwo, the dots in the $xy$-plane lie almost entirely in the region where $M_1(1) < M_1(0)$, indicating as expected that scrubbers predominantly decrease \SOTwo\, emissions.  Associative effects for \SOTwo\, are indicated by the downward slope of the surface in the direction of decreasing $M_1(1) - M_1(0)$ (towards the left of the viewer), indicating that larger decreases (increases) in \SOTwo\, are associated with larger decreases (increases) in \PMTwo.  

The analogous surfaces for \NOx\, and \COTwo\, appear in Figures \ref{fig6}\subref{nox} and \ref{fig6}\subref{co2}, respectively.  In contrast to the surface for \SOTwo, the dots in the $xy$-plane fall more closely and symmetrically around the line $M_k(1) = M_k(0)$, reflecting that scrubbers do not affect these emissions, on average.  The surface for \NOx\, exhibits some evidence of associative effects in the opposite direction of those for \SOTwo; there is some downward slope of the surface in the direction of increasing $M_k(1) - M_k(0)$ (towards the right of the viewer), indicating that larger increases (decreases) in these emissions are associated with larger decreases (increases) in \PMTwo.  

Table \ref{tab:ps} lists posterior mean and standard deviation of EDE, EAE$^-$, and EAE$^+$ for all possible $\mathcal{K}$.  
\begin{table}[h]
\centering 
\caption{Posterior means (standard deviations) for expected associative and dissociative effects of \SOTwo\, scrubbers.}\label{tab:ps}
\resizebox{\textwidth}{!}{  
\begin{tabular}{c|c|ccccccc}
& & \SOTwo & \NOx & \COTwo & \SOTwo\, \& \NOx & \SOTwo\, \& \COTwo & \NOx\, \& \COTwo & \SOTwo\, \& \NOx\, \& \COTwo \\
\hline\hline
\multirow{2}{*}{EAE$^-$}& Mean &  -1.19 & -0.77 & -1.14 & -0.84 & -1.18 &  -0.90 & -0.94 \\
& SD & (0.46) & (0.59) & (0.56) & (0.59) & (0.57) & (0.67) & (0.68)\\
\hline
\multirow{2}{*}{EDE}& Mean &  -0.32 & -0.69 & -0.82 & -0.09 & -0.31 & -0.48 & -0.15 \\
& SD & (0.57) & (0.54) & (0.49) & (0.71) & (0.68) & (0.69) & (0.86)\\
\hline
\multirow{2}{*}{EAE$^+$}& Mean & 0.60 & -1.68 & -1.08 & 0.38 & 1.28 & -1.63 &  0.69 \\
& SD & (2.52) & (0.74) & (0.75) & (3.67) & (3.78) & (1.04) & (4.68)\\
\hline
\end{tabular}}
\end{table}
Estimates of EDE for all $\mathcal{K}$ indicate little to no reduction in \PMTwo\, among plants where emissions were not affected in excess of $\mathcal{C}^D_\mathcal{K}$, with the exception of some pronounced estimates of EDE for $\mathcal{K} = \{$\NOx$\}$ and $\mathcal{K} = \{$\COTwo$\}$.  Estimates of EAE$^-$  and EAE$^+$ tend to be less than zero.  The most pronounced estimate of EAE$^-_\mathcal{K} =$ -1.19 (0.46) for $\mathcal{K} = \{$\SOTwo$\}$ suggests that \PMTwo\, was reduced among power plants where \SOTwo\, emissions were substantially reduced, which corresponds to the contour of the surface in Figure \ref{fig6}\subref{so2} and is consistent with the anticipated causal pathway whereby scrubbers reduce \PMTwo\, through reducing \SOTwo\, emissions.  In accordance with the opposite sloping surface in Figures \ref{fig6}\subref{nox}, the estimate of EAE$^+_\mathcal{K}$ is most pronounced for $\mathcal{K} = \{$\NOx$\}$, and $\{$\NOx, \COTwo$\}$, indicating that ambient \PMTwo\, is decreased among plants with substantial {\it increases} in \NOx\, emissions.  

\begin{figure}[t]
\centering
\scalebox{0.5}
{\includegraphics{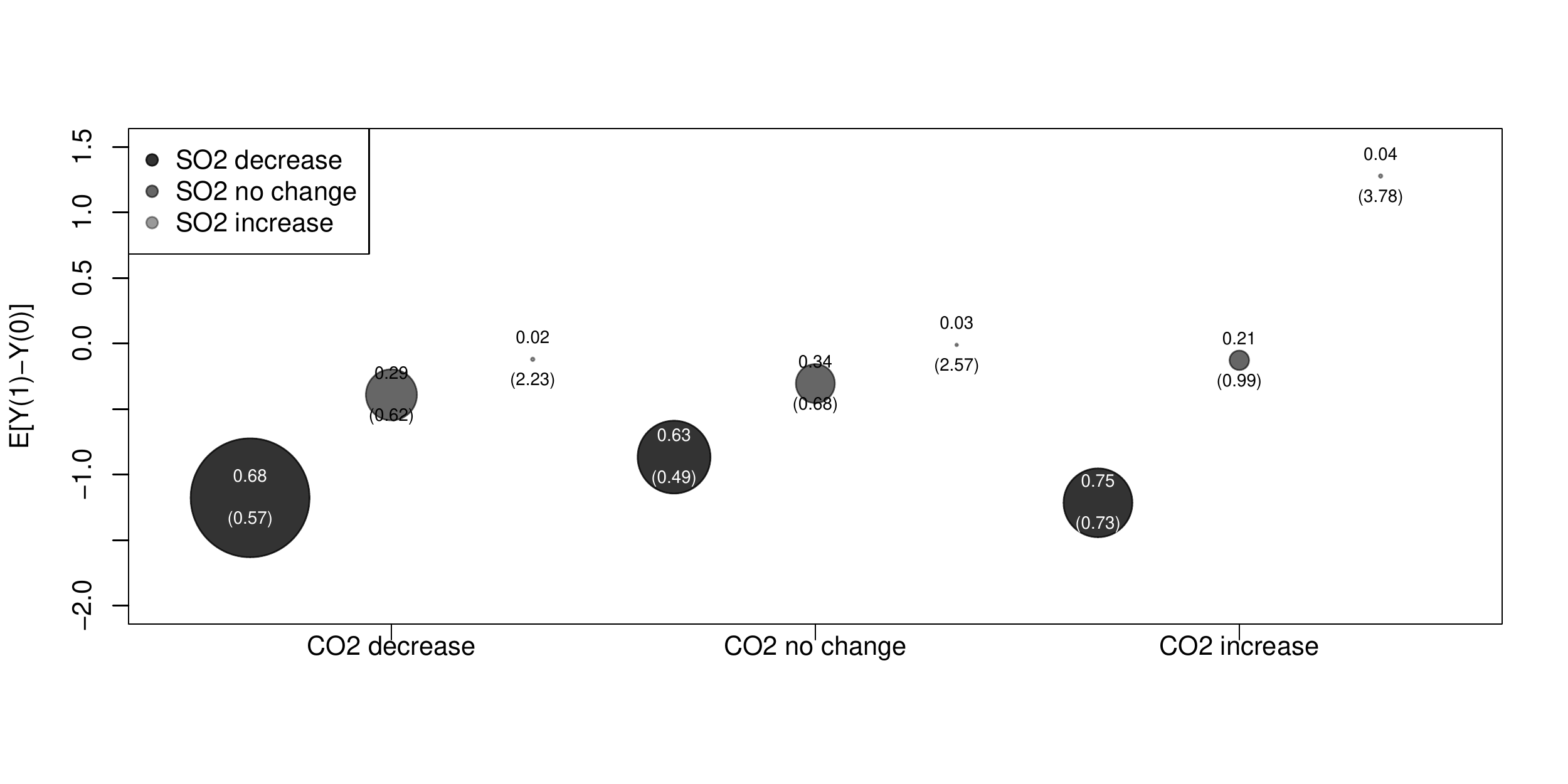}} 
\caption{Posterior mean estimates of principal effects for strata defined by cross-classifying changes in \COTwo\, ($x$-axis) and changes in \SOTwo\, (colored circles).  Size of circle symbolizes the proportion of each \COTwo\, stratum falling in the corresponding \SOTwo\, category, and number (and number in parentheses) listed is posterior mean proportion (and posterior standard deviation).}
\label{fig:ps}
\end{figure}

Recall that the estimates in Table \ref{tab:ps} represent average principal effects over only a subset of principal strata, in particular those where changes in multiple emissions are concordant (i.e., all decreasing, all increasing, or none changing). Other strata may be of interest.  Figure \ref{fig:ps} provides estimates of principal effects in a cross-classification of strata defined by changes in \COTwo\, and \SOTwo, with changes defined as increases, decreases, or no change in reference to $C_{\mathcal{K}}^D$ and $C_{\mathcal{K}}^A$. For example, the third column of Figure \ref{fig:ps} subdivides the stratum defined by causal increases in \COTwo\, into three substrata: those where \COTwo\, increases and \SOTwo\, (1) decreases (in excess of $C_{\mathcal{K}}^A$); (2) does not substantially change (beyond $C_{\mathcal{K}}^D$ ) ; or (3) increases (in excess of $C_{\mathcal{K}}^A$).  Principal causal effect estimates for these three substrata appear along with their relative proportion among the stratum defined by \COTwo\, increases, indicated by the size of the plotting symbol.  The light grey dot  corresponds to EAE$_\mathcal{K}^+$ for $\mathcal{K} = \{$\SOTwo, \COTwo$\}$ as reported in Table \ref{tab:ps}, but note that only 4\% of the \COTwo-increase stratum exhibits \SOTwo\, increases.  The dark grey dot corresponds to the principal effect among the 21\% of the \COTwo-increase stratum in substratum (2) where \SOTwo\, does not change, with a principal effect estimate of -0.13 (0.99). The remaining proportion (75\%) of the \COTwo-increase stratum belongs to substratum (3) where the plants exhibiting decreases in \SOTwo\, and a corresponding principal effect estimate of -1.21 (0.73).  Thus, for $\mathcal{K}=\{$\COTwo$\}$, the negative estimate of EAE$_\mathcal{K}^+$ from Table \ref{tab:ps} is revealed to be generated in large part by strata where \SOTwo\, decreases and there is a pronounced negative effect on \PMTwo.  Analogously, the second column of Figure \ref{fig:ps} considering the stratum where \COTwo\, emissions do not substantially change (used to estimate EDE) reveals that 63\% of this strata exhibited causal reduction in \SOTwo\, and a causal reduction in \PMTwo\, of -0.87 (0.49), explaining in large part the negative estimate of EDE$_\mathcal{K}$ for $\mathcal{K}=\{$\COTwo$\}$ in Table \ref{tab:ps}.   Analogous cross-classification of strata by changes in \NOx\, and \SOTwo\, appears very similar to Figure \ref{fig:ps} and is not presented.

The main conclusions from the principal stratification analysis are that 1) scrubbers reduce \SOTwo\, on average, but not \NOx\, or \COTwo, 2) there is some evidence of a nonzero dissociative effect for \SOTwo, 3) associative effects for \SOTwo\, are more pronounced than dissociative effects, with \PMTwo\, reduced more around plants where scrubbers cause large reductions in \SOTwo; 4) associative effects for \NOx\, and \COTwo\, are more pronounced than dissociative effects, with \PMTwo\, reduced more around plants where scrubbers cause larger {\it increases} in these emissions; but that 5) strata defined by increases (or no change) in \NOx\, and/or \COTwo\, are comprised in large part by substrata where \SOTwo\, and \PMTwo\, were causally reduced. This analysis points towards (but cannot confirm) the conclusion that scrubbers affect \PMTwo\, among plants where emissions are not changed, and that scrubber effects on \PMTwo\, are mediated in part through effects on \SOTwo, with less evidence of a mediating role of \NOx\, and \COTwo.

\subsection{Mediation Effects}
To estimate direct and indirect effects, we augment the principal stratification analysis with Assumption 3 in Section \ref{sec:assumptions_med} about {\it a priori} counterfactuals.  Figure 1 (top) in the Web Appendix depicts boxplots of the posterior distributions of TE, NDE, JNIE$_{123}$, JNIE$_{12}$, JNIE$_{23}$,  JNIE$_{13}$, and the individual NIEs. The estimated NDE, representing the direct effect of a scrubber on ambient \PMTwo\, that is not mediated through any emissions changes, is  -0.53 (-1.51, 0.39) $\mu g /m^3$, indicating no evidence of a direct effect of scrubbers on \PMTwo\, that is not mediated through \SOTwo, \NOx, or \COTwo. 
The NIEs for \NOx\, (NIE$_2$) and \COTwo\, (NIE$_3$) are estimated to be very close to 0, -0.02 (-0.26, 0.21) and -0.04 (-0.33, 0.23), respectively.  The estimated NIE for \SOTwo\, (NIE$_1$) is -0.54 (-1.20, -0.01), indicating a significant indirect effect.  The joint natural indirect effects involving \SOTwo\, are all similar in magnitude to NIE$_1$, with estimates of JNIE$_{12}$, JNIE$_{13}$, and JNIE$_{123}$ of -0.56 (-1.23, -0.01), -0.58 (-1.25, -0.02), and -0.59 (-1.27, -0.02), respectively.  The estimated JNIE$_{23}$ is -0.03 (-0.31, 0.23).


As discussed in Section \ref{sec:mediationeffects}, a benefit of the proposed approach is the accommodation of overlap between NIEs, and the opportunity to examine the extent of overlap. We evaluate the relationship between the joint effects $\text{JNIE}_{jk}$ and the mediator-specific effects $\text{NIE}_1$, $\text{NIE}_2$,
$\text{NIE}_3$ through 
$(\text{NIE}_1+\text{NIE}_2) - \text{JNIE}_{12}   =  -0.01 \,(-0.18, 0.16)$, 
$(\text{NIE}_1+\text{NIE}_3) - \text{JNIE}_{13}   =  0.01 \,(-0.22, 0.23)$ and 
$(\text{NIE}_2+\text{NIE}_3) - \text{JNIE}_{23}   =  0.00 \,(-0.19, 0.15)$,
which give no evidence of overlap between NIEs. That is,
the effect of a scrubber on ambient \PMTwo\, that is
mediated through emissions changes appears to be described by indirect effects that act additively and do not exhibit any apparent synergy that would lead to overlapping effects. The lack of overlapping indirect effects, combined with the fact that a) all indirect effects involving \SOTwo\, (NIE$_1$, JNIE$_{12}$, JNIE$_{13}$, and JNIE$_{123}$) are similar in magnitude and b) all indirect effects not involving \SOTwo\, (NIE$_2$, NIE$_3$, JNIE$_{23}$) are estimated to be zero, provides strong evidence that the effect of scrubbers on \PMTwo\, is primarily driven by effects on \SOTwo.

In the Web Appendix, we also conduct inference using flat priors on plausible values of the partially-identifiable parameters, and the estimates for the effects are similar to those in the main analysis. 

The conclusions of the causal mediation analysis are clear and mostly consistent with those from the principal stratification analysis: scrubber effects on ambient \PMTwo\, are almost entirely mediated through reductions in \SOTwo\, emissions.  Combining reductions in \SOTwo\, with reductions of \NOx\, and \COTwo\, does not significantly change the mediated effect.  In fact, \NOx\, and \COTwo\, appear to play no role in the causal effect of scrubbers on \PMTwo.  

\subsection{Results from Alternative Analyses}
We conduct two simpler analyses for comparison.  First, we implement separate single-mediator analyses using the methods described above with $K=1$.   Results are largely consistent with the multiple mediator analysis, as suggested by the apparent absence of overlapping effects.  For \SOTwo\, emissions, the total, indirect and direct effects are estimated to be -1.28 (-2.25, -0.62), -0.70 (-1.51, -0.04) and -0.58 (-1.35, 0.37), respectively. For \NOx\, emissions, the total, indirect and direct effects are estimated to be -1.21 (-2.05, -0.40), -0.04 (-0.32, 0.28) and -1.17 (-1.99, -0.32), respectively. With \COTwo\, emissions, the total, indirect and direct effects are estimated to be -1.22 (-1.98, -0.29), 0.03 (-0.26, 0.33) and -1.25 (-2.05, -0.30), respectively. Note that significant estimated direct effects for \NOx\, and \COTwo\, suggest pathways that are not through \NOx\, and \COTwo\, (i.e., the pathway through \SOTwo). 

  
For a second comparison, we conduct a multiple mediator analysis using a traditional regression approach to mediation with the same model in Section \ref{sec:sim}. The mediation effects are estimated to be $NIE_1 = \alpha_{11}\beta_2 =  -0.39 (95\% C.I. -1.11, 0.25), NIE_2 = \alpha_{12}\beta_3 =  -0.09 (95\% C.I. -0.44, 0.22), NIE_3 = \alpha_{13}\beta_4 = 0.08 (95\% C.I. -0.08, 0.35), NDE = \beta_1 = -0.18 (95\% C.I. -2.56, 0.11)$.  Thus, while these results are on average consistent with the results from the proposed methods, the estimate of the $NIE_1$ is not significant.  Note that this analysis explicitly assumes that the mediators do not interact with each other in the outcome model, implying an estimate of the joint indirect effect of all three mediators that is the sum of all three indirect effect (i.e., $JNIE_{123} = -0.40 (95\% C.I. -1.15, 0.34)$) which is also not significant. The discrepancy between the results of the traditional regression approach and ours is due to our flexible modeling strategy using Bayesian nonparametric methods (Dirichlet process mixtures) that even in presence of additivity, allows for nonlinearities and non-normal errors. 

\section{Discussion}\label{sec:discussion}
We have developed flexible Bayesian methods for principal stratification and causal mediation analysis in the presence of multiple mediating variables.  To accommodate the setting of multiple pollutants that are emitted contemporaneously and possibly interact with one another, we have developed methods to accommodate multiple contemporaneous and non-independent mediators. Bayesian nonparametric modeling approaches provided flexible models for the observed data (marginal distribution for each mediator and conditional distribution for the outcome under each intervention $z=0,1$), and linked observed data distributions to joint distributions of potential mediators using explicit and transparent assumptions about both observable and {\it a priori} counterfactuals.

A key feature of our approach is the integration of principal stratification and causal mediation analysis in a manner that relies on the same models for the observed data.  
Deployment of these methods in the power plant analysis represents, to our knowledge, the most comprehensive consideration of these two approaches and the implications of the results in the context of a single analysis. We use Assumption 3 to relate {\it a priori} counterfactual outcomes to observed outcomes, and show that this assumption implies homogeneity across principal strata, which aids interpretation.  This assumption also has close ties to that of sequential ignorability \citep{Imai:2010b}. Benefits of formulating Assumption 3 as done here include facilitation of a sensitivity analysis to this assumption following the general approach of \cite{Dani:Roy:Kim:Hoga:Perr:2012} and the aided interpretation implied by the relationship to homogeneity assumptions.  While a version of sequential ignorability relevant to the setting of multiple contemporaneous mediators with interactions and that can be used to identify each mediator-specific effect has not been previously formulated, Web Appendix E explores the relationship between our Assumption 3 and sequential ignorability in the case of a single mediator.  In this case, implications of these two assumptions are identical for the types of estimands considered here, although one assumption does not generally imply the other. 

The results of the principal stratification and causal mediation analyses should be interpreted jointly and are, in this case study, largely consistent with one another.  Principal stratification indicated that scrubbers tended to decrease ambient \PMTwo\, around plants where scrubbers substantially reduced \SOTwo\, emissions, a result consistent with the estimated natural indirect effects from the mediation analysis.  Jointly interpreting results related to other emissions proved more subtle, and highlighted the difficulty involved in interpreting principal effects as mediated effects, in particular when there are multiple mediators.  A finer examination of principal strata defined by cross-classification of \SOTwo\, changes and changes in \COTwo\, (or \NOx\,) revealed the dominating role of scrubber effects on \SOTwo\, that was corroborated by the results of the mediation analysis. This cross-classification also reconciled the lack of evidence for a natural direct effect with the apparent evidence of dissociative effects pertaining to \NOx\, and \COTwo\, that were revealed to be driven primarily by changes in \SOTwo. The evidence of nonzero dissociative effects for \SOTwo\, is likely explained by the negative expected direct effect. The relative magnitudes of principal effects and mediation effects are consistent with the well-known result that, in general, associative effects are a mixture of direct and indirect effects. Overall, these results are largely consistent with expectations: scrubbers appear to causally reduce \SOTwo\, emissions but not those of \NOx\, or \COTwo; scrubbers causally reduce ambient \PMTwo\, (within 150km); the effect on \PMTwo\, is primarily mediated by causal reductions in \SOTwo\, emissions and not \NOx\, or \COTwo\, emissions; and there appears to be direct effect of scrubbers on \PMTwo.

The results of this case study should be interpreted in light of several important limitations.  First is the relative simplicity with which we linked power plants to monitors.  Specifically, our strategy links power plants to all of the ambient monitors within 150km.  Thus, our analysis is of the causal effects of scrubbers on average \PMTwo\, measured within 150km. This likely does not reflect the full effect of emissions changes on ambient air quality, which are expected to have implications at distances greater than 150km.  A related limitation is the assumption that there is no interference between observations.  If the effect of a scrubber on ambient \PMTwo\, extends far enough beyond 150km so that a scrubber at a given power plant causally affects ambient \PMTwo\, surrounding \textit{other} power plants, then this assumption would be violated.   More sophisticated strategies for causal inference in the presence of interference and for linking ambient monitors to power plants based on features such as atmospheric conditions and weather patterns are warranted.  Nonetheless analysis presented here represents an important approximation that still yields valuable conclusions, especially with respect to quantifying causal pathways.  Another important limitation of this analysis is that it assumes that the factors listed in Table \ref{Data} are sufficient to control for confounding, which in this case would consist of differences between power plants or other features related to ambient \PMTwo\, that are also associated with whether a power plant had scrubbers installed in 2005. Our approach is not readily extended to categorical mediators. We save this as potential future research. Despite these limitations, we have developed new statistical methodology and leveraged an unprecedented linked data base to provide the first empirical evaluation of the presumed causal relationships that motivate a variety of regulations for improving ambient air quality and, ultimately, human health. 

\section*{Acknowledgements}
This work was supported by NIH R01ES026217, NIH R01CA183854, NIH R01GM112327, NIH R01GM111339, NCI P01 CA134294, HEI 4909, HEI 4953, and USEPA RD-83587201. Its contents are solely the responsibility of the grantee and do not necessarily represent the official views of the USEPA.  Further, USEPA does not endorse the purchase of any commercial products or services mentioned in the publication.  

\section*{supplement}
Web Appendices A-J, tables and figures are provided as supplementary materials.

\bibliographystyle{agsm}
\bibliography{multiple,HEIAccountabilityGrant}

\section*{A. Observed Data Models and Prior Specification} 

We specify Dirichlet process mixtures of truncated normals (bounded from below at 0) for the
distribution of  each mediator \citep{muller_bayesian_1996}.  Specifically, for each
intervention $z=0,1$, $k=1,2,3$ and baseline covariates $\boldsymbol{X}=\boldsymbol{x}$, the conditional distribution of the $k$-th observed
mediator is specified as
\begin{eqnarray*}
M_{k,i} | Z_i=z, \boldsymbol{X}_i=\boldsymbol{x}_i & \sim &
N(\beta^z_{k0,i}+\boldsymbol{x}_i^\top \boldsymbol{\beta}^z_{k1},\,\,
\tau^z_{k,i}), \qquad M_{k,i \geq 0};\, i=1,\cdots, n^z\\
\beta^z_{k0, i}, \tau^z_{k,i} & \sim & F_{k}^z,\\
F_{k}^z & \sim & DP (\lambda_{k}^z, \,\,\mathcal{F}_{k}^z),
\end{eqnarray*}
where the subscript $i$ indexes the observation, $k$ indicates the $k$-th mediator, and the superscript $z$ indicates the intervention
arm.  For example, $\beta^z_{k0,i}$ and $\tau^z_{k,i}$
denote the intercept and precision parameters for the $k$-th emission at the $i$-th power plant that received intervention $z$. Here, $DP$ denotes the Dirichlet process with two parameters, a mass
parameter ($\lambda_k^z$) and a base measure ($\mathcal{F}_k^z$) for
each mediator $k$ and intervention $z$. To not overly complicate the model we only `mixed' over the
intercept parameters and precisions in the conditional distributions,
$\beta^z_{k0, i}$ and $\tau^z_{k,i}$. The base distribution $\mathcal{F}_k^z$ is taken to be the normal-Gamma distribution,
$N(\mu_k^z, S_k^z) G(a_k^z,b_k^z)$, where $S_k^z$ is the precision parameters and the Gamma is
parametrized as the mean to be $a_k^z / b_k^z$ and we set a
Gamma prior $G(1,1)$ on the mass parameter
$\lambda_k^z$. For the hyper-priors,
we follow the specification from \cite{Tadd:2008} such
that $\mu_k^z \sim (\mu_k^{z\star}, S_k^{z\star}), S_k^z \sim G(a_k^{z\star},
b_k^{z\star})$ and $a_k^z = b_k^z  = 1$. $S_k^{z\star}$ is set to $2/\hat{\Sigma}_k^z$ and $\mu_k^{z\star}$ is
set to the mean of the data. And $a_k^{z\star} \sim \text{Unif}(1, 5), b_k^{z\star}  = a_k^z \times
\hat{\Sigma}_k^z/2$. From these specifications, $E(\tau_{k,i}^{z}) =E(S_k^{z}) =
S_k^{z\star} = 2/\hat{\Sigma}_k^z$ (i.e., the expected variance
components are an attenuated value of the MLE of the variance of the
data). These observed-data models can be represented using
the stick-breaking construction \citep{sethuraman_constructive_1994}
which can be approximated by a finite mixture of normals such that,
for example, the conditional distribution of $M_1$ under intervention $z=1$ can be
represented as
\[f_{M_1}(m|z=1, \boldsymbol{x}) = \sum_{k=1}^K \theta_{k} N(m\, ;\,
\beta_{10,k}^{z=1} + \boldsymbol{x}^\top \boldsymbol{\beta}_{11}^{z=1}, \tau_{1,k}^{z=1}  ), \]
where $\theta_k = \theta^\prime_{k} \prod_{h<k} (1-\theta^\prime_k),
\theta^\prime_k \sim \text{Beta}(1, \lambda_1^{z=1})$, and $(\beta_{10,k}^{z=1},
\tau_{1,k}^{z=1}) \overset{iid}{\sim} \mathcal{F}_1^{z=1}$ and $K$ is a
maximum number of clusters. We use the stick-breaking construction for posterior samplings.

To model the distributions of the potential outcomes for each $z=0,1$ conditional on all potential mediators and covariates, we use a locally weighted mixture of normal regression model \citep{muller_bayesian_1996} that is induced by specifying a DP mixture of normals for the joint distribution of the outcome, all mediators and covariates. Let $Z_i = (Y_i, \bM(0,0,0), \bM(1,1,1), \boldsymbol{X}_i)$. The model for the joint distribution of $Z_i$ for each $z=0,1$ is as follows:
\begin{eqnarray*}
Z_i|\mu_i^z, \Sigma_i^z &\sim& N(\boldsymbol{\mu}_i^z, \Sigma_i^z), \qquad i=1, \cdots, n_z\\
(\boldsymbol{\mu}_i^z, \Sigma_i^z) | G^z &\sim& G^z\\
G^z | \alpha^z G_0^z &\sim & DP (\alpha^z G_0^z)
\end{eqnarray*}
where $G_0^z = N(\boldsymbol{\mu}^z | m_1, (1/k_0) \Sigma^z) IW (\Sigma^z | 25, \psi_1^z)$. The following hyperpriors are also specified: $\alpha^z \sim Gamma(10,1)$, $m_1^z \sim N(m_2^z, s_2)$, $k_0 \sim Gamma(6.01/2, 2.01/2)$ and $\psi_1^z \sim IW(25, \psi_2^z)$ where $m_2^z = \text{mean}(Z_i) \text{ for } i=1, \cdots, n_z$, $s_2 = 0.5 \text{cov}(Z_i) \text{ for } i=1, \cdots, n_z$ and $\psi_2^z = s_2$.
This joint distribution induces the following conditional distribution model of the outcome for each intervention $z=0,1$:
\begin{eqnarray*}
\lefteqn{f(y_i | \mathbf{m}_i(0,0,0), \mathbf{m}_i(1,1,1), \boldsymbol{x}_i,Z_i=z})\\
& = & \sum_{l=1}^{\infty} \omega_l^z
  N(y_i,\mathbf{m}_i(0,0,0),\mathbf{m}_i(1,1,1),\boldsymbol{x}_i \,|\,\boldsymbol{\mu}_l^z, \Sigma_l^z)
\end{eqnarray*}
where $\omega_l^z = \gamma_l^z / (\sum_{j=1}^\infty \gamma_j^z N(\mathbf{m}_i(0,0,0),\mathbf{m}_i(1,1,1),\boldsymbol{x}_i \,|\,\boldsymbol{\mu}_{j,\setminus 1}^z, \Sigma_{j, (\setminus 1, \setminus 1)}^z))$ and $\boldsymbol{\mu}_{j,\setminus 1}^z$ denotes all elements of mean parameters $\boldsymbol{\mu}_j^z$ except for $Y_i$. Similarly, $\Sigma_{j, (\setminus 1, \setminus 1)}^z$ denotes
a submatrix of covariance matrix $\Sigma_{j}^z$ formed by deleting the the first row and the first column. The weight consists of the parameter with this prior $\gamma_j^{\prime,z} \sim \text{Beta}(1, \alpha^z)$ where $\gamma_j^z = \gamma_j^{\prime,z}\prod_{h<j (1-\gamma_h^{\prime,z})}$. In the MCMC computation, we use the R package, \verb|DPpackage| \citep{Jara:Hans:Quin:Mull:2011}, for the subroutine to fit the joint model in each iteration.

\section*{B. Specification of Correlations in Assumption 2} 
To better understand the specification of the correlation structure in
Assumption 2, we provide a table based on two mediators (e.g., $K=2$). Let $r_{jk}(z)$ be the correlation between $M_j(z)$ and $M_k(z)$. 

\begin{table}[h]
\centering
\resizebox{\textwidth}{!}{  
\begin{tabular}{c|cc|cc}
 & $M_1(0)$ & $M_2(0)$  & $M_1(1)$ & $M_2(1)$ \\\hline\hline
$M_1(0)$ & \cellcolor{blue!25} $r_{11}(0)=1$ &\cellcolor{blue!25} $r_{12}(0)$ & $\frac{1}{2} \cdot [r_{11}(0)+r_{11}(1)]\cdot \rho$
&$\frac{1}{2}\cdot [r_{12}(0)+r_{12}(1)]\cdot \rho$ \\
$M_2(0)$ &\cellcolor{blue!25} & \cellcolor{blue!25}$r_{22}(0) =1$&$\frac{1}{2}\cdot
[r_{12}(0)+r_{12}(1)]\cdot \rho$ & $\frac{1}{2}\cdot
[r_{22}(0)+r_{22}(1)]\cdot \rho$ \\ \hline
$M_1(1)$ & & &\cellcolor{blue!25} $r_{11}(1)=1$&\cellcolor{blue!25} $r_{12}(1)$ \\
$M_2(1)$ & & &\cellcolor{blue!25} &\cellcolor{blue!25} $r_{22}(1)=1$\\
\hline
\end{tabular}}
\caption{Colored cells represent correlations identified from
  the observed data. }
\end{table}

\section*{C. Estimation of the Causal Effects}
In the following, we show that Assumptions 1-3 are sufficient to estimate the principal causal effects and the NDE, JNIE's and NIE's.

\subsection*{C.1. Estimation of the principal causal effects}
\begin{small}
\begin{eqnarray}
\lefteqn{\text{EDE}_\mathcal{K}}\\
 & = & E[Y(1;\bM(1,1,1)) - Y(0;\bM(0,0,0)) \,\big\vert\, |(\bM(1,1,1) - \bM(0,0,0)) |_\mathcal{K} < C^D_\mathcal{K}] \nonumber\\
& = & E[Y(1) - Y(0) \,\big\vert\, |(\bM(1,1,1) - \bM(0,0,0)) |_\mathcal{K} < C^D_\mathcal{K}] \nonumber\\
& = & E[Y(1) \,\big\vert\, |(\bM(1,1,1) - \bM(0,0,0)) |_\mathcal{K} < C^D_\mathcal{K}] \nonumber\\
& & - E[Y(0) \,\big\vert\, |(\bM(1,1,1) - \bM(0,0,0)) |_\mathcal{K} < C^D_\mathcal{K}] \nonumber\\
& = & \int_{|(\bM(1,1,1) - \bM(0,0,0)) |_\mathcal{K} < C^D_\mathcal{K}} \int_y \,y \,\, dF_{Y(1)  | \bM(1,1,1),\bM(0,0,0)}(y) \,dF_{\bM(1,1,1), \bM(0,0,0)}(\mathbf{m}_1, \mathbf{m}_0) \nonumber\\
&  & - \int_{|(\bM(1,1,1) - \bM(0,0,0)) |_\mathcal{K} < C^D_\mathcal{K}} \int_y \,y \,\, dF_{Y(0)  | \bM(1,1,1),\bM(0,0,0)}(y) \,dF_{\bM(1,1,1), \bM(0,0,0)}(\mathbf{m}_1, \mathbf{m}_0) \nonumber
\end{eqnarray}
\end{small}
where $F_{Y(1)  | \bM(1,1,1),\bM(0,0,0)}(y)$, $F_{Y(1)  | \bM(1,1,1),\bM(0,0,0)}(y)$, and $F_{\bM(1,1,1), \bM(0,0,0)}(\mathbf{m}_1, \mathbf{m}_0)$ denote the conditional distributions of the outcomes and the joint distribution of the mediators, all of which are estimated from the observed data with Assumptions 1 and 2 (and the conditional outcome model).  Similarly, we estimate $\text{EAE}_\mathcal{K}$.

\subsection*{C.2. Estimation of the NDE, JNIE's and NIE's}
The NDE, JNIE's and NIE's conditional on covariates $\boldsymbol{X} = \boldsymbol{x}$ are estimated with
\begin{eqnarray}
\lefteqn{NDE(\boldsymbol{x})}\nonumber\\
 & = & E[Y(1;\bM(0,0,0)) - Y(0;\bM(0,0,0)) | \boldsymbol{X}=\boldsymbol{x}]\nonumber\\
& = & \int E[Y(1;\bM(0,0,0)) | \bM(1,1,1)=\mathbf{m}_1, \bM(0,0,0)=\mathbf{m}_0,\boldsymbol{X}=\boldsymbol{x}] \nonumber\\
& & \times dF_{\bM(0,0,0), \bM(1,1,1) | \boldsymbol{X}= \boldsymbol{x}} (\mathbf{m}_0,\mathbf{m}_1) -  E[Y(0;\bM(0,0,0)) | \boldsymbol{X}=\boldsymbol{x}] \nonumber\\
& = &  \int E[Y(1;\bM(1,1,1)) | \bM(1,1,1)=\mathbf{m}_0, \boldsymbol{X}=\boldsymbol{x}] dF_{\bM(0,0,0) | \boldsymbol{X}= \boldsymbol{x}} (\mathbf{m}_0)\nonumber \\
& & - E[Y(0;\bM(0,0,0)) | \boldsymbol{X}=\boldsymbol{x}] \nonumber \qquad \qquad \qquad \text{by Assumption 3}\\
& = &   \int E[Y(1) | \bM(1,1,1)=\mathbf{m}_0, \boldsymbol{X}=\boldsymbol{x}] dF_{\bM(0,0,0) | \boldsymbol{X}= \boldsymbol{x}} (\mathbf{m}_0) -  E[Y(0) | \boldsymbol{X}=\boldsymbol{x}]\nonumber, 
\end{eqnarray}
where the second term is estimated by the observed data model under Assumption 1 and all elements in the first term are estimated from the observed data with Assumptions 1 and 2 (and the conditional outcome model). For the JNIE, we can estimate with
\begin{eqnarray}
\lefteqn{JNIE(\boldsymbol{x})}\nonumber\\
 & = & E[Y(1;\bM(1,1,1)) - Y(1;\bM(0,0,0)) | \boldsymbol{X}=\boldsymbol{x}]\nonumber\\
& = &  E[Y(1;\bM(1,1,1)) | \boldsymbol{X}=\boldsymbol{x}] \nonumber\\
& & - \int E[Y(1;\bM(0,0,0)) | \bM(1,1,1)=\mathbf{m}_1, \bM(0,0,0)=\mathbf{m}_0,\boldsymbol{X}=\boldsymbol{x}] \nonumber\\
& & \times dF_{\bM(0,0,0), \bM(1,1,1) | \boldsymbol{X}= \boldsymbol{x}} (\mathbf{m}_0,\mathbf{m}_1)\nonumber\\
& = &  E[Y(1;\bM(1,1,1)) | \boldsymbol{X}=\boldsymbol{x}] \nonumber\\
& & - \int E[Y(1;\bM(1,1,1)) | \bM(1,1,1)=\mathbf{m}_0, \boldsymbol{X}=\boldsymbol{x}] dF_{\bM(0,0,0) | \boldsymbol{X}= \boldsymbol{x}} (\mathbf{m}_0)\nonumber\\
& &  \qquad\qquad \text{by Assumption 3}\nonumber\\
& = &  E[Y(1) | \boldsymbol{X}=\boldsymbol{x}]  - \int E[Y(1) | \bM(1,1,1)=\mathbf{m}_0, \boldsymbol{X}=\boldsymbol{x}] dF_{\bM(0,0,0) | \boldsymbol{X}= \boldsymbol{x}} (\mathbf{m}_0)\nonumber, 
\end{eqnarray}
where the first term is estimated by the observed data model under Assumption 1 and all elements in the second term are estimated from the observed data with Assumptions 1 and 2 (and the conditional outcome model). Also, 
\begin{eqnarray}
\lefteqn{NIE_1(\boldsymbol{x})}\nonumber\\
 & = & E[Y(1;\bM(1,1,1)) - Y(1;\bM(0,1,1)) | \boldsymbol{X}=\boldsymbol{x}]\nonumber\\
& = &  E[Y(1;\bM(1,1,1)) | \boldsymbol{X}=\boldsymbol{x}] \nonumber\\
& & - \int E[Y(1;\bM(0,1,1)) | \bM(1,1,1)=\mathbf{m}_1, \bM(0,0,0)=\mathbf{m}_0,\boldsymbol{X}=\boldsymbol{x}] \nonumber\\
& & \times dF_{\bM(0,0,0), \bM(1,1,1) | \boldsymbol{X}= \boldsymbol{x}} (\mathbf{m}_0,\mathbf{m}_1)\nonumber\\
& = &  E[Y(1;\bM(1,1,1)) | \boldsymbol{X}=\boldsymbol{x}] \nonumber\\
& & - \int E[Y(1;\bM(1,1,1)) | \bM(1,1,1)=\mathbf{m}_{011}, \boldsymbol{X}=\boldsymbol{x}] dF_{\bM(0,1,1) | \boldsymbol{X}= \boldsymbol{x}} (\mathbf{m}_{011})\nonumber \\
& & \quad \text{by Assumption 3}\nonumber\\
& = &  E[Y(1) | \boldsymbol{X}=\boldsymbol{x}]  - \int E[Y(1) | \bM(1,1,1)=\mathbf{m}_{011}, \boldsymbol{X}=\boldsymbol{x}] dF_{\bM(0,1,1) | \boldsymbol{X}= \boldsymbol{x}} (\mathbf{m}_{011})\nonumber, 
\end{eqnarray}
where $\mathbf{m}_{011}$ denotes a vector of values for the mediators $M_1, M_2, M_3$ under interventions $0, 1, 1$, respectively.
The first term in the last equation is estimated by the observed data model under Assumption 1 and all elements in the second term are estimated from the observed data with Assumptions 1 and 2 (and the conditional outcome model). For the remaining mediation effects, we can estimate them analogously.

\section*{D. Assumption 3 and the Assumptions of Homogeneity Across Principal Strata}

The homogeneity assumption in \cite{Forastiere:2016} applied to both intervention arms and extended to the case of multiple mediators can be written as: 
\begin{equation}
Y(1;(m_1,m_2,m_3)) \independent \bM(1-z,1-z,1-z) \,|\, \bM(z,z,z) = (m_1,m_2,m_3), \boldsymbol{X}=\boldsymbol{x}, \label{h3}
\end{equation}
where $m_1, m_2, m_3$ are realized emissions values for the first, second and third mediators and the treatment is denoted as $z \in \{0,1\}$.   For {\it a priori} counterfactuals defined based on values of $M(z,z,z)$, Assumption 3 implies this assumption: 

Invoking SUTVA, Assumption 3 can be represented as:
\begin{eqnarray}
\lefteqn{f(Y(1;(m_1, m_2, m_3))|\bM(0,0,0)=(m_1, m_2, m_3), \bM(1,1,1),\boldsymbol{x})}\nonumber\\
& =& f(Y(1;(m_1, m_2, m_3))|\bM(0,0,0), \bM(1,1,1)=(m_1, m_2, m_3),\boldsymbol{x}). \label{A3}
\end{eqnarray}
Since Equation (\ref{A3}) holds regardless of the values for $\bM(1,1,1)$ in the LHS, it implies 
\begin{equation}
Y(1;(m_1,m_2,m_3)) \independent \bM(1,1,1) \,|\, \bM(0,0,0) = (m_1,m_2,m_3), \boldsymbol{X}=\boldsymbol{x}. \label{A3a}
\end{equation}
Similarly, since Equation (\ref{A3}) holds regardless of the values for $\bM(0,0,0)$ in the RHS, it implies 
\begin{equation}
Y(1;(m_1,m_2,m_3)) \independent \bM(0,0,0) \,|\, \bM(1,1,1) = (m_1,m_2,m_3), \boldsymbol{X}=\boldsymbol{x}, \label{A3b}
\end{equation}
which together imply (\ref{h3}).

However, this homogeneity only applies to {\it a priori} counterfactuals relying on $M(z,z,z)$.  Those of the more general form $M(z_1, z_2, z_3)$ cannot be assumed homogeneous across principal strata since {\it a priori} counterfactual mediator values defined as simultaneously subject to different interventions do not appear in the definition of principal strata. Thus, Assumption 3 and the extended homogeneity assumptions in (\ref{h3}) are equivalent in the case of a single mediator (see Section \ref{sec:single} for details), but with multiple mediators Assumption 3 entails additional assumptions about {\it a priori} counterfactuals used in the decomposition of the overall natural indirect effect into indirect effects attributable to subsets of the mediators.

\subsection*{D.1. Single Mediatior Case}\label{sec:single}
With a single mediator, Assumption 3 and the extended homogeneity assumptions are equivalent. First, Assumption 3 implies the following two homogeneity assumptions: \begin{eqnarray}
Y(1;m) &\perp& M(1) \,|\, M(0)=m, \boldsymbol{X}=\boldsymbol{x} \label{eqn1}\\
Y(1;m) &\perp& M(0) \,|\, M(1)=m, \boldsymbol{X}=\boldsymbol{x}\label{eqn2}.
\end{eqnarray}
The statement in (\ref{eqn1}) implies that the distribution of the {\it a priori} counterfactual $Y(1; M(0))$ is homogenous across all principal strata with $M(0)=m$, regardless of the value of $M(1)$.  The statement in (\ref{eqn2}) implies that the distribution of the observable counterfactual $Y(1;M(1))$ is homogeneous across principal strata with $M(1)=m$. Estimation of the distribution of {\it a priori} counterfactual $Y(1;M(0))$ with $M(0)=m$ follows using observed values of $Y(1;M(1))$ among observations with $Z=1$ and $M(1)=m$. 

To show that the extended homogeneity assumptions, 
\begin{equation}
Y(1; m) \perp M(z) \,|\, M(1-z)=m, \boldsymbol{X}=\boldsymbol{x} \quad \text{for } z=0, 1,\label{H}
\end{equation}
imply Assumption 3, recall that $Y(1; (m)) \perp M(0) | M(1) = m, \boldsymbol{X}=\boldsymbol{x}$ implies that the distribution of potential outcomes $Y(1;M(1))$ with mediator $M(1)=m$ are the same for all values of $M(0)$, which can be represented as 
\begin{eqnarray}
\lefteqn{f(Y(1;M(1)) |  M(0)=m^a, M(1)=m, \boldsymbol{X}=\boldsymbol{x})}\label{H4}\\
& =& f(Y(1;M(1)) |  M(0)=m^b, M(1)=m, \boldsymbol{X}=\boldsymbol{x}),\nonumber
\end{eqnarray}
for all $m^a$ and $m^b$.
Also, $Y(1; (m)) \perp M(1) | M(0) = m, \boldsymbol{X}=\boldsymbol{x}$ implies that the distribution of potential outcomes $Y(1;M(0))$ with mediator $M(0)=m$ are the same for all values of $M(1)$, which can be represented as 
\begin{eqnarray}
\lefteqn{f(Y(1;M(0)) |  M(0)=m, M(1)=m^a, \boldsymbol{X}=\boldsymbol{x}) }\label{H5}\\
&=& f(Y(1;M(1)) |  M(0)=m, M(1)=m^b, \boldsymbol{X}=\boldsymbol{x}), \nonumber
\end{eqnarray}
for all $m^a$ and $m^b$.
Combining (\ref{H4}) and (\ref{H5}), we have Assumption 3 as follows
\begin{eqnarray*}
\lefteqn{ f(Y(1;M(0)) |  M(0)=m, M(1), \boldsymbol{X}=\boldsymbol{x})}\\
& = & f(Y(1;M(0)) |  M(0)=m, M(1)=m, \boldsymbol{X}=\boldsymbol{x}) \qquad \text{by (\ref{H4})}\\
& = & f(Y(1;m) |  M(0)=m, M(1)=m, \boldsymbol{X}=\boldsymbol{x})  \qquad \text{by (SUTVA)}\\
& = & f(Y(1;M(1)) |  M(0)=m, M(1)=m, \boldsymbol{X}=\boldsymbol{x}) \qquad \text{by (SUTVA)}\\
& = & f(Y(1;M(1)) |  M(0), M(1)=m, \boldsymbol{X}=\boldsymbol{x}) \qquad \text{by (\ref{H5})}.
\end{eqnarray*}
Therefore, in the setting of a continuous single mediator, Assumption 3 and the extended homogeneity assumptions (\ref{H}) are equivalent.

\section*{E. Assumption 3 and the Sequential Ignorability Assumption}
Even though Assumption 3 and the sequential ignorability assumption do not imply one another, they are closely related to each other.
Assumption 3 implies a consequence of the sequential ignorability (S.I.) assumption \citep{Imai:2010b} in the setting of $K=1$ mediator. That is, Assumption 3 ($K=1$) and the SI have the same implication from an inferential perspective.
Specifically, S.I. implies 
\[f(Y(1;{M}(0)) \,|\, {M}(0) = {m},\boldsymbol{X}=\boldsymbol{x}) = f(Y(1;M(1)) \,|\,  M(1)=m, \boldsymbol{X}=\boldsymbol{x}),\]
which is also implied by a single mediator version of our Assumption 3,
\begin{eqnarray*}
\lefteqn{f(Y(1;{M}(0)) \,|\, {M}(0) = {m},M(1),\boldsymbol{X}=\boldsymbol{x})}\\
 &=& f(Y(1;M(1)) \,|\,  M(0), M(1)=m, \boldsymbol{X}=\boldsymbol{x}).
\end{eqnarray*}
The difference is that Assumption 3 states that this relationship holds for all values of $M(1)$ in the LHS and all values of $M(0)$ in the RHS while the consequence of the S.I. only holds when $M(1)$ and $M(0)$ are marginalized over in the LHS and the RHS, respectively.

\section*{F. Posterior Inference}
Full Bayesian inference for the principal causal effects and the natural direct and indirect effects is based on posterior samples of the parameters, $\boldsymbol{\theta}$, of the observed-data models since all the causal effects of interest are functions of these parameters. We ran 10,000 MCMC iterations for the observed-data posterior and used the last 7,500 with thinning of 5 to obtain $N=1500$ posterior samples of the parameters.  Since it is not easy to check convergence of cluster specific parameters due to the label switching issue, it is suggested that it is sufficient to monitor convergence of the posterior samples for the global parameters $\boldsymbol{\beta}_j$ for all $j$ \citep{Moli:2010}.  Visual inspection of time-series plots provided no indication against convergence. R codes used to perform the MCMC parameter estimation and post-processing, are available in GitHub (https://github.com/lit777/MultipleMediators).

Posterior sampling for our approach consists of four major steps: (1) sample from the marginal distributions of the mediators, (2) sample from the correlation matrix of the Gaussian copula, (3) impute missing mediators from the joint distribution, and (4) sample from the conditional distribution of the outcome given each set of posterior samples of the mediators and observed covariates. In this paper, we use the efficient Bayesian approach for sampling from Gaussian copula models outlined in Pitt et al. (2006)\nocite{Pitt:2006}.
For notational simplicity, denote $T_{1i} = M_{1,i}(0),T_{2i} = M_{2,i}(0),T_{3i} = M_{3,i}(0),T_{4i} = M_{1,i}(1), T_{5i} = M_{2,i}(1), T_{6i} = M_{3,i}(1)$ and $H_{ji} = \Phi^{-1}\{F_j(T_{ji}; \boldsymbol{\theta}_j, \boldsymbol{X}_i)\}$ where $\boldsymbol{\theta}_j$ is a vector of parameters in the $j$-th marginal and $\Phi^{-1}$ is the inverse univariate standard normal CDF.

In the first step, for $j=1,\cdots, 6$, the parameters $\boldsymbol{\theta}_j$ are sampled and $H_{ji}$ is updated; then, in the second step, the correlation matrix $\boldsymbol{R}$ is sampled. Missing mediator values are imputed based on the joint distribution of all potential mediators and (observed) outcomes. Finally, we sample $\boldsymbol{\xi}_z$, the vector of parameters in the outcome model for $Z=z$, all potential mediators and covariates.

The likelihood of $\boldsymbol{\theta}$ and $\boldsymbol{R}$ in the Gaussian Copula model has the form
\begin{eqnarray}
f(\boldsymbol{T}, \boldsymbol{X} \,|\,  \boldsymbol{\theta}, \boldsymbol{R}) &=& \prod_{i=1}^n f(\boldsymbol{T}_{\LargerCdot i} \,|\,\boldsymbol{X}_i, \boldsymbol{\theta}, \boldsymbol{R}) f(\boldsymbol{X}_i)\nonumber\\
& = & \prod_{i=1}^n f(T_{1i}, T_{2i}, T_{3i}, T_{4i}, T_{5i}, T_{6i} \,|\,\boldsymbol{X}_i, \boldsymbol{\theta}, \boldsymbol{R} ) f(\boldsymbol{X}_i)\nonumber\\
& = & |\boldsymbol{R}|/^{-n/2} \prod_{i=1}^n \exp \left\{\frac{1}{2}\boldsymbol{H}_{\LargerCdot i}^\top (I-\boldsymbol{R}^{-1}) \boldsymbol{H}_{\LargerCdot i} \right\} \prod_{j=1}^6 f(T_{ji} \,|\, \boldsymbol{X}_i, \boldsymbol{\theta}_j)f(\boldsymbol{X}_i), \label{likelihood}
\end{eqnarray}
where $\boldsymbol{T}_{\LargerCdot i} = \{ T_{1i}, T_{2i}, T_{3i}, T_{4i}, T_{5i}, T_{6i}\}^\top$, $\boldsymbol{H}_{\LargerCdot i} = \{H_{1i}, H_{2i}, H_{3i}, H_{4i}, \allowbreak H_{5i}, H_{6i}\}^\top$ and $f(\boldsymbol{X}_i)$ denotes the empirical PDF of covariates $\boldsymbol{X}_i$. Assuming independent priors $p(\boldsymbol{\theta}) = \prod_{j=1}^6 p(\boldsymbol{\theta}_j)$, $p(R)$ and $p(\boldsymbol{\xi}) = p(\boldsymbol{\xi}_0)p(\boldsymbol{\xi}_1)$, the posterior distribution of $[\boldsymbol{\theta},\boldsymbol{\xi}, \boldsymbol{R}]$ can be represented as follows (\cite{imbens1997bayesian}, \cite{jin2008principal} and \cite{schwartz2011bayesian}),
\begin{small}
\begin{eqnarray*}
\lefteqn{f(\boldsymbol{\theta}, \boldsymbol{\xi}, \boldsymbol{R}|\boldsymbol{T}^\text{obs}, \boldsymbol{Y}^\text{obs}, \boldsymbol{Z},\boldsymbol{X})}\\ 
& \propto & p(\boldsymbol{\theta})p(\boldsymbol{\xi})p(\boldsymbol{R})\int \prod_{i=1}^n f(T_{1i}, T_{2i}, T_{3i}, T_{4i}, T_{5i}, T_{6i}, Y(1), Y(0) \,|\,\boldsymbol{X}_i,\boldsymbol{\theta}, \boldsymbol{\xi},\boldsymbol{R} ) \,\, d\boldsymbol{T}_{\LargerCdot i}^\text{mis} d\boldsymbol{Y}_i^\text{mis}.
\end{eqnarray*}
\end{small}
We can obtain the joint distribution, $f(\boldsymbol{\theta}, \boldsymbol{\xi},\boldsymbol{R}, \boldsymbol{T}^\text{mis}|\boldsymbol{T}^\text{obs}, \boldsymbol{Y}^\text{obs}, \boldsymbol{Z},\boldsymbol{X})$ by sampling iteratively from $f(\boldsymbol{\theta}|\boldsymbol{T}^\text{obs},\boldsymbol{T}^\text{mis},\boldsymbol{Y}^\text{obs},\boldsymbol{X},\boldsymbol{\xi},\boldsymbol{R})$, $f(\boldsymbol{T}^\text{mis}|\boldsymbol{T}^\text{obs},\boldsymbol{Y}^\text{obs},\boldsymbol{X},\boldsymbol{\theta},\boldsymbol{\xi},\boldsymbol{R})$, $f(\boldsymbol{R}|\boldsymbol{T}^\text{obs},\allowbreak \boldsymbol{T}^\text{mis},\boldsymbol{Y}^\text{obs}\boldsymbol{X},\boldsymbol{\theta},\boldsymbol{\xi})$ and $f(\boldsymbol{\xi}|\boldsymbol{T}^\text{obs},\boldsymbol{T}^\text{mis},\boldsymbol{Y}^\text{obs},\boldsymbol{X},\boldsymbol{\theta},\boldsymbol{R})$.

Specifically, we can use four steps:
\begin{itemize}
\item Step 1. For $j = 1, \cdots, 6$, sample $\boldsymbol{\theta}_j$ from $f(\boldsymbol{\theta}_j \,|\, T_{j \LargerCdot}^\text{obs},T_{j \LargerCdot}^\text{mis}, H_{\setminus j 
\LargerCdot}, \boldsymbol{X}, R)$ where $H_{\setminus j \LargerCdot}$ denotes all elements of $H$ except $\{H_{j1}, H_{j2}, \cdots, H_{jn}\}$  and $T_{j \LargerCdot}^\text{obs}$  denotes all observed entries of $T_{ji}$ for $i=1,\cdots,n$. Then, update  $H_{ji} = \Phi^{-1}\{F_j(T_{ji}; \boldsymbol{\theta}_j, \boldsymbol{X}_i)\}$ for all $i \in \{1, \cdots, n\}$.

\item Step 2. Sample $\boldsymbol{R}$ from $f(\boldsymbol{R}|\boldsymbol{T}^\text{obs},\boldsymbol{T}^\text{mis},\boldsymbol{X},\boldsymbol{\theta})$.

\item Step 3. For each $j=1, \cdots, 6$, impute $T_{j\LargerCdot}^{\text{mis}}$ from $f(T_{j\LargerCdot}^{\text{mis}} | T_{j\LargerCdot}^{\text{obs}}, T_{\setminus j \LargerCdot}, \boldsymbol{Y}^\text{obs}, \boldsymbol{X}, \boldsymbol{\theta},  \boldsymbol{\xi},  \boldsymbol{R})$ and update $H_{j\LargerCdot}^{\text{mis}} = \Phi^{-1}\{F_j(T_{j\LargerCdot}^{\text{mis}}; \boldsymbol{\theta}_j, \boldsymbol{X})\}$

\item Step 4. Sample $\boldsymbol{\xi}$ from $f(\boldsymbol{\xi}|\boldsymbol{T}^\text{obs},\boldsymbol{T}^\text{mis},\boldsymbol{Y}^\text{obs},\boldsymbol{X},\boldsymbol{\theta},\boldsymbol{R})$ which is proportional to $p(\boldsymbol{\xi}) \prod_{i=1}^n f(Y_i(0)|\boldsymbol{T}_{\LargerCdot i}^\text{mis}, \boldsymbol{T}_{\LargerCdot i}^\text{obs}, \boldsymbol{X}_i; \boldsymbol{\xi})^{(1-Z_i)}f(Y_i(1)|\boldsymbol{T}_{\LargerCdot i}^\text{mis}, \boldsymbol{T}_{\LargerCdot i}^\text{obs}, \boldsymbol{X}_i; \boldsymbol{\xi})^{(Z_i)}f(\boldsymbol{T}_{\LargerCdot i}, \boldsymbol{X}_i \,|\,  \boldsymbol{\theta}, \boldsymbol{R})$.

\end{itemize}
Note that the first step represents sampling from $\boldsymbol{\theta}_j \sim f(\boldsymbol{\theta}_j \,|\, T_{j, \LargerCdot}, \boldsymbol{X}, \boldsymbol{\theta}_{\setminus j}, \boldsymbol{R})$ for $j=1, \cdots, 6$ where $\boldsymbol{\theta}_{\setminus j}=\{\boldsymbol{\theta}_1, \boldsymbol{\theta}_2, \cdots, \boldsymbol{\theta}_6\} \setminus \boldsymbol{\theta}_j$ since  $f(\boldsymbol{\theta}_j \,|\, T_{j \LargerCdot}, \boldsymbol{X}, \boldsymbol{\theta}_{\setminus j}, \boldsymbol{R})$ is equivalent to $f(\boldsymbol{\theta}_j \,|\, T_{j \LargerCdot}, H_{\setminus j \LargerCdot}, \boldsymbol{X}, \boldsymbol{R})$ (because $\boldsymbol{\theta}_{\setminus j}$ affects $\boldsymbol{\theta}_j$ only through $H_{\setminus j \LargerCdot}$ in (\ref{likelihood})). We obtain
\begin{eqnarray}
\lefteqn{\log f(\boldsymbol{\theta}_j | T_{j \LargerCdot}, H_{\setminus j \LargerCdot}, \boldsymbol{X},\boldsymbol{R})} \nonumber\\
& = &  \text{const} + \frac{1}{2} (1-\boldsymbol{R}^{-1}_{jj})\sum_{i=1}^n H_{ji}^2 - \sum_{i=1}^n \sum_{k=1, k\neq j}^6 (\boldsymbol{R}^{-1})_{jk} H_{ji}H_{ki} \label{conditional}\\
& & + \sum_{i=1}^n \log f(T_{ji}| \boldsymbol{X}_i, \boldsymbol{\theta}_j)+\log p(\boldsymbol{\theta}_j). \nonumber
\end{eqnarray}
Since $H_{ji} = \Phi^{-1}\{F_j(T_{ji}; \boldsymbol{\theta}_j, \boldsymbol{X}_i)\}$, we have $H_{ji}$ in the expression (\ref{conditional}) despite conditioning only on $H_{\setminus j \LargerCdot}$. 

\medskip

\noindent {\bf Step 1}

We specify a Bayesian nonparametric model (Dirichlet process mixtures of truncated normals; DPM) for $f(T_{ji}| \boldsymbol{X}_i, \theta_j)$ which makes it difficult to generate $\boldsymbol{\theta}_j$ from the nonstandard conditional density of $\boldsymbol{\theta}_j$. Also, the number of covariates can be large, which makes it hard to sample directly from the conditional density. To overcome these issues, we use a block Metropolis algorithm. As we mentioned earlier in Section A, we use the finite stick-breaking approximation of the DPM specification \citep{sethuraman_constructive_1994} for the marginal distribution,
\[f(T_{ji}| \boldsymbol{X}_i, \theta_j) = \sum_{k=1}^K \omega_k N(T_{ji}\,;\, \beta_{j0,(k)}+\boldsymbol{X}_i\boldsymbol{\beta}_j, \tau_{j,(k)}),\]
where $(\beta_{j0,(k)}, \tau_{j,(k)}) \overset{\text{iid}}{\sim} N(\mu_j, S_j) G(a_j, b_j)$ and $\omega_k = \omega^\prime_k \prod_{h<k} (1-\omega_h^\prime), \omega_h^\prime \sim \text{Beta}(1, \lambda_j)$ and $K$ is a maximum number of clusters; we set $8$ because the smallest cluster of each marginal model has a sufficiently small weight, $\text{min} (\omega_k) < 0.005$. The hyper-priors are specified as $\lambda_j \sim G(1,1)$, $\mu_j \sim N(\mu_j^\star, S_j^\star)$, $S_j \sim G(a_j^\star, b_j^\star)$, $a_j = b_j = 1$,  $a_j^\star \sim \text{Unif}(1, 5)$ and $b_j^\star = 100 a_j^\star$ (see Section A for details of the specification). At every iteration of the MCMC scheme for Step 1, we need to sample values $\lambda_j, \mu_j, a_j^\star, S_j,  \boldsymbol{\beta}_j$ and $\omega_k, \beta_{j0,(k)}, \tau_{j,(k)}$ for $k=1, \cdots, K$ using a block Metropolis algorithm. Specifically, for each $j=1, \cdots, 6$,
\begin{itemize}
\item[Step 1.a] $\Omega=\{\omega_1, \cdots, \omega_K\}, \lambda_j$ :  Denote the latent cluster indicator variable for each subject $i$ as $\mathcal{Z}_i$ which can take a value in $\{1,2,\cdots, K\}$. Also, denote the values of $\beta_{j0,(k)}, \allowbreak \boldsymbol{\beta}_j, \tau_{j,(k)}, \omega$ at the $t$-th iteration as $ \beta_{j0,(k)}(t), \allowbreak \boldsymbol{\beta}_j(t), \tau_{j,(k)}(t), \omega_k (t)$. Draw $\mathcal{Z}_i$ from $\mathcal{Z}_i \sim \text{Categorical}(p_1, p_2, \cdots, p_K)$ where $p_k = N(T_{ji}\,;\, \beta_{j0,(k)}(t-1)+\boldsymbol{X}_i\boldsymbol{\beta}_j(t-1),  \tau_{j,(k)}(t-1)) \times \omega_k (t-1)$.
Draw $\omega^\prime_k \sim \text{Beta}(1+n_k, \lambda_j+\sum_{q=h+1}^{K-1} n_q)$ for $k=1, \cdots K-1$ where $n_k$ denotes the number of subjects having $\mathcal{Z}_i = k$. Then, update $\Omega=\{\omega_1, \cdots, \omega_K\}$ via $\omega_k = \omega^\prime_k \prod_{h<k} (1-\omega_h^\prime)$. Then, draw $\lambda_j$ from $\text{Gamma}(1+K-1, 1-\sum_{k=1}^K (\log (1-\omega_k)))$.
\item[Step 1.b] $a_j^\star,  \mu_j, S_j$ : Denote the values at the $t$-th iteration as $ a_j^\star(t), \allowbreak \mu_j(t), S_j(t)$. We propose values independently from  $a_{j(prop)}^\star \sim \text{Unif}(1,5)$, $\mu_{j(prop)} \sim N(\mu_j(t-1),  2/\Sigma_j)$,  $S_{j(prop)} \sim \text{Unif}(S_{j}(t-1)-0.1, S_j(t-1)+0.1)$ and denote this joint distribution as $q\{\Psi_{prop}; \Psi(t-1)\}$ where $\Psi_{prop} = \{a^\star_{j(prop)}, \allowbreak\mu_{j(prop)}, S_{j(prop)}\}$ and $\Psi(t) = \{a^\star_{j}(t), \mu_{j}(t), S_{j}(t)\} $. Then, calculate the acceptance probability of the proposed values
\[AR_1 = \text{min}\left\{1, \frac{f(\boldsymbol{\theta}_{j (prop)} | T_{j \LargerCdot}, H_{\setminus j \LargerCdot}, \boldsymbol{X},\boldsymbol{R})q\{\Psi(t-1)\,;\, \Psi_{prop}\}}{f(\boldsymbol{\theta}_j (t-1) | T_{j \LargerCdot}, H_{\setminus j \LargerCdot}, \boldsymbol{X},\boldsymbol{R})q\{\Psi_{prop} \,;\, \Psi(t-1)\}}\right\},\]
where $f(\boldsymbol{\theta}_{j(prop)} | T_{j \LargerCdot}, H_{\setminus j \LargerCdot}, \boldsymbol{X},R)$ denotes the conditional distribution in (\ref{conditional}) before log-transformation and set $\lambda_j, a_j, a_j^\star, \mu_j, S_j$ to their proposed values and other parameters are set to their current values. Similarly, for $f(\boldsymbol{\theta}_{j}(t-1) | T_{j \LargerCdot}, H_{\setminus j \LargerCdot}, \boldsymbol{X},R)$, we plug in the values from the $(t-1)$-th iteration instead. Then, we accept the proposed value with probability $AR_1$.
\item[Step 1.c] $\boldsymbol{\beta}_{j0} = \{\beta_{j0,(1)}, \beta_{j0,(2)}, \cdots, \beta_{j0,(K)}\}$ : propose values from  $\beta_{j0,(k),(prop)} \sim N(\beta_{j0,(k)}(t-1), 0.1)$ and use the same Metropolis algorithm as in the previous sub-step.
\item[Step 1.d] $\boldsymbol{\beta}_{j} = \{\beta_{j,(1)}, \beta_{j,(2)}, \cdots, \beta_{j,(P)}\}$ for the coefficients of $P$ covariates: propose values using adaptive Metropolis \citep{rosenthal2011} from $\boldsymbol{\beta}_{j0, (prop)} \sim MVN(\boldsymbol{\beta}_{j0} (t-1), \Sigma^\dagger)$ where $\Sigma^\dagger = (2.38)^2/(2P) \Sigma_{t-1} + 0.1^2/(2P) I_p $ with the empirical covariance matrix $\Sigma_t$ of $\boldsymbol{\beta}_{j}(1), \boldsymbol{\beta}_{j}(2), \cdots, \boldsymbol{\beta}_{j}(t-1)$ and use the same Metropolis algorithm as in the previous sub-step.
\item[Step 1.e] $\boldsymbol{\tau}_j = \{\tau_{j, (1)}, \cdots, \tau_{j,(K)}\}$ : propose values from $1/\tau_{j,(k)} \sim \text{Gamma}(0, 1/\tau_{j,(k)}(t-1) \times c, c )$ for $k=1, \cdots, K$ where $c$ is a large constant to concentrate the probability around $1/\tau_{j,(k)}(t-1)$.
\end{itemize}


\medskip

\noindent {\bf Step 2}

In Step 2, we conduct the analysis using 2 different specifications: (1) we specify uniform priors on all the association parameters in $\boldsymbol{R}$ where their intervals are restricted to give the positive definite matrix; (2) we give restrictions on the (partially-identifiable) association parameters using a parameter $\rho$ (see Section 5.2.3 in the main paper) and put $\text{Unif}(0,1)$ priors on this parameter and uniform priors on other remaining association parameters. However, for computational efficiency, we propose a value for each association parameter $r_{(prop)}$ from $\text{Unif}(r_L, r_U)$ where $r_L$ and $r_U$ are determined to give the positive definite matrix $\boldsymbol{R}$. See \cite{Barn:McCu:Meng:2000} for the computation details of calculating these intervals.
 Then, calculate the acceptance probability of the proposed values
\[AR_2 = \text{min}\left\{1, \frac{f(\boldsymbol{R}_{r(prop)} | \boldsymbol{T},\boldsymbol{X},\boldsymbol{\theta} )q(r(t-1))}{f(\boldsymbol{R}_r(t-1) | \boldsymbol{T},\boldsymbol{X},\boldsymbol{\theta})q(r_{(prop)} )} \right\},\]
where $\boldsymbol{R}_{r(prop)}$ denotes the correlation matrix with the $r$-th element set to the proposed value and other entries are set to their current values. Similarly, $\boldsymbol{R}_{r}(t-1)$ denotes the correlation matrix where $r$-th element is set to the value from the $(t-1)$-th iteration and other entries are set to their current values.  We accept the proposed value with probability $AR_2$.

\medskip

\noindent {\bf Step 3}

In Step 3, we impute $\boldsymbol{T}^{\text{mis}}$. Specifically, for each $j=1, \cdots, 6$, draw $T_{j(prop)} \sim N(T_{j}(t-1), \sigma_T)$ and calculate the acceptance probability of the proposed values
\[AR_3 = \min\left\{1, \frac{f(T_{j(prop)},T_{j \LargerCdot}^\text{obs}, T_{\setminus j \LargerCdot}, \boldsymbol{X}|\boldsymbol{\theta},\boldsymbol{R})f(\boldsymbol{Y}^\text{obs}|T_{j(prop)},T_{j \LargerCdot}^\text{obs}, T_{\setminus j \LargerCdot}, \boldsymbol{X},\boldsymbol{\theta},\boldsymbol{\xi})q(T_{j}(t-1) )}{f(T_{j}(t-1),T_{j \LargerCdot}^\text{obs}, T_{\setminus j \LargerCdot}, \boldsymbol{X}|\boldsymbol{\theta},\boldsymbol{R})f(\boldsymbol{Y}^\text{obs}|T_{j}(t-1),T_{j \LargerCdot}^\text{obs}, T_{\setminus j \LargerCdot}, \boldsymbol{X},\boldsymbol{\theta},\boldsymbol{\xi})q(T_{j(prop)} )}\right\} \]
Then, we accept the proposed value with probability $AR$.

\medskip

\noindent {\bf Step 4}

Finally, in Step 4, once we obtain the missing components of the potential mediators ($\boldsymbol{m}(0,0,0)$ for subjects with $Z=0$ and $\boldsymbol{m}(1,1,1)$ for subjects with $Z=1$), we specify locally weighted mixture normal regression outcome models that are induced by specifying DP mixtures of normals for the joint distributions of $(Y(0), \bM(1,1,1), \bM(0,0,0), \boldsymbol{X})$ for $Z=0$ or $(Y(1), \bM(1,1,1), \bM(0,0,0), \boldsymbol{X})$ for $Z=1$. For this subroutine, we use \verb|DPcdensity| function in the R package \verb|DPpackage| to obtain posterior samples in each iteration. Here, we use the default hyper-parameter specifications described in Section A. The detail of the posterior computation can be found in \cite{Jara:Hans:Quin:Mull:2011}. 

\section*{G. Simulation Setup}
\begin{itemize}
\item Confounders : \[X_1 \sim N(1.5, 0.3^2),\,\, X_2 \sim N(-1.5, 0.3^2),\,\, X_3 \sim N(2,0.1^2)\]
\item Potential Mediators : \[\mathbf{M}(z, z, z) \sim MVN \left( \begin{array}{r}
         2+0.4 z +0.5 x_1 + 0.4 x_2 + 0.5 x_3\\
         1+0.4 z -0.4 x_1 + 0.4 x_2 - 0.5 x_3\\ 
         -0.5 -0.4 z +0.5 x_1 + 0.4 x_2 + 0.5 x_3\\
        \end{array} , \quad \Sigma_z \right)\]
\item Potential Outcome : \[Y(z; \mathbf{M}(z_1,z_2,z_3)) \sim N(1-1 z+ 0.8 M_1(z_1) + 0.8 M_2(z_2) + 0.8 M_3(z_3) + h(M_1,M_2,M_3)+x_1 + x_2+0.8 x_3, 0.1^2)\]       
\end{itemize}
where $h()$ is a function for `interaction' term(s). We consider two cases: (1) Case 1: $h() = 1.5M_1(z_1)\times M_2(z_2)$ and (2) Case 2: and $h() = 1.5M_1(z_1) \times M_2(z_2) + 0.6 M_2(z_2) \times M_3(z_3)$. Simultaneously, we test the model for `correlated mediators' assuming two cases \[\text{Case A} :  \Sigma = \left( \begin{array}{ccc}
0.64 & 0 & 0 \\
0 & 0.64 & 0 \\
0 & 0 & 0.64 \end{array} \right) \text{ for } z=1; \Sigma = \left( \begin{array}{ccc}
0.04 & 0 & 0 \\
0 & 0.04 & 0 \\
0 & 0 & 0.04 \end{array} \right) \text{ for } z=0 \]
\[ \text{Case B} :  \Sigma = \left( \begin{array}{ccc}
0.64 & 0.128 & 0.128 \\
0.128 & 0.64 & 0.128 \\
0.128 & 0.128 & 0.64 \end{array} \right) \text{ for } z=1; \Sigma = \left( \begin{array}{ccc}
0.04 & 0.01 & 0.01 \\
0.01 & 0.04 & 0.01 \\
0.01 & 0.01 & 0.04 \end{array} \right) \text{ for } z=0 \]

\section*{H. Uniform Priors on the Correlation Parameters}
As mentioned in Section 5.2, we conduct an additional analysis with uniform priors on entries of the correlation matrix, $\boldsymbol{R}$, in the Copula model. 
On average, the estimates for the principal causal effects and mediation effects are similar to those in the main paper. However, credible intervals are wider in this case.
\begin{table}[h]
\centering 
\caption{Posterior means (95\% C.I.s) for expected associative and dissociative effects of \SOTwo\, scrubbers.}\label{tab:ps}
\resizebox{\textwidth}{!}{  
\begin{tabular}{c|c|ccccccc}
& & \SOTwo & \NOx & \COTwo & \SOTwo\, \& \NOx & \SOTwo\, \& \COTwo & \NOx\, \& \COTwo & \SOTwo\, \& \NOx\, \& \COTwo \\
\hline\hline
\multirow{2}{*}{EAE$^-$}& Mean &  -1.77 & -1.62 & -1.68 & -1.66 & -1.71 & -1.64 & -1.67 \\
& SD & (0.53) & (0.71) & (0.64) & (0.72) & (0.65) & (0.78) & (0.79)\\
\hline
\multirow{2}{*}{EDE}& Mean &  -1.42 & -1.48 & -1.42 & -1.31 & -1.14 & -1.17 & -1.06 \\
& SD & (0.72) & (0.64) & (0.59) & (0.90) & (0.85) & (0.76) & (1.04)\\
\hline
\multirow{2}{*}{EAE$^+$}& Mean & -0.21 & -1.91 & -1.84 &  -0.59 & 0.13 & -2.03 & -0.47\\
& SD & (3.50) & (0.76) & (0.87) & (4.93) & (5.26) & (1.09) & (5.78)\\
\hline
\end{tabular}}
\end{table}
\begin{figure}[p]
\centering
  \includegraphics[scale=0.5]{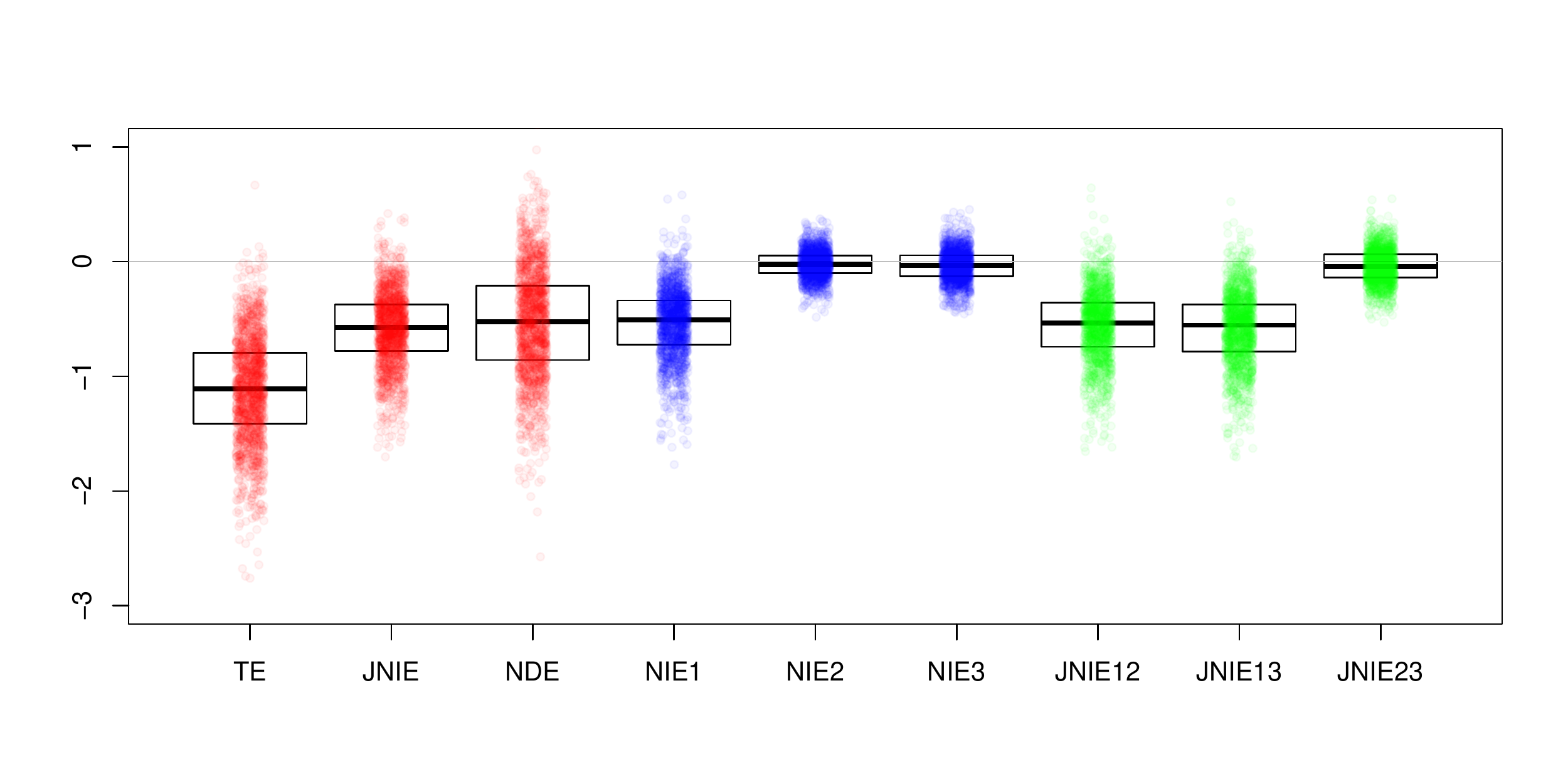}
  \includegraphics[scale=0.5]{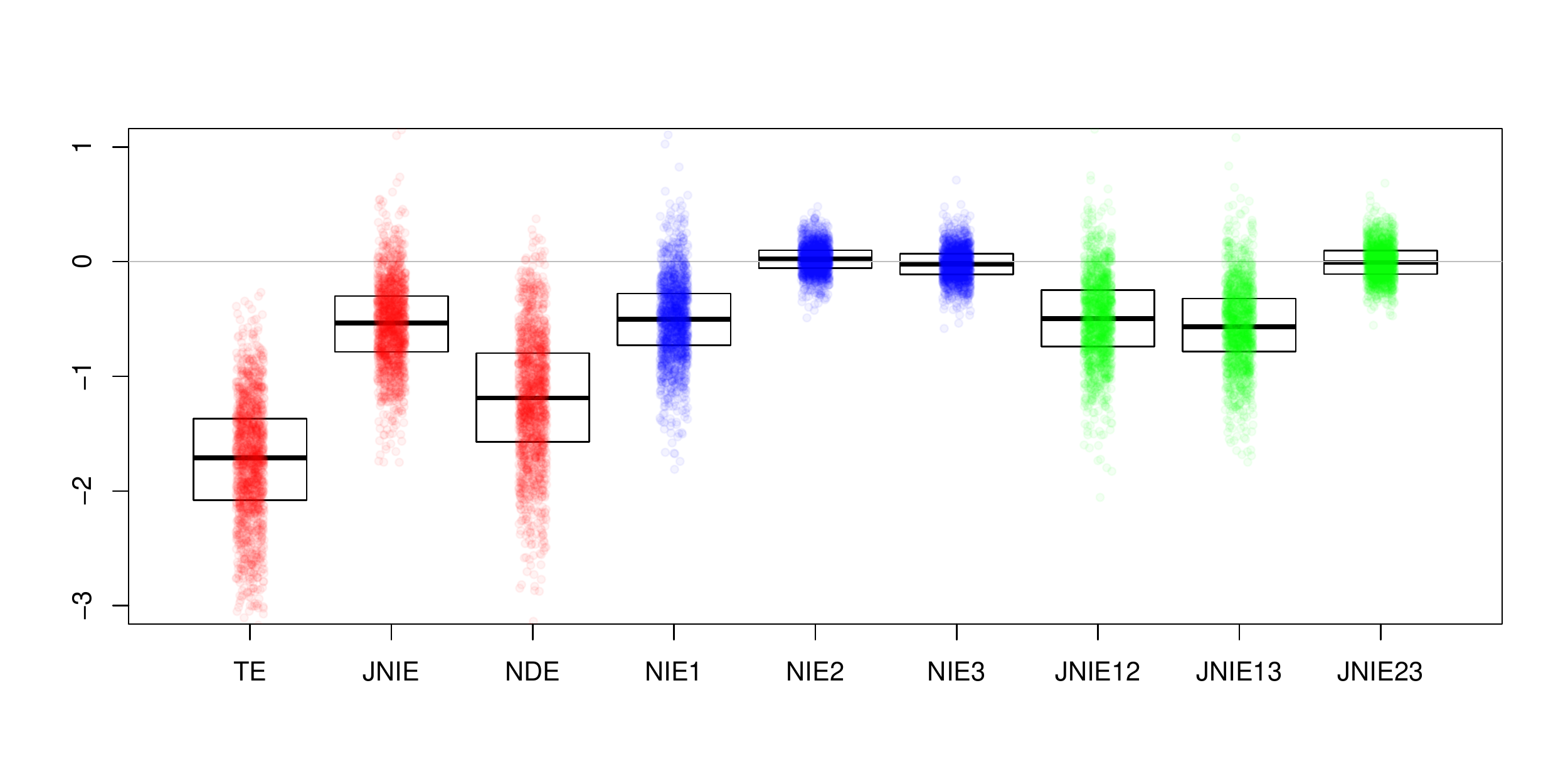}
\caption{Posterior distributions (means and SDs) of the mediation effects in the analysis of \SOTwo\, controls for the main analysis with the optional assumption in Section 5.2.3 (Top) and for the analysis with uniform priors over the correlations (Bottom).}
\label{fig7}
\end{figure}

\section*{I. Model Fit}
\begin{figure}[pt]
\centering
  \includegraphics[scale=0.6]{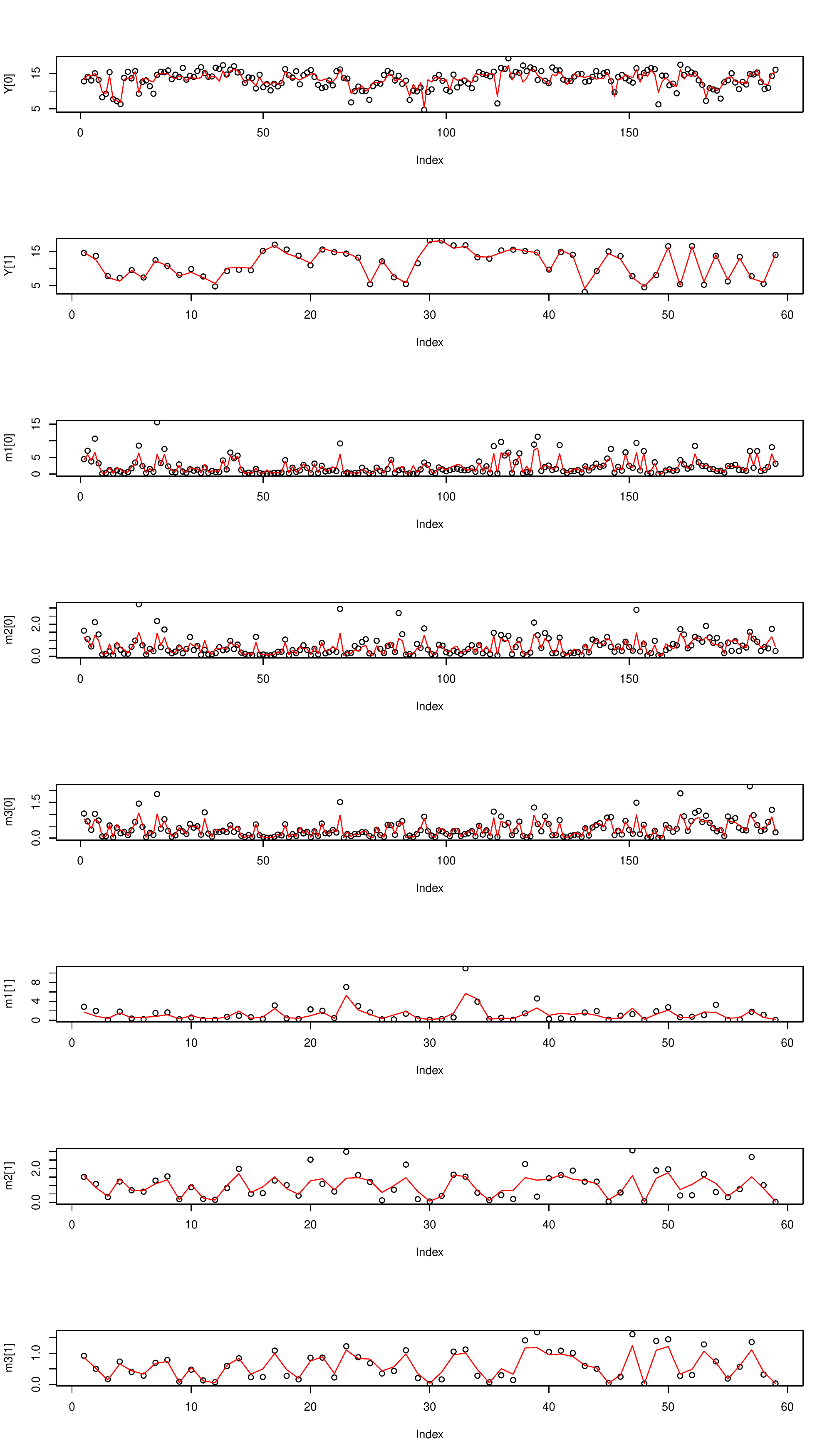}
\caption{Posterior predictive means (redlines) of emissions and ambient \PMTwo\, under each intervention $z=0,1$ and the observed data points (black dots) for each case.}
\label{predictive}
\end{figure}

\begin{figure}[pt]
\centering
  \includegraphics[scale=0.6]{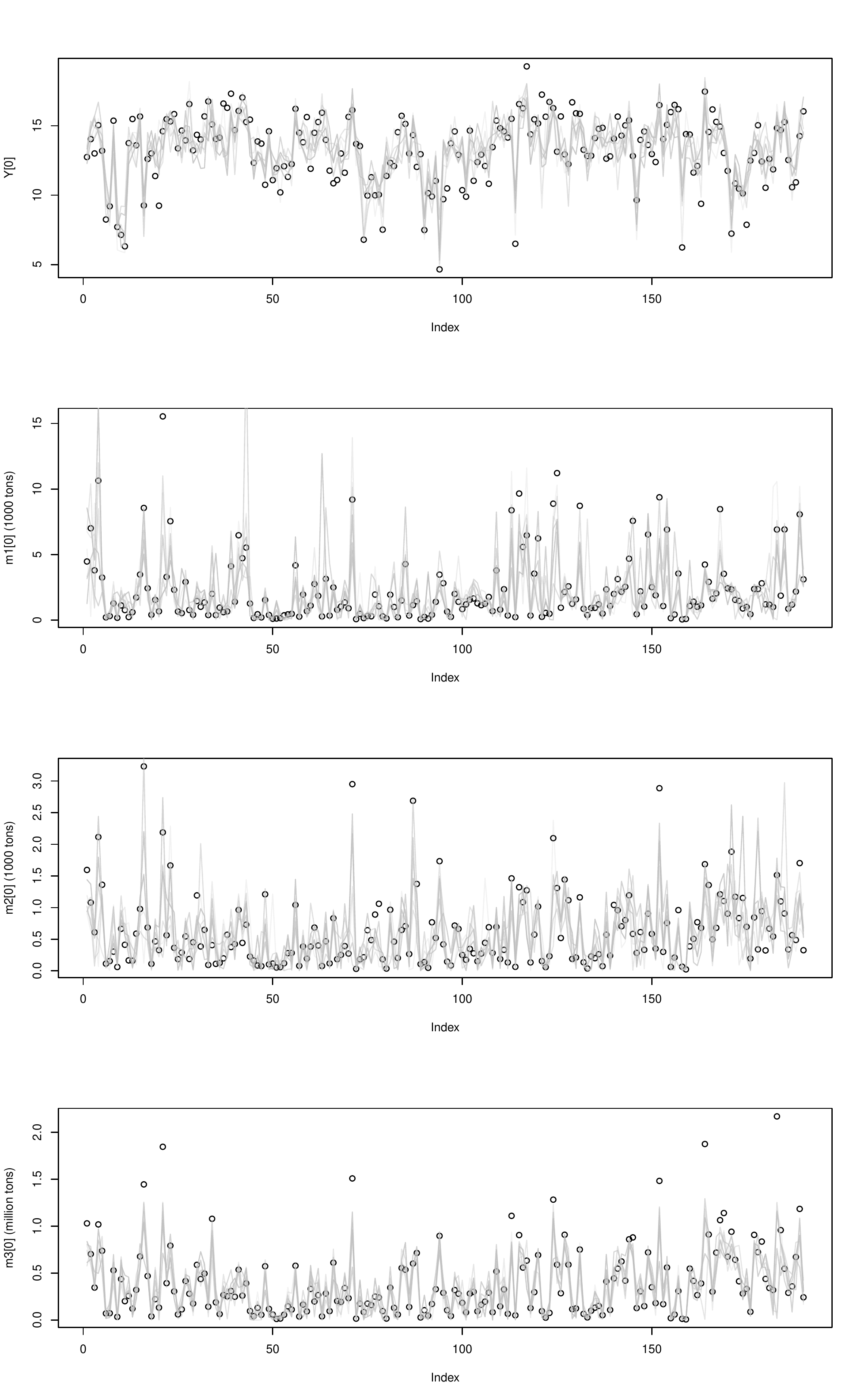}
\caption{Posterior predictive replications (7 gray lines) of emissions and ambient \PMTwo\, under $z=0$ and the observed data points (black dots) for each case.}
\label{predictive1}
\end{figure}

\begin{figure}[pt]
\centering
  \includegraphics[scale=0.6]{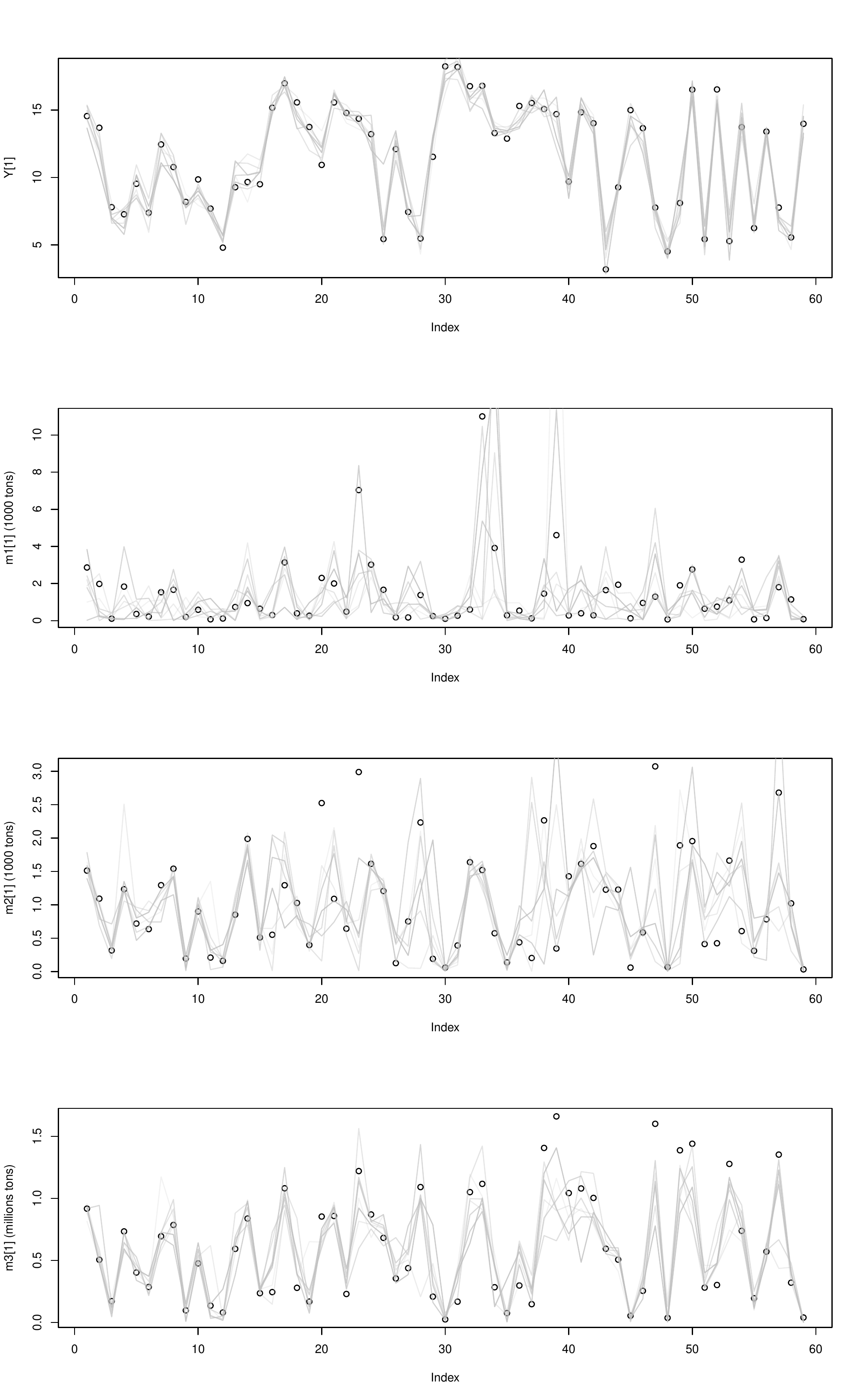}
\caption{Posterior predictive replications (7 gray lines)  of emissions and ambient \PMTwo\, under $z=1$ and the observed data points (black dots) for each case.}
\label{predictive2}
\end{figure}

We compared the fit of parametric and nonparametric models for the power plant data. Particularly, we assessed relative fit using a variation of DICs \citep{celeux2006deviance} for mixture models since our nonparametric model can be specified as a mixture model using the stick-breaking construction \citep{Seth:1994}
\[DIC_3 = -4E_\theta[\log f(\boldsymbol{T}|\theta)| \boldsymbol{T}]+
2\log \hat{f}(\boldsymbol{T}),\]
where $\hat{f}(\boldsymbol{T}) = \prod_{i=1}^n
\hat{f}(T_{i} )$ 
with the MCMC predictive density which is defined as \[\hat{f}(T_{i}) = \frac{1}{m}\sum_{l=1}^m \sum_{k=1}^K \omega_k^{(l)}
N(T_{i};\boldsymbol{\mu}_k^{(l)},\boldsymbol{\Sigma}_k^{(l)});\]
$m$ denotes the number of MCMC
samples and $(\boldsymbol{\mu}_k^{(l)},
\boldsymbol{\Sigma}_k^{(l)}, \omega_k^{(l)})_{1 \leq k \leq K}$ are the
values at iteration $l$. An `equivalent' parametric model for a marginal distribution was specified as a linear regression model with the same set of covariates and non-informative priors. Table \ref{table:dic} shows that the nonparametric model has lower DIC's for all marginal distributions, which supports the more complex model for the power plant data.
\begin{table}
\centering
\caption{DIC's for the parametric and nonparametric models based on the Power Plant data. Lower is
better.}\label{table:dic}
  \begin{tabular}{ c || c | c }
    \hline
Observed Data & Parametric model & Nonparametric model \\
\hline\hline
    $M_1(1)$ & 31.3 & -144.7 \\ 
    $M_2(1)$ & 158.7 & 82.5 \\ 
    $M_3(1)$ & -257.4 & -285.7\\ \hline
    $M_1(0)$ & -144.7 & -462.8 \\
    $M_2(0)$ & 181.2 & -22.5 \\
    $M_3(0)$ & -877.7 & -1236.9 \\
    \hline
  \end{tabular}
\end{table}
We also assessed the posterior predictive means and replications of emissions and ambient \PMTwo\, simulated under the model conditional on the observed covariates. Figure \ref{predictive} illustrates that the observed data points (black circle) and the corresponding posterior predictive means (red lines). Figure \ref{predictive1} \& \ref{predictive2} illustrate 4 posterior predictive replications for each case. They all suggest reasonable fit of the nonparametric model.

\section*{J. Sensitivity Analysis}

\subsection*{J.1 Sensitivity Analysis for Assumption 3}\label{sec:assumption1}
To assess sensitivity of
Assumption 3, we divide the assumption into 2 sub-assumptions (3.a and
3.b) with sensitivity parameters $\epsilon$ and $\chi$. 
Let 
$S_k$ be the variance of difference between the $k^{th}$ mediator under
opposite interventions, $[M_k(1)-M_k(0)|\boldsymbol{X}]$, for
$k=1,2,3$ which can be estimated by Assumption 2 and the observed data.  Let random
variable $D_k$ be the Mahalanobis distance, $D_k = \sqrt{[M_k(1)-M_k(0)|\boldsymbol{X}]^2/S_k}$,
which quantifies the treatment effect on the $k^{th}$ mediator. For
notational simplicity, we suppress notation for covariates $\boldsymbol{X}$ in the following conditional distributions.

\bigskip

 \noindent{\bf Assumption 3.a } {\it For a fixed $\epsilon$, the following three equalities hold
\begin{eqnarray*}
\lefteqn{f_{1,\bM(0,1,1)}(y|\mathbf{M}(0,1,1)=\mathbf{m}, \mathbf{M}(1,0,0), D_1< \epsilon)}\\
 &=&f_{1,\bM(1,1,1)}(y|\mathbf{M}(1,1,1)=\mathbf{m}, \mathbf{M}(0,0,0), D_1<\epsilon), \\
\lefteqn{f_{1,\bM(1,0,1)}(y|\mathbf{M}(1,0,1)=\mathbf{m}, \mathbf{M}(0,1,0), D_2< \epsilon)}\\
 &=&f_{1,\bM(1,1,1)}(y|\mathbf{M}(1,1,1)=\mathbf{m}, \mathbf{M}(0,0,0), D_2<\epsilon), \\
\lefteqn{f_{1,\bM(1,1,0)}(y|\mathbf{M}(1,1,0)=\mathbf{m}, \mathbf{M}(0,0,1), D_3< \epsilon)}\\
 &=&f_{1,\bM(1,1,1)}(y|\mathbf{M}(1,1,1)=\mathbf{m}, \mathbf{M}(0,0,0), D_3<\epsilon), 
\end{eqnarray*}
where a vector $\mathbf{m}$ is a vector of realized values of three mediators, $\bM(z_1,z_2,z_3)\equiv \{M_1(z_1), M_2(z_2), M_3(z_3)\}$.}

The idea behind this assumption is among subject for whom the treatment
effect on the $k^{th}$ mediator ($D_k$) is small (as quantified by
$\epsilon$), the distribution of the outcome is the same whether the
mediator value was induced by $Z = 1$ or $Z = 0$. Here, $\epsilon$
corresponds to the size of the change in terms of the number of
standard deviations. We set a single sensitivity parameter $\epsilon$
instead of setting $\epsilon_k$ for $k=1,2,3$ since differences are already standardized by $S_k$. 

For conditional distributions of potential outcomes with any two mediators under different interventions (e.g., $Y(1; \bM(1,0,0)), Y(1; \bM(0,1,0)), Y(1; \bM(0,0,1))$), we assume similar equalities hold. To be consistent with $\epsilon$ which measures the size of the change in terms of the number of standard deviations, the treatment effect on the $j^{th}$ and $k^{th}$ mediators ($D_{jk}$) is quantified by \begin{eqnarray}\sqrt{\left( \begin{array}{cc}
\epsilon & \epsilon \end{array} \right)\left( \begin{array}{cc}
1 & r_{jk}  \\
r_{jk} & 1 \end{array} \right)\left( \begin{array}{ccc}
\epsilon \\
\epsilon \end{array} \right)}=\sqrt{\epsilon^2(2+2 r_{jk})},\label{epsilon22}\end{eqnarray} where the correlation, $r_{jk}$, is 
\begin{eqnarray*}
r_{jk}&  = & \text{Cor}(M_j(1)-M_j(0), M_k(1)-M_k(0)) \\
& = & \frac{\text{Cov}(M_j(1),
  M_k(1))+\text{Cov}(M_j(0), M_k(0)) -
  \text{Cov}(M_j(0), M_k(1)) - \text{Cov}(M_j(1),
  M_k(0))}{\sqrt{\text{Var}(M_j(1)-M_j(0))
    \times \text{Var}(M_k(1)-M_k(0))}}.
\end{eqnarray*}
 Here, the correlations $r_{jk}$'s are estimable based on Assumption 2. Then, for the potential outcomes $Y(1; \bM(1,0,0))$, $Y(1; \bM(0,1,0))$, and $Y(1; \bM(0,0,1))$,  we assume the following equalities,
\begin{eqnarray*}
\lefteqn{f_{1,\bM(1,0,0)}\Big(y|\mathbf{M}(1,0,0)=\mathbf{m}, \mathbf{M}(0,1,1), D_{23}< \sqrt{\epsilon^2(2+2 r_{23})} \Big)}\\
& = & f_{1,\bM(1,1,1)}\Big(y|\mathbf{M}(1,1,1)=\mathbf{m}, \mathbf{M}(0,0,0), D_{23}<\sqrt{\epsilon^2(2+2 r_{23})}\Big)\\
\lefteqn{f_{1,\bM(0,1,0)}\Big(y|\mathbf{M}(0,1,0)=\mathbf{m}, \mathbf{M}(1,0,1), D_{13}< \sqrt{\epsilon^2(2+2 r_{13})}\Big)}\\
& = & f_{1,\bM(1,1,1)}\Big(y|\mathbf{M}(1,1,1)=\mathbf{m}, \mathbf{M}(0,0,0), D_{13}<\sqrt{\epsilon^2(2+2 r_{13})}\Big)\\
\lefteqn{f_{1,\bM(0,0,1)}\Big(y|\mathbf{M}(0,0,1)=\mathbf{m}, \mathbf{M}(1,1,0), D_{12}< \sqrt{\epsilon^2(2+2 r_{12})}\Big)}\\
& = & f_{1,\bM(1,1,1)}\Big(y|\mathbf{M}(1,1,1)=\mathbf{m}, \mathbf{M}(0,0,0), D_{12}<\sqrt{\epsilon^2(2+2 r_{12})}\Big)
\end{eqnarray*}
where $D_{jk} = \sqrt{\bW_{jk}^{T}S_{jk}^{-1}\bW_{jk}}$ and
$\bW_{jk}^T = [M_j(1)-M_j(0), M_k(1)-M_k(0)|\boldsymbol{X}]$ and $S_{jk}$ is the
covariance matrix of $\bW_{jk}$. Then, $D_{jk}$ is an average standardized treatment effect on
the $j^{th}$ and $k^{th}$ mediators. In this way, we quantify (approximately) the
treatment effect on the $j^{th}$ and $k^{th}$ mediators ($D_{jk}$) by the
size of the change in terms of the number of standard deviations in
(\ref{epsilon22}).  

In the same manner, for the conditional distribution of $Y(1; \bM(0,0,0))$,
\begin{eqnarray*}
\lefteqn{f_{1,\bM(0,0,0)}\Big(y|\mathbf{M}(0,0,0)=\mathbf{m}, \mathbf{M}(1,1,1), D_{123}< \sqrt{\epsilon^2 (3+2r_{12}+2r_{13} + 2r_{23}) }\,\,\Big) }\\
&=&f_{1,\bM(1,1,1)}\Big(y|\mathbf{M}(1,1,1)=\mathbf{m}, \mathbf{M}(0,0,0), D_{123}<\sqrt{\epsilon^2 (3+2r_{12}+2r_{13} + 2r_{23}) }\,\,\Big), 
\end{eqnarray*}
where $D_{123} = \sqrt{\bW_{123}^{T}S_{123}^{-1}\bW_{123}}$ and
$\bW_{123}^T = [M_1(1)-M_1(0), M_2(1)-M_2(0), M_3(1)-M_3(0)|\boldsymbol{X}]$ and
$S_{123}$ is the covariance matrix of $\bW_{123}$.  

\bigskip

\noindent {\bf Assumption 3.b } {\it The second part of the assumption is for the subgroup of subjects for whom the intervention has a greater than $\epsilon$ effect on the $k$-th mediator in terms of $D_k$.  For potential outcomes with one mediator under different intervention (e.g., $Y(1; \bM(0,1,1)), Y(1; \bM(1,0,1)), Y(1; \bM(1,1,0))$), let $k$ indicate which element of $\{z_1, z_2, z_3\}$ is set to $0$ (e.g., if $z_1=0, z_2=1, z_3=1$, then $k=1$).  Then, for a fixed $\epsilon$ and $\chi_k$ for $k=1,2,3$, we assume
\begin{eqnarray}
\lefteqn{f_{1,\bM(0,1,1)}\Big(y|\mathbf{M}(0,1,1)=\mathbf{m},
  \mathbf{M}(0,1,1), D_1 \geq \epsilon\Big)}\nonumber\\
 & = &e^{ \tilde{y} (\log
   (\chi_1))}f_{1,\bM(1,1,1)}\Big(y|\mathbf{M}(1,1,1)=\mathbf{m},
 \mathbf{M}(0,0,0), D_1\geq\epsilon\Big), \label{epsilon11}\\
\lefteqn{f_{1,\bM(1,0,1)}\Big(y|\mathbf{M}(1,0,1)=\mathbf{m},
  \mathbf{M}(1,0,1), D_2 \geq \epsilon\Big)}\nonumber\\
 & = &e^{ \tilde{y} (\log
   (\chi_2))}f_{1,\bM(1,1,1)}\Big(y|\mathbf{M}(1,1,1)=\mathbf{m},
 \mathbf{M}(0,0,0), D_2\geq\epsilon\Big), \nonumber\\
 \lefteqn{f_{1,\bM(1,1,0)}\Big(y|\mathbf{M}(1,1,0)=\mathbf{m},
  \mathbf{M}(1,1,0), D_3 \geq \epsilon\Big)}\nonumber\\
 & = &e^{ \tilde{y} (\log
   (\chi_3))}f_{1,\bM(1,1,1)}\Big(y|\mathbf{M}(1,1,1)=\mathbf{m},
 \mathbf{M}(0,0,0), D_3\geq\epsilon\Big), \nonumber
\end{eqnarray}
where $\tilde{y}=(y-\bar{y}_{1,obs})/\text{s.d.}(y_{1,obs})$, the standardized outcome with respect to the observed outcome under intervention $Z=1$ (e.g., $y_{1,obs}$).}  

This assumption states that among subjects for whom intervention
$Z=1$ has a greater than $\epsilon$ (number of standard deviation)
effect on the mediator (measured via $D_k$),  the conditional
distribution of the outcome under intervention $Z=1$ with the
$k^{th}$ mediator set to its value under the opposite intervention ($Z=0$) is equal to the
conditional distribution of the outcome under intervention $Z=1$
with all mediators set to their values under intervention $Z=1$ through an exponential tilt
relationship. Note that the standardized outcome is used in the exponential tilt for numerical stability especially when the observed outcomes have large values.

For potential outcomes with any two mediators under different interventions (e.g., $Y(1; \bM(0,0,1))$, $Y(1; \bM(0,1,0))$, $Y(1; \bM(1,0,0))$), we assume
\begin{eqnarray*}
\lefteqn{f_{1,\bM(0,0, 1)}\Big(y|\mathbf{M}(0,0,1)=\mathbf{m},
  \mathbf{M}(1,1,0), D_{12}\geq
  \sqrt{\epsilon^2(2+2 r_{12})}\,\,\Big)}\\ & = &e^{\tilde{y}(\log (\chi_1)+\log(\chi_2))} f_{1,\bM(1,1,1)}\Big(y|\mathbf{M}(1,1,1)=\mathbf{m}, \mathbf{M}(0,0,0), D_{12}\geq\sqrt{\epsilon^2(2+2 r_{12}}\,\,\Big), \\
  \lefteqn{f_{1,\bM(0,1, 0)}\Big(y|\mathbf{M}(0,1,0)=\mathbf{m},
  \mathbf{M}(1,0,1), D_{13}\geq
  \sqrt{\epsilon^2(2+2 r_{13})}\,\,\Big)}\\ & = &e^{\tilde{y}(\log (\chi_1)+\log(\chi_3))} f_{1,\bM(1,1,1)}\Big(y|\mathbf{M}(1,1,1)=\mathbf{m}, \mathbf{M}(0,0,0), D_{13}\geq\sqrt{\epsilon^2(2+2 r_{13})}\,\,\Big), \\
\lefteqn{f_{1,\bM(1,0, 0)}\Big(y|\mathbf{M}(1,0,0)=\mathbf{m},
  \mathbf{M}(1,0,0), D_{23}\geq
  \sqrt{\epsilon^2(2+2 r_{23})}\,\,\Big)}\\ & = &e^{\tilde{y}(\log (\chi_2)+\log(\chi_3))} f_{1,\bM(1,1,1)}\Big(y|\mathbf{M}(1,1,1)=\mathbf{m}, \mathbf{M}(0,0,0), D_{23}\geq\sqrt{\epsilon^2(2+2 r_{23})}\,\,\Big), 
\end{eqnarray*}
 where $D_{jk} =
 \sqrt{\bW_{jk}^{T}S_{jk}^{-1}\bW_{jk}}$ with $\bW_{jk}^T =
 [M_j(1)-M_j(0), M_k(1)-M_k(0)|\boldsymbol{X}]$ and the covariance matrix $S_{jk}$ of
 $\bW_{jk}$. The difference is quantified by the same quantity as in (\ref{epsilon22}). 

Here, we implicitly assume additivity of sensitivity parameter
$\chi$'s on a log scale, which in turn implies multiplicativity of
sensitivity parameter $\chi$'s. The idea
behind this assumption is two
conditional distributions are proportional to each other by $\chi_j
\times \chi_k$ which
is less than $\text{min}(\chi_j, \chi_k)$ if $0<\chi_j, \chi_k <1$ or
larger than $\text{max}(\chi_j, \chi_k)$ if $\chi_j, \chi_k > 1$ while two conditional distributions in the case
(\ref{epsilon11}) are proportional to each other by $\chi_k$. Thus, under this setting, a conditional distribution of {\it a priori} counterfactual is more deviated from the conditional distribution of the observed outcome as more mediators are set to values under the other intervention (e.g., $Z=0$). 

Similarly, for the conditional distribution of $Y(1; \bM(0,0,0))$ where all mediators are under intervention $Z=0$, we assume
\begin{eqnarray*}
\lefteqn{f_{1,\bM(0,0,0)}\Big(y|\mathbf{M}(0,0,0)=\mathbf{m},
  \mathbf{M}(1,1,1), D_{123}\geq
  \sqrt{\epsilon^2 (3+2r_{12}+2r_{13} + 2r_{23}) }\,\,\Big)}\\
 & = &e^{\tilde{y} (\sum_k \log (\chi_k))} f_{1,\bM(1,1,1)}\Big(y|\mathbf{M}(1,1,1)=\mathbf{m}, \mathbf{M}(0,0,0), D_{123}\geq\sqrt{\epsilon^2 (3+2r_{12}+2r_{13} + 2r_{23}) }\,\,\Big), 
\end{eqnarray*}
where $D_{123} = \sqrt{\bW_{123}^{T}S_{123}^{-1}\bW_{123}}$ with
$\bW_{123}^T = [M_1(1)-M_1(0), M_2(1)-M_2(0), M_3(1)-M_3(0)|\boldsymbol{X}]$ and
$S_{123}$ is the covariance matrix of $\bW_{123}$. 

Assumption 3.a and 3.b together differentiate the population into those for which the intervention has a large effect on the mediators versus those for which the intervention has a small effect on the mediators. 
Based on this framework, we can assess sensitivity of inferences to
Assumption 4. Note that two assumptions correspond to the Assumption 4
when we set $\epsilon=\infty$ or $\chi_k=1$ for all $k=1,2,3$. 

\subsection*{J.2. Sensitivity Parameters}\label{sec:sensitivity}

Assumption 3.a and 3.b contain sensitivity parameters, $(\epsilon,
\chi_k)$. In this section, we discuss strategies for eliciting values of each sensitivity parameter.

For $\epsilon$, we can consider this as the size of the change in terms of the number of standard deviation. A default approach can be $\epsilon \in [0.5, 2]$, from half to 2 standard deviations.

To better understand plausible ranges of $\chi_k$ for $k=1,2,3$ in Assumption 3.b, for the
subgroup of subjects for whom the intervention has a greater than
$\epsilon$ effect on the first mediator ($M_1$) in terms of $D_{1}$, we
assume the negative effect of the intervention through the first
mediator 
(i.e., the negative $\text{NIE}_1$), then the following equality is likely to hold for
some value of the outcome such as $y^\star\equiv\bar{y}_{obs,1}+sd(\bar{y}_{obs,1})$:
\begin{eqnarray}
\lefteqn{f_{1,\bM(1,1,1)}(y^\star|\mathbf{M}(1,1,1)=\mathbf{m},
  \mathbf{M}(0,0,0), D_{1}\geq \epsilon\,\,)} \nonumber\\
 & \leq &f_{1,\bM(0,1,1)}(y^\star|\mathbf{M}(0,1,1)=\mathbf{m},
  \mathbf{M}(1,0,0), D_{1}\geq \epsilon\,\,).  \label{sens11}
\end{eqnarray}
where the RHS in (\ref{sens11}) is equal to
\begin{equation}
e^{ \log (\chi_1)}f_{1,\bM(1,1,1)}(y^\star|\mathbf{M}(1,1,1)=\mathbf{m},
  \mathbf{M}(0,0,0), D_{1}\geq \epsilon  \,\,) \label{sens22}
\end{equation}
by Assumption 3.b. Thus, we equate (\ref{sens11}) to (\ref{sens22}) and
have the following equality
\begin{eqnarray}
1 & \leq & e^{
  \log(\chi_1)} \, = \, \chi_1, \label{sens33}
\end{eqnarray}
which gives a possible range of $\chi_1$ provided that we have a prior
expectation about the negative $\text{NIE}_1$. Analogously, we can elicit
possible ranges of $\chi_2$ and $\chi_3$ assuming negative
mediator specific effects for $M_2$ and $M_3$, 
\begin{equation}
1 \leq
\chi_2 , \qquad 1 \leq \chi_3. \label{sens335}\end{equation}

Additionally, if we assume the negative direct effect of the
intervention, then the following equality is likely to hold for some value of
the outcome such as $y^\star\equiv\bar{y}_{obs,1}+sd(\bar{y}_{obs,1})$:
\begin{eqnarray}
\lefteqn{f_{0,\bM(0,0,0)}(y^\star|\mathbf{M}(0,0,0)=\mathbf{m},
  \mathbf{M}(1,1,1), D_{123}\geq \epsilon\,\,)} \nonumber\\
 & \geq &f_{1,\bM(0,0,0)}(y^\star|\mathbf{M}(0,0,0)=\mathbf{m},
  \mathbf{M}(1,1,1), D_{1}\geq \epsilon\,\,),  \label{sens44}
\end{eqnarray}
where the RHS in (\ref{sens44}) is equal to 
\begin{equation}
e^{(\log(\chi_1)+\log(\chi_2)+\log(\chi_3))}f_{1,\bM(1,1,1)}(y^\star|\mathbf{M}(1,1,1)=\mathbf{m},
  \mathbf{M}(0,0,0), D_{123}\geq \epsilon  \,\,) \label{sens55}
\end{equation}
by Assumption 3.b. Then, we equate (\ref{sens44}) to (\ref{sens55}) to
have the following equality
\begin{eqnarray}
\frac{f_{0,\bM(0,0,0)}(y^\star|\mathbf{M}(0,0,0)=\mathbf{m},
  \mathbf{M}(1,1,1), D_{123}\geq \epsilon\,\,)}{f_{1,\bM(1,1,1)}(y^\star|\mathbf{M}(1,1,1)=\mathbf{m},
  \mathbf{M}(0,0,0), D_{123}\geq \epsilon  \,\,)} 
 & \geq & e^{
  (\log(\chi_1)+\log(\chi_2)+\log(\chi_3))} \nonumber\\
 & = & \chi_1 \times
 \chi_2 \times \chi_3, \label{sens66}
\end{eqnarray}
where the LHS in (\ref{sens66}) acts as an upper bound which is likely to be larger than 1
under assuming the negative causal effect of the treatment. Thus, we have 
\[ 1 <\chi_1 \times \chi_2 \times \chi_3 < \text{some upper bound specified in }(\ref{sens66}).\]
With the inequalities in (\ref{sens335}) we can set possible ranges for
$\chi_k$ for $k=1,2,3$.

\begin{table}[ht]
\centering
\resizebox{\textwidth}{!}{  
\begin{tabular}{c|c|cccccc}
 & $D_1$ & $D_2$ & $D_3$ & $D_{12}$ & $D_{23}$ & $D_{13}$ & $D_{123}$ 
\\ \hline
Mean (S.D.) & 0.96 (0.06) & 0.78 (0.02) & 0.77 (0.03) & 0.98 (0.04) & 0.87 (0.02) & 0.97 (0.04) & 0.98 (0.03)\\
\hline
\end{tabular}}
\caption{Posterior means (and standard deviations) of the treatment effects on the $D$'s.}\label{tab:differences}
\end{table}

\begin{figure}[h]
\centering
\scalebox{0.075}
{\includegraphics{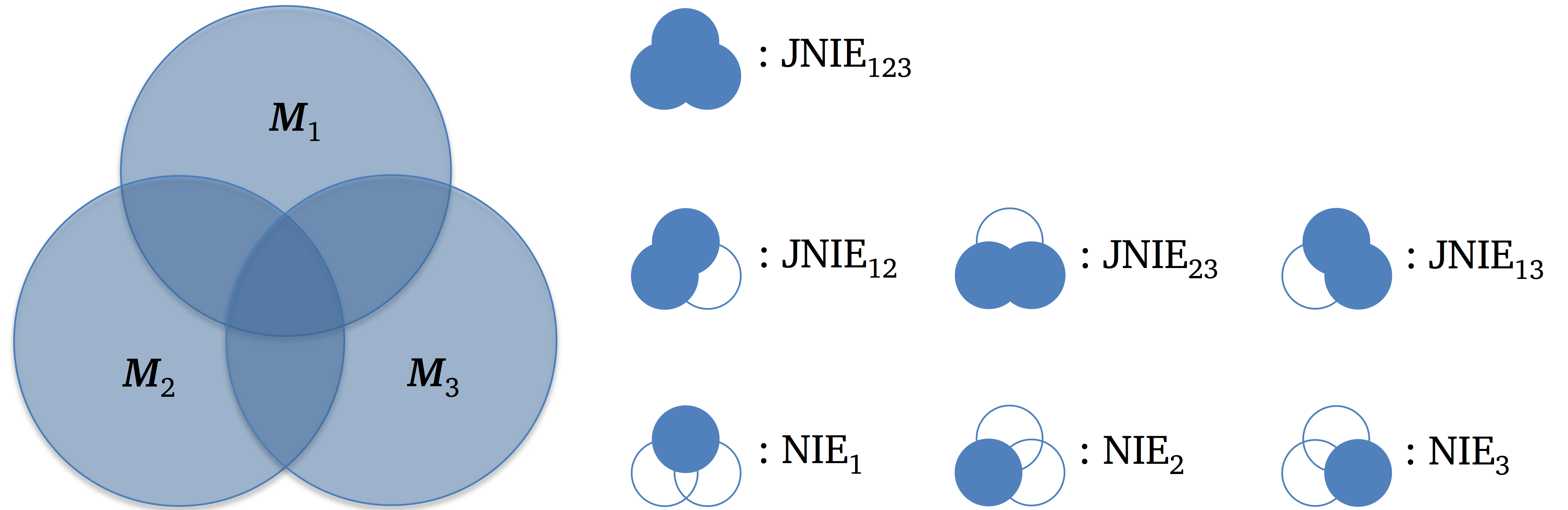}} 
\caption{Graphical representation of the decomposition of the $\text{JNIE}_{123}$ into mediator-specific NIEs
and the joint effects of all possible pairs of the 
candidate mediators for the case $K=3$.}
\label{fig1}
\end{figure}

\end{document}